\documentclass[%
 reprint,
 superscriptaddress,
 amsmath,amssymb,
 aps,
]{revtex4-2}
\usepackage{graphicx}% Include figure files
\usepackage{dcolumn}% Align table columns on decimal point
\usepackage{bm}% bold math
\usepackage{subcaption}

\usepackage[utf8]{inputenc}
\usepackage{mwe}

\usepackage{xcolor}
\usepackage{mathtools}

\usepackage{physics}
\usepackage[normalem]{ulem}
\usepackage{lineno}
\usepackage{soul} % Allows \st strikethrough
\usepackage{float} % Make figures place with H

\newcommand{\rev}[1]{{\color{black} #1}}  %colors revisions as red, switch to black for clean copy ** but watch out for \sout

\usepackage{hyperref} 
\hypersetup{colorlinks=true, citecolor=blue}

\newcommand{\RsXs}{%
  \mathrel{%
    \vcenter{\offinterlineskip
      \ialign{##\cr$R_s$\cr\noalign{\kern+1.5pt}$X_s$\cr}%
    }%
  }%
}

\begin{document}

\preprint{APS/123-QED}

%\title{Electrodynamic properties and low energy excitations of UTe$_2$}
%\title{Isotropic power-law for penetration depth in UTe$_2$ weaker than point nodes}
%\title{Weaker than point-nodal, isotropic sub-gap excitations in UTe$_2$}
%\title{Revealing isotropic sub-gap excitations in UTe$_2$ through complex microwave surface impedance}
\title{Revealing isotropic abundant low-energy excitations in UTe$_2$ through complex microwave surface impedance}

\author{Arthur Carlton-Jones}
\affiliation{Maryland Quantum Materials Center, Physics Department,\\
University of Maryland, College Park, MD 20742-4111  USA}
\author{Alonso Suarez}
\affiliation{Maryland Quantum Materials Center, Physics Department,\\
University of Maryland, College Park, MD 20742-4111  USA}
\author{Yun-Suk Eo}
\affiliation{Texas Tech University, Department of Physics \& Astronomy,\\
1200 Memorial Circle, MS 41051, Lubbock, TX 79409-1051 USA}
\author{Ian M. Hayes}
\affiliation{Maryland Quantum Materials Center, Physics Department,\\
University of Maryland, College Park, MD 20742-4111  USA}
\author{Shanta R. Saha}
\affiliation{Maryland Quantum Materials Center, Physics Department,\\
University of Maryland, College Park, MD 20742-4111  USA}
\author{Johnpierre Paglione}
\affiliation{Maryland Quantum Materials Center, Physics Department,\\
University of Maryland, College Park, MD 20742-4111  USA}
\affiliation{Canadian Institute for Advanced Research, Toronto, Ontario M5G 1Z8, Canada}
\author{Nicholas P. Butch}
\affiliation{Maryland Quantum Materials Center, Physics Department,\\
University of Maryland, College Park, MD 20742-4111  USA}
\affiliation{NIST Center for Neutron Research, National Institute of Standards and Technology,\\
Gaithersburg, Maryland 20899, USA}
\author{Steven M. Anlage}
\affiliation{Maryland Quantum Materials Center, Physics Department,\\
University of Maryland, College Park, MD 20742-4111  USA}
\date{February, 2025}

\begin{abstract}
The complex surface impedance is a well-established tool to study the super- and normal-fluid responses of superconductors. Fundamental properties of the superconductor, such as the pairing mechanism, Fermi surface, and topological properties, also influence the surface impedance. We explore the microwave surface impedance of spin-triplet UTe$_2$ single crystals as a function of temperature using resonant cavity perturbation measurements employing a novel multi-modal analysis to gain insight into these properties. We determine a composite surface impedance of the crystal for each mode using resonance data combined with the independently measured normal state dc resistivity tensor. The normal state surface impedance reveals the weighting of current flow directions in the crystal of each resonant mode. For UTe$_2$, we find an isotropic $\Delta \lambda(T) \sim T^\alpha$ power-law temperature dependence for the magnetic penetration depth for $T\le T_c/3$ with $\alpha < 2$, which is inconsistent with a single pair of point nodes on the Fermi surface under weak scattering. We also find a similar power-law temperature dependence for the low-temperature surface resistance $R_s(T) \sim T^{\alpha_R}$ with $\alpha_R < 2$. We observe a strong anisotropy of the residual microwave loss across these modes, with some modes showing loss below the universal line-nodal value, to those showing substantially more. We compare to predictions for topological Weyl superconductivity in the context of the observed isotropic power-laws, and anisotropy of the residual loss.

\end{abstract}

\maketitle

\section{Introduction}
%\paragraph{Introduction}
UTe$_2$ is a remarkable heavy Fermion superconductor near the edge of ferromagnetism, which shows evidence of spin-triplet pairing \cite{Ran19}, including extraordinarily high critical fields with multiple superconducting phases, as well as re-entrant and orphan superconducting phases \cite{Knebel19,Ran19_2}. UTe$_2$ has an orthorhombic crystal structure with D$_{2h}$ point group symmetry. Early generation samples (CVT1, T$_c$ $\sim$ 1.6 K), grown with a chemical vapor transport (CVT) method \cite{Ran21_CVT}, show evidence of time reversal invariance (TRI) breaking in the superconducting state.  Two transitions in specific heat and a non-zero polar Kerr effect which onsets near T$_c$ have been observed \cite{Hayes21}. Electrodynamic properties of CVT1 samples are consistent with point nodal behavior of the superconducting gap \cite{Metz19,Seokjin21,Ishihara23}. Similarly, thermodynamic properties are also consistent with nodal behavior in UTe$_2$ \cite{Ran19,Hayes21,Hayes24}.

The CVT growth method has been improved to produce samples with higher transition temperatures (CVT2, $T_c \lesssim 2$ K). These samples show only one superconducting transition \cite{Rosa22}. There is now a generation of molten salt flux (MSF) grown samples with high RRR $\sim100$ (MSF1, $T_c\approx 2.1$ K), which likewise, shows only one superconducting transition \cite{Sakai22,Aoki24}. Polar Kerr measurements have also been performed on single transition samples grown with both CVT2 and MSF1 methods, which do not show evidence of TRI breaking in the superconducting state \cite{Ajeesh23}. Muon spin relaxation ($\mu$SR) measurements also show no evidence of TRI breaking for MSF grown samples \cite{Azari23}. For these CVT2 and MSF1 crystals, $T_c$ increases with RRR, but saturates for RRR $\gtrsim 50$ \cite{Rosa22,Sakai22,Aoki24}. If UTe$_2$ is a simple p-wave superconductor, then ordinary disorder should suppress T$_c$; hence, higher-T$_c$ crystals should imply higher quality (less disordered) materials. In actuality, this is not the case as samples of the same $T_c$ can have a variety of RRR values \cite{Rosa22,Sakai22,Aoki24}.

Numerous studies have been performed on the CVT2 and MSF1 samples to determine the specific single component order parameter of UTe$_2$; however, there is no agreement. Thermal conductivity has been measured vs.~temperature and applied magnetic field for both CVT and MSF grown UTe$_2$ samples \cite{Hayes24,Suetsugu24}. Hayes \textit{et. al.} find that the extrapolated zero temperature thermal conductivity to temperature ratio shows a rapid increase with magnetic field for very low fields, which is indicative of nodal superconductivity. They conclude that UTe$_2$ is specifically a $B_{1u}$ (nodes along the $k_z$ direction) or $B_{2u}$ (nodes along the $k_y$ direction) pairing symmetry. Suetsugu \textit{et. al.}, however, do not observe this behavior when performing the same measurements, concluding an $A_u$ (fully gapped) pairing symmetry. A $B_{1u}$ pairing symmetry has also been concluded from measurements of the Josephson coupling between UTe$_2$ and an s-wave superconductor \cite{Li23}, and a $B_{2u}$ pairing symmetry has also been concluded from ultrasound measurements \cite{Theuss23}.

A multi-component order parameter has also been proposed to interpret some magnetic screening measurements \cite{Ishihara23,Iguchi23}. Iguchi \textit{et al.} determine the local superfluid response of CVT grown UTe$_2$ samples, which they compare to a theoretical model of the possible single component order parameters on three different Fermi surfaces \cite{Iguchi23}. They find the best agreement with $B_{3u}$ (nodes along the $k_x$ directions) or $A_u$ or a multi-component combination of these for a single-sheet, cylindrical Fermi surface along the $k_z$-direction. A two-sheet cylindrical Fermi surface has been observed in quantum oscillation measurements on MSF grown samples \cite{Eaton24,Broyles23}. Broyles \textit{et al.} additionally observe a spherical pocket in this Fermi surface \cite{Broyles23}. Similarly, a $B_{3u}+i A_u$ order parameter has been proposed by Ishihara \textit{et al.} for penetration depth measurements of two CVT1 samples as well as a MSF1 sample \cite{Ishihara23}.

In this paper, we measure electrodynamic properties of two CVT2 UTe$_2$ single crystals, including the complex surface impedance and magnetic penetration depth. We study the low energy excitations out of the ground state in the form of the low temperature power-law temperature dependence of the penetration depth. Additionally, the large microwave residual loss which we observe is also consistent with nodal superconductivity. We observe an isotropic, sub-$T^2$ power-law for the penetration depth temperature dependence, which is inconsistent with a single pair of point nodes on the Fermi surface for weak scattering. We discuss other possible interpretations of these results.

\subsection{Experiment}
\label{sec:experiment}
We have access to high-$T_c$ $\text{UTe}_2$ samples grown in the Quantum Material Center of the University of Maryland by Shanta Saha using the CVT2 technique \cite{Ran21_CVT,Rosa22}. These samples have a $T_c$ near 2 K  as opposed to the CVT1 samples which had $T_c=1.6-1.8$ K.

The UTe$_2$ samples used in this experiment are prepared in a controlled atmosphere with a coarse and a fine polishing, as discussed in \S\ref{sec:SM_sample_prep}. The prepared UTe$_2$ samples have side lengths of roughly $0.5-1$ mm.

\begin{figure*}[t!]  %---------------------------------
\hspace*{-2cm}\includegraphics[scale=0.4,width=1.2\textwidth]{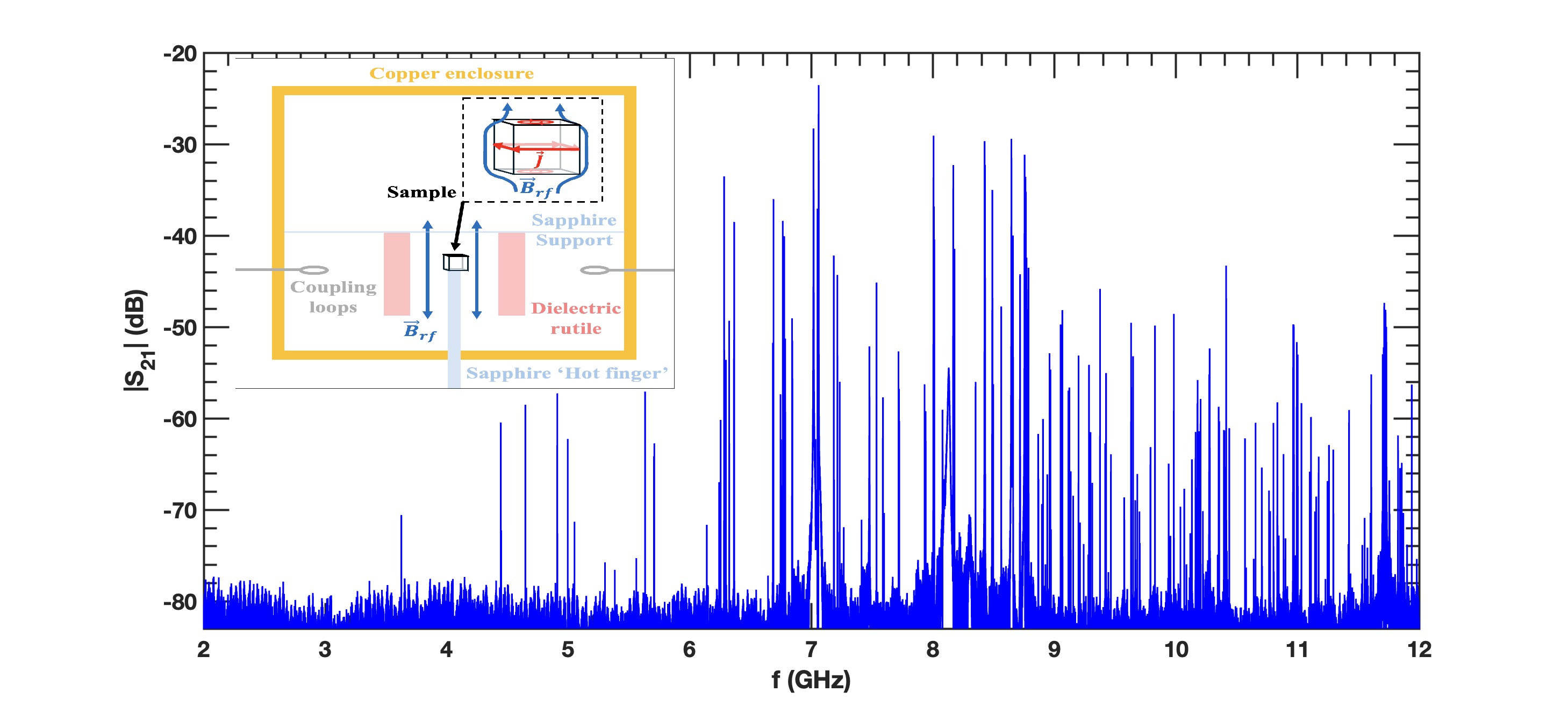}
\caption{Plot of the transmission magnitude $|S_{21}|$ through the cylindrical, hollow dielectric resonator as a function of frequency with no sample present. Roughly 300 resonant modes appear above the noise floor (-80 dB) in the 2-12 GHz measurement. The inset shows a schematic diagram (not to scale) of the rf magnetic field in a mode of the resonator as well as the screening current response of the sample.}
\label{fig:full_spectrum}
\end{figure*}

We perform cavity perturbation experiments with these samples in a cylindrical, hollow rutile dielectric resonator. A microwave signal is sent through the cavity, and the transmission is measured, which is shown in Fig.~\ref{fig:full_spectrum}. The sample perturbs the resonances of the cavity revealing the electrodynamic properties through changes in the resonant frequency $f_0$ and quality factor $Q$ of the modes as functions of sample temperature. Resonator perturbation experiments like this have been used to study the electrodynamics of a diverse variety of superconductors for many years \cite{Waldron67,Sridhar88,Rubin88,Bonn91,Hardy93,Mao95,Trunin98,Ormeno02,Seokjin21}.  A novel aspect of this work is that numerous modes in the spectrum shown in Fig.~\ref{fig:full_spectrum} are utilized to make multiple independent measurements of the sample surface impedance.  Another novelty of this work is the use of the anisotropic normal state resistivity tensor to determine the two parameters ($G$ and $X_0$) that convert the raw data into the absolute surface impedance and magnetic penetration depth. \rev{Both of these novel features are discussed in detail in \S\ref{sec:App_general} and \S\ref{sec:App_imp_fitting}.}

	%---------------------------------

\subsection{Resonator}
\label{sec:resonator}
The sample is introduced into a hollow cylindrical, rutile dielectric resonator \cite{Seokjin21} to perturb its resonances. The sample is attached to the end of a sapphire-rod `hot finger' \cite{Sridhar88,Rubin88,Mao95,Trunin98} using Apiezon N-grease, which holds it on the symmetry axis in the middle of the resonator and allows heat to be conducted from a heater outside the resonator cavity directly to the sample over a range from 100 mK to 20 K, while isolating it from the copper walls of the cavity and the rutile, which are nominally held at 100 mK. The microwave transmission through the resonator is measured for frequencies between 2 and 12 GHz, and the resonant frequency $f_0$ and quality factor $Q$ of many modes are determined by fitting \cite{Petersan98}. The resonant properties are measured both with \rev{$(f_{0,\text{tot.}}, Q_{\text{tot.}})$} and without \rev{$(f_{0,r}, Q_{r})$} the sample present in the cavity, which allows us to isolate an effective $\Delta f_{0\rev{,\text{sample}}}$ and $Q_{\rev{\text{sample}}}$ arising from the sample alone \cite{Mao95,Seokjin21}, where $\Delta f_0$ denotes the change in $f_0$ from its value at the lowest measurement temperature $T_0\sim 150$ mK. See \S\ref{sec:SM_mode_selection} for \rev{further} details.

\section{Data analysis (utilizing minimal assumptions)}
\label{sec:analysis_min}
%\paragraph{Data analysis}
The first electrodynamic property of interest is the complex surface impedance $Z_s=R_s+i X_s$, made up of the surface resistance $R_s(T)$ and surface reactance $X_s(T)$. Note that the surface impedance which we measure is a \textit{composite} of multiple anisotropic responses\rev{, namely the 3 components of the diagonal surface impedance tensor}. Our model of this is shown in Eq.~\ref{eq:Zs_comp}, and will be discussed in detail in \S\ref{sec:analysis_max_imp_fitting} below. We recover the real and imaginary components of the surface impedance from the sample resonance data as functions of temperature \cite{Mao95,Seokjin21},
% \begin{equation}
% \begin{split}
%     R_s=&\frac{G}{Q}\\
%     X_s=&-2G\frac{\Delta f_0}{f_{0,\text{tot.}}(T_0)} + X_0.
% \end{split}
% \label{eq:f0Q->Zs}
% \end{equation}
\begin{equation}
    Z_s=\frac{G}{Q_{\rev{\text{sample}}}} + i\left[ -2G\frac{\Delta f_{0\rev{,\text{sample}}}}{f_{0,\text{tot.}}(T_0)} + X_0 \right].
\label{eq:f0Q->Zs}
\end{equation}
Here $X_0$ is the minimum-temperature surface reactance, which sets the starting point for $X_s$ since this measurement is only directly sensitive to relative changes in $X_s$. The geometry factor $G$ (Eq. \ref{eq:Gdefinition}) has dimensions of resistance and relates the field structure at the sample location to that of the rest of the resonator for each mode. Here we assume that the sample creates a predominantly magnetic perturbation to the resonant modes considered. See \S\ref{sec:SM_mode_selection} for a discussion of the perturbation due to the sample. To fully relate the measured $f_{0\rev{,\text{sample}}}$ and $Q_{\rev{\text{sample}}}$ to $Z_s$, we must determine $G$ and $X_0$, which are unique to each mode of the resonator, as well as the sample and its orientation. 

Importantly, though, the overall temperature dependence of $Z_s$ and several other key electrodynamic properties can still be \textit{extracted directly from the data} with $G$ and $X_0$ undetermined. 
%The insets of Fig.~\ref{fig:lambda_pow_laws} demonstrate the effect of changing $X_0/G$ on the dimension-less impedance ($R_s(T)/G$, $X_s(T)/G$). While $R_s(T)/G$ is independent of $X_0/G$, $X_s(T)/G$ is shifted upward by a uniform amount by increasing $X_0/G$ (see color bar in Fig.~\ref{fig:lambda_pow_laws}). 
We proceed as follows.  The minimum value of $X_0/G$ is dictated by the data through the constraint that $X_s\ge R_s$ for all measured temperatures, which is a general property of the normal state electrodynamics \cite{Hein01}. The maximum value of $X_0$ is constrained in the non-local limit by $X_s \le \sqrt{3} R_s$ \cite{Reuter48}. See \S\ref{sec:SM_Xs_constraints} for further discussion of these constraints. We will use the dimension-less quantity $Z_s(T)/G$ for various $X_0/G$ to analyze the complex conductivity and  magnetic penetration depth as functions of $X_0/G$. 

\begin{figure}[!ht]	%---------------------------------
\begin{subfigure}[b]{0.4\textwidth}
\caption{}
\includegraphics[scale=0.23,width=1.0\textwidth]{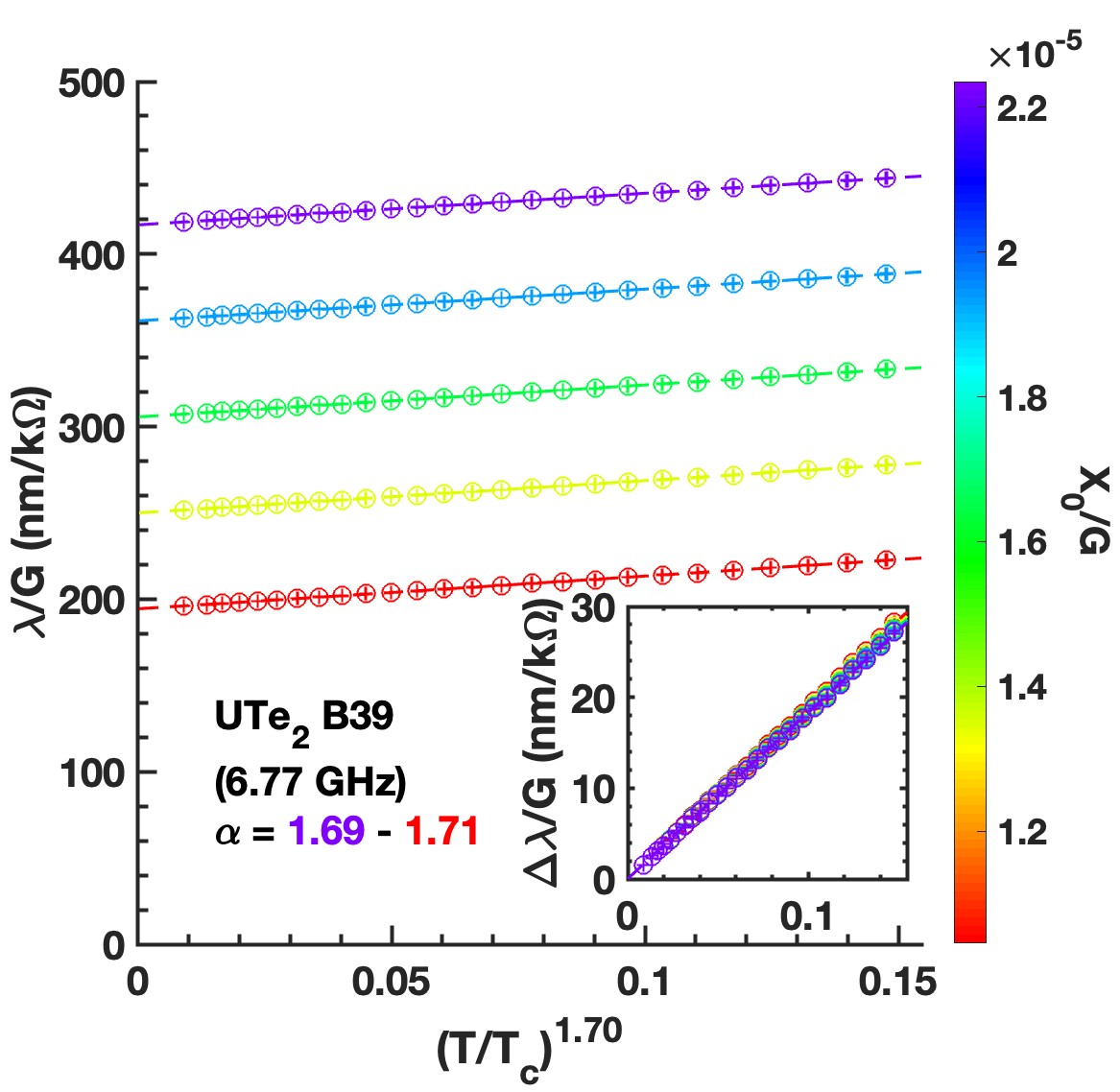}
\label{fig:lambda_pow_law_B39}
\end{subfigure}
\begin{subfigure}[b]{0.4\textwidth}
\caption{}
\includegraphics[scale=0.23,width=1.0\textwidth]{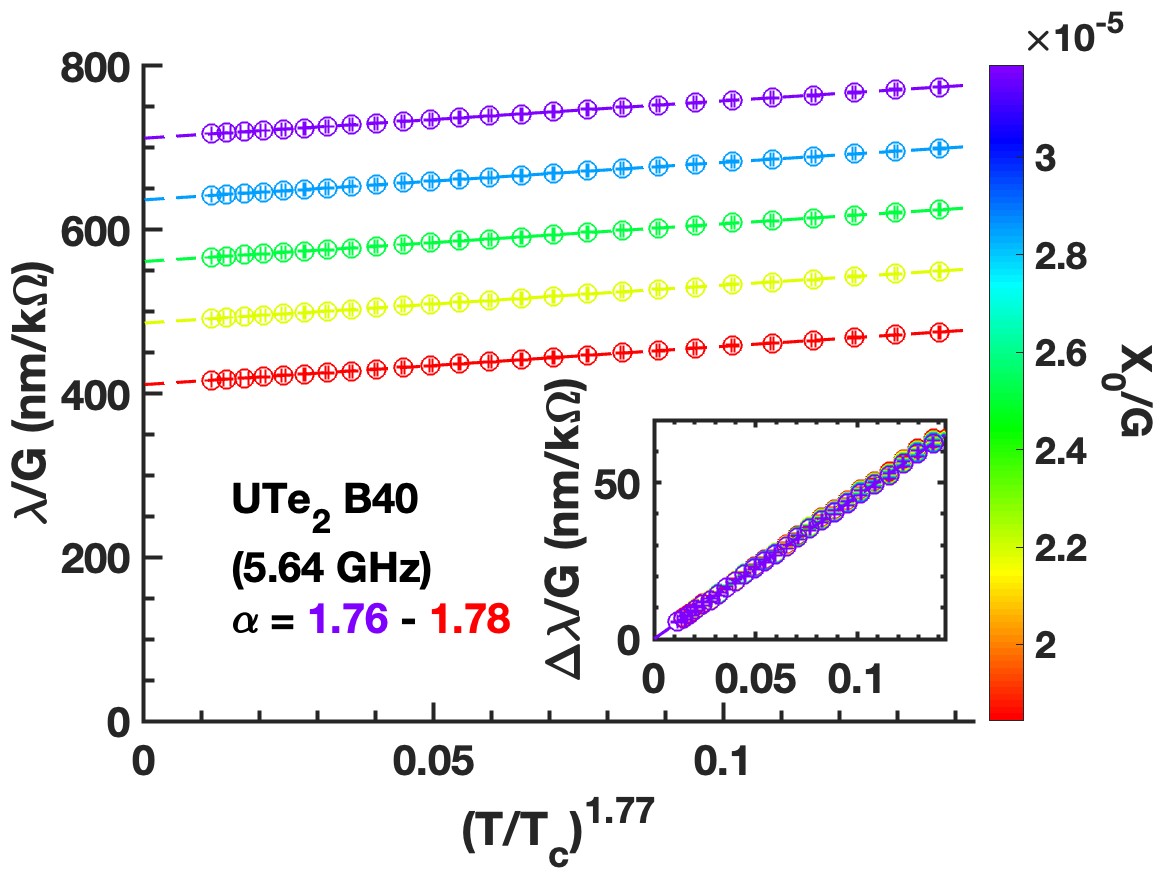}
\label{fig:lambda_pow_law_B40}
\end{subfigure}
\caption{\rev{Plots of reduced penetration depth $\lambda/G$ vs. an intermediate exponent of temperature from power-law fits (shown as dashed lines) for $T\le T_c/3$ for (a) sample B39 and (b) sample B40. The different colors correspond to different values of $X_0/G$ over its allowed range, from minimum (red) to maximum (purple), for which the power-law $\alpha$ slightly changes. The insets show the change in penetration depth $\Delta\lambda/G = (\lambda-\lambda_0)/G$ with the same respective x-axes. These plots largely overlap for each value of $X_0/G$.}}
\label{fig:lambda_pow_laws}
\end{figure}	%---------------------------------

In the local limit, the surface impedance is related to the complex conductivity $\sigma=\sigma_1-i\sigma_2$ by $Z_s=\sqrt{\frac{i\mu_0\omega}{\sigma}}$, where $\omega=2\pi f$. We discuss our results for the complex conductivity in \S\ref{sec:analysis_max_sigma} below.
%in \S\ref{sec:SM_sigma}. 
When $\sigma_2 \gg \sigma_1$, the penetration depth is given by, $\lambda=1/\sqrt{\mu_0\omega\sigma_2}$ \cite{Seokjin21,Anlage92}. 
\rev{Both $\sigma$ and $\lambda$ calculated from composite $Z_s$ in this way are also composite quantities.}
%Fig.~\ref{fig:lambda_pow_laws} shows the temperature dependence of the change in the penetration depth $\Delta\lambda(T)/G=(\lambda(T)-\lambda_0)/G$ from its zero temperature value $\lambda_0$ at low temperature $T\le T_c/3$, where $T_c$ is the microwave critical temperature, for two UTe$_2$ samples, using only the raw $f_0(T)$ and $Q(T)$ data for multiple possible values of $X_0/G$.
Fig.~\ref{fig:lambda_pow_laws} shows the temperature dependence well below the microwave critical temperature, $T_c$, of the penetration depth, for two UTe$_2$ samples, using only the raw $\rev{\Delta}f_{0\rev{,\text{sample}}}(T)$ and $Q_{\rev{\text{sample}}}(T)$ data for multiple possible values of $X_0/G$. Importantly, the $\Delta\lambda(T)/G$ data\rev{, shown in the insets of Fig.~\ref{fig:lambda_pow_laws},} demonstrates an independence on $X_0/G$ in this temperature range. 

A power-law temperature dependence for the penetration depth at low temperatures \rev{($T\le T_c/3$)} can be indicative of nodal superconductivity. The form,
\begin{equation}
    \frac{\lambda(T)-\lambda_0}{\lambda_0} = \eta \left(\frac{T}{T_c}\right)^{\alpha},
    \label{eq:lambda_pow_law}
\end{equation}
has theoretical predictions for the dimensionless parameters $\alpha$ and $\eta$ depending on the pairing symmetry, direction of current flow, and scattering conditions present in the sample \cite{Gross86,Klemm88}. For both UTe$_2$ B39 and UTe$_2$ B40, we estimate $T_c=1.96$ K. We used data from base temperature to $T_c/3$ to do the power law fitting in Eq.~\ref{eq:lambda_pow_law} with $\lambda_0/G$, $\alpha$, and $\eta$ as the fitting parameters for the composite $\lambda/G$ data. \rev{These power-law fits are included in Fig.~\ref{fig:lambda_pow_laws} as dashed lines. See \S\ref{sec:lambda_pow_law_fits} for a detailed discussion of this fitting, as well as Table~\ref{tab:fitting_uncerts} which summarizes the uncertainties of the parameters in this fit and others.} We also compare this low-temperature $\lambda(T)$ for UTe$_2$ with that of s-wave superconductors Nb and NbSe$_2$ measured in the same apparatus in \S\ref{sec:SM_lambda_exp_fit}.

For the range of possible $X_0/G$ values we find $\alpha=1.69-1.71$ for sample B39 and $\alpha=1.76-1.78$ for sample B40 for the modes in Fig.~\ref{fig:lambda_pow_laws}. Note that both $X_0$ and $G$ are different for these different modes and samples. The behavior of $\alpha$, $\eta$, and $\lambda_0$ with respect to $X_0/G$ is further discussed in \S\ref{sec:lambda_pow_law_fits}. Overall, we find no systematic variation of $\alpha$ between modes, which is discussed further in \S\ref{sec:discussion} below. We therefore take the average and standard deviation among all the modes to find $\alpha=1.70\pm0.08$ for sample B39 and $\alpha=1.82\pm0.25$ for sample B40. These averages use a fit to fix $G$ and $X_0$ for each mode, which is discussed in \S\ref{sec:analysis_max_imp_fitting} below. The variation of $\alpha$ due to changing the value of $X_0/G$ is then much smaller than its variation between modes, so the particular choice of fitting method to determine $G$ and $X_0$ will not alter the conclusion. The aggregate results of these power-law fittings are summarized in Table~(\ref{tab:lambda_and_loss}).

%Since this power-law dependence for both UTe$_2$ samples is robust to variations in the choice of $X_0$, the particular choice of fitting method to determine $G$ and $X_0$ will not alter the conclusion, so we will now discuss ways to fix these parameters to determine residual loss values and estimate the contribution of each crystallographic direction to the measured effective surface impedance of each mode.

\begin{figure}	%---------------------------------
\hspace*{-0.5cm}\includegraphics[scale=0.23,width=0.5\textwidth]{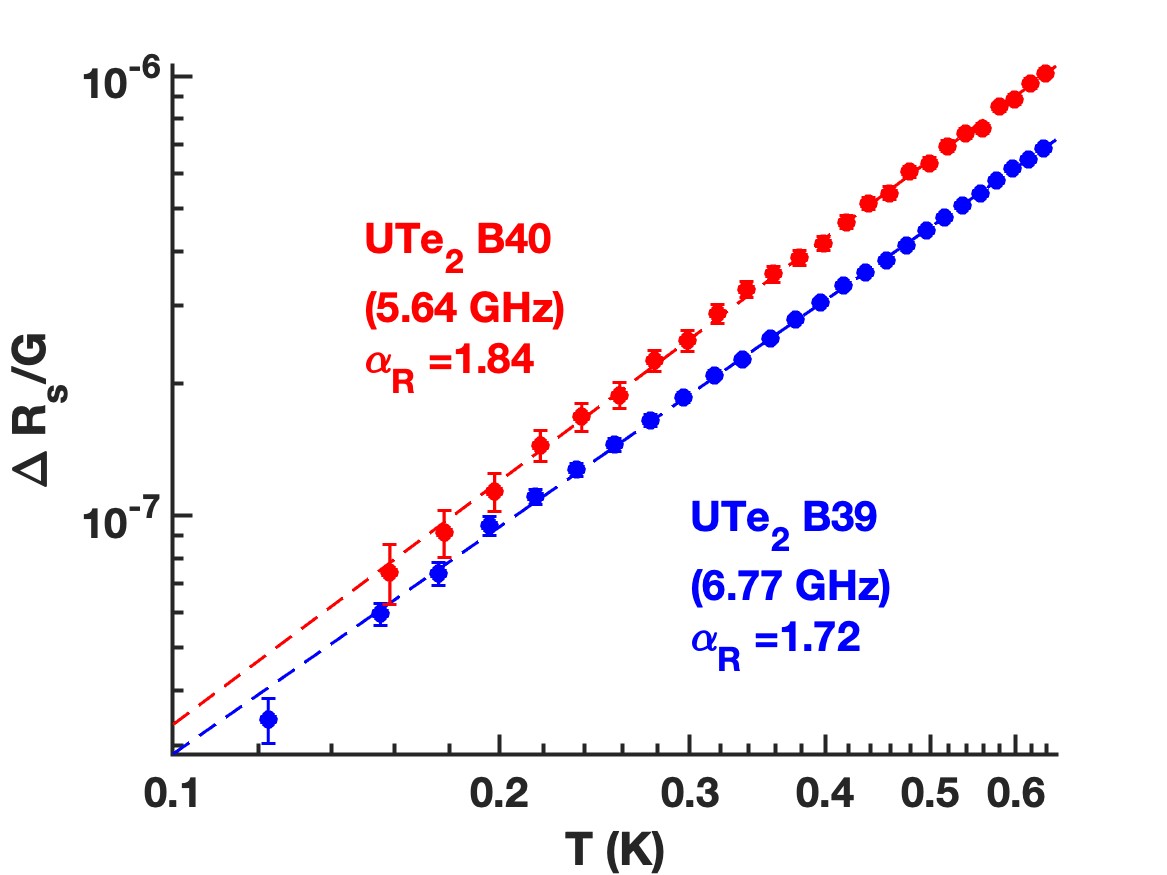}
\caption{Plots of change in dimensionless surface resistance $\Delta R_s/G = (R_s-R_0)/G$ vs. temperature and power-law fits for $T\le T_c/3$. UTe$_2$ B39 is shown in blue and UTe$_2$ B40 is shown in red. Shown are examples of the power-law fits to $R_s(T)$ for a mode of sample B39 ($f_0=6.77$ GHz) and sample B40 ($f_0=5.64$ GHz), for $T<T_c/3$.}
\label{fig:Rs_pow_law_sup_mat}
\end{figure}	%---------------------------------

We also observe a power-law temperature dependence for the dimensionless surface resistance as, 
\begin{equation}
    \frac{R_s(T)-R_0}{R_0} = \eta_R \left(\frac{T}{T_c}\right)^{\alpha_R},
    \label{eq:Rs_pow_law}
\end{equation}
for $T\le T_c/3$, which we use to determine the zero-temperature surface resistance $R_0$ for each mode. In fact, the determination of $\alpha_R$, $\eta_R$, and $R_0/G$ is explicitly independent of $G$ and $X_0$. See \S\ref{sec:SM_resistance_pow_law} for a \rev{detailed} discussion of this fitting\rev{, as well as Table~\ref{tab:fitting_uncerts} which summarizes the uncertainties of the parameters in this fit and others}. Similarly to the penetration depth power-law exponents, the surface resistance power-law exponents, $\alpha_R$, show no systematic variation between modes. These give average results of $\alpha_R=1.70\pm 0.20$ for UTe$_2$ B39 and $\alpha_R=1.86\pm 0.20$ for UTe$_2$ B40, which are effectively the same as the penetration depth power-law exponents considering the uncertainties. Table~\ref{tab:lambda_and_loss} also includes a summary of $\alpha_R$ and the nominally frequency-independent residual loss $R_0/\omega^2$.  The significance of this power-law behavior is not straightforward to interpret; see \S\ref{sec:SM_resistance_pow_law} for more discussion.  We note in passing that a power-law behavior of $R_s(T)$ with $\alpha_R\approx 1$ independent of frequency was observed in CeCoIn$_5$ \cite{Ormeno02}.

\begin{table}
    \begin{center}
    \caption{A comparison of several composite and axis-resolved electrodynamic properties for various UTe$_2$ samples.}
    \begin{tabular}{ | c | c | c | c | c | c | c | c | }
     Batch/ & Growth & $T_c$ & $\alpha$ & $\eta$ & $\lambda_0$  & $\alpha_R$ & $R_0/\omega^2$ \\ 
      citation & method & (K) &  &  & $(\mu m)$ &  & $(\Omega\ ps^2)$ \\  
     \hline
     B39 & CVT2 & 1.96 & $1.70$ & $0.58$ & $0.64$ & $1.70$ & $0.10$ \\ % our data
      &  &  & $\pm0.08$ & $\pm0.23$ & $-2.01$ & $\pm0.20$ & $-5.41$ \\ 
     B40 & CVT2 & 1.96 & $1.82$ & $0.70$ & $1.00$ & $1.86$ & $0.14$ \\ 
      &  &  & $\pm0.25$ & $\pm0.28$ & $-2.42$ & $\pm0.20$ & $-22.4$ \\ 
     \hline
     \cite{Seokjin21} & CVT1 & 1.57 & $2.11$ &  & $0.791$ &  & $3.0$ \\ % Seokjin
     \hline
     \cite{Ishihara23} & CVT1 & 1.65 & (2.01, &  &  &  & \\ % Ishihara
      &  &  & 1.90, &  &  &  & \\ 
      &  &  & 1.60) &  &  &  & \\ 
      & CVT1 & 1.75 & (2.08, &  &  &  & \\
      &  &  & 2.06, &  &  &  & \\ 
      &  &  & 1.84) &  &  &  & \\ 
      & MSF1 & 2.1 & (2.07, &  & (1.42, &  & \\
      &  &  & 2.16, &  & 0.710, &  & \\ 
      &  &  & 2.04) &  & 2.75) &  & \\ 
     \label{tab:lambda_and_loss}
    \end{tabular}
    \end{center}

\end{table}

\section{Data analysis (requiring more fittings and assumptions)}
\label{sec:analysis_max}
In the normal state, we assume that $\sigma$ is given by a single Drude peak \cite{Ormeno06}, which is consistent with THz spectroscopy data on UTe$_2$,\cite{Arm22}, though a second, much weaker Drude has also been observed at THz frequencies, which should still be weak at the GHz frequencies used in our experiment. The main Drude peak is then given by $\sigma=\frac{1}{\rho_{\text{dc}} (1+i\omega\tau)}$, where $\rho_{\text{dc}}$ is the DC resistivity and $\tau$ is the quasiparticle scattering time. Here we assume a single, composite function for the scattering time $\tau(T)$ despite the anisotropic nature of the sample.

\subsection{Scattering time}
\label{sec:analysis_max_omega_tau}
\begin{figure}[ht!]  %---------------------------------
\begin{subfigure}[b]{0.4\textwidth}
\caption{}
\hspace*{-0.5cm}\includegraphics[scale=0.23,width=1.0\textwidth]{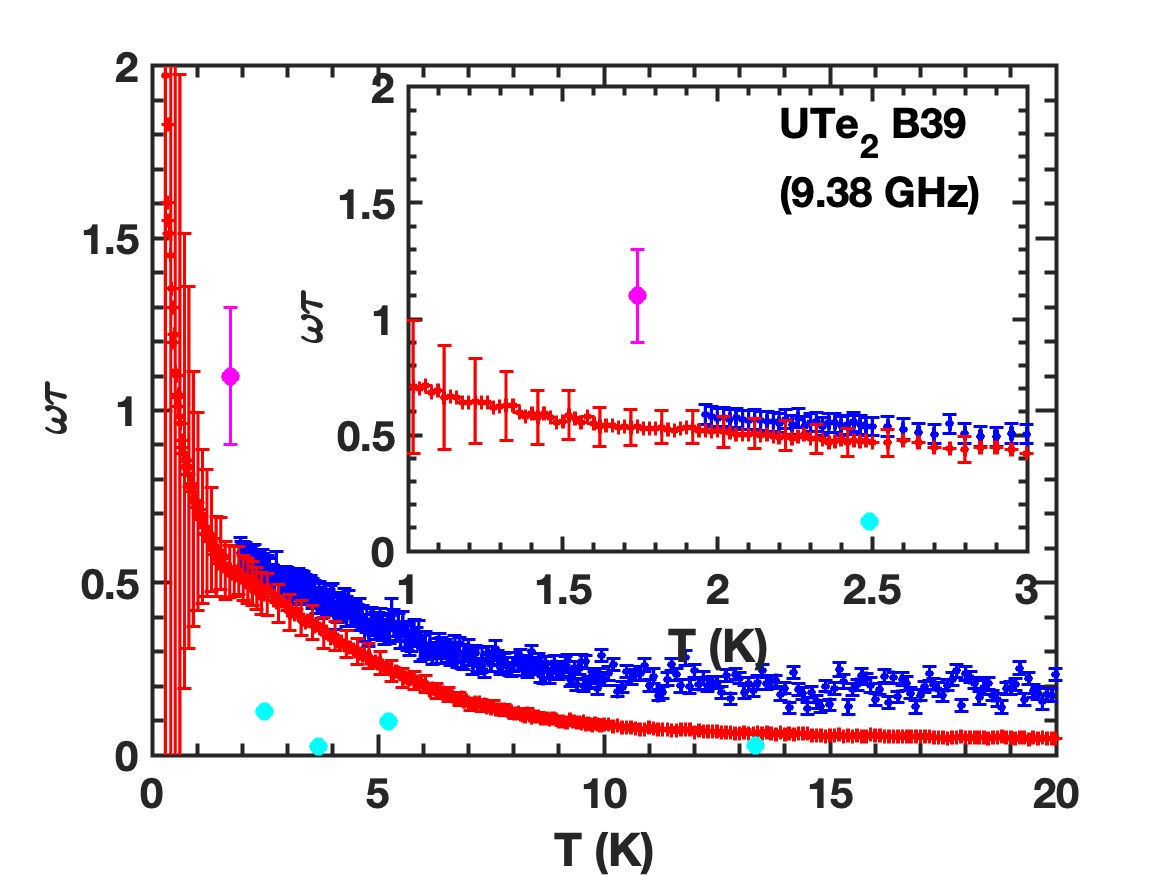}
\label{fig:omega_tau_comparison}
\end{subfigure}
\begin{subfigure}[b]{0.4\textwidth}
\caption{}
\hspace*{-0.5cm}\includegraphics[scale=0.6,width=1.0\textwidth]{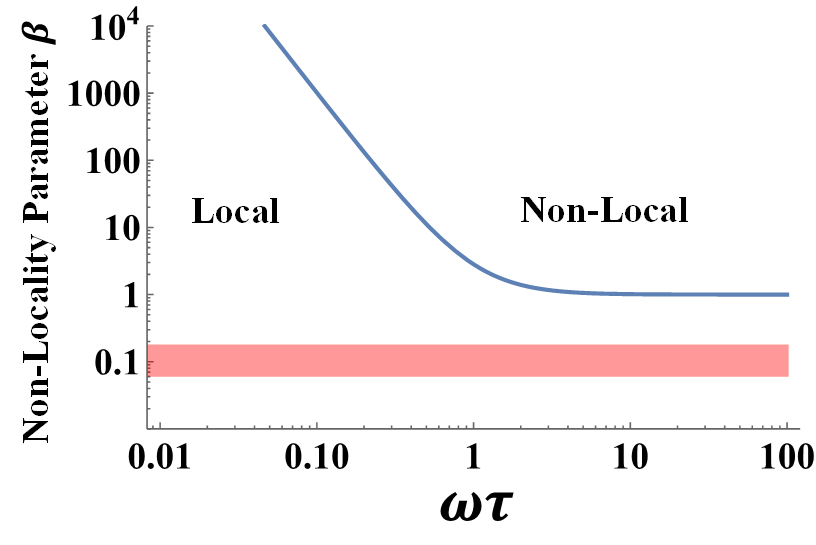}
\label{fig:BetavsOmegaTau}
\end{subfigure}
\caption{(a) Comparison of multiple independent determinations of composite scattering time $\omega\tau(T)$ for the UTe$_2$ crystals. The blue and red curves show $\omega \tau$ estimated from the Drude and two-fluid model results, respectively, see \S\ref{sec:2f_model}. The cyan points show $\omega\tau$ from the normal state impedance slope, $\Xi(T)$ in the local-limit, see \S\ref{sec:OmTauAnalysis}. The magenta point is the mean value of $\omega\tau_{eff} (>\omega\tau)$ from the reactance peak for all the measured B39 modes, see \S\ref{sec:SM_reactance_peak}. The inset shows temperatures near $T_c$. (b) Log-log plot of non-locality parameter $\beta$ vs. $\omega\tau$.  The region above (below) the blue line ($\beta_c$) is the non-local (local) limit.  The estimated range of $\beta$ values for the UTe$_2$ samples measured at the frequencies 
 in this study are shown in the red box.}
\label{fig:omega_tau_and_beta}
\end{figure}

We consider multiple methods to determine $\omega\tau$ for UTe$_2$, which are compared in Fig.~\ref{fig:omega_tau_comparison}. We estimate $\omega\tau \lesssim 1$ just below $T_c$ for all of our data from a minimum assumption analysis of the reactance peak $X_s(T)$ for both UTe$_2$ samples \cite{Hein01b,Ormeno06}. We discuss this analysis in \S\ref{sec:SM_reactance_peak}. In the normal state, we obtain $\omega\tau\sim 0.1$ from a minimal assumption analysis of the variation of the data in the complex $Z_s$ plane (see Fig.~\ref{fig:Zs_contours_sup_mat} to visualize this geometry) \cite{Hein01b}. We discuss this analysis in \S\ref{sec:OmTauAnalysis}. Both of these analyses do not require knowledge of $G$ or $X_0$. Our analysis with an isotropic Drude model gives $\omega\tau(T)=(X_s^2(T)-R_s^2(T))/2R_s(T) X_s(T)$ in the normal state, which requires only raw data and the determination $X_0/G$. We can extend this analysis to the superconducting state using the two-fluid model, as discussed in \S\ref{sec:2f_model}, which requires determination of both $X_0/G$ and a composite (reduced) London penetration depth $\lambda_L/G$. Near $T_c$, we find good agreement of $\omega\tau$ between the Drude and the two-fluid models, as shown in Fig.~\ref{fig:omega_tau_comparison}. 

We will now discuss determination of the degree of nonlocality of the electrodynamics in our experiment to justify our assumption of the local limit. We introduce the frequency-dependent dimensionless nonlocality parameter $\beta = \frac{v_F}{\omega \lambda_L} = \frac{\ell_{mfp}/\lambda_L}{\omega\tau}$, where $v_F$ is the Fermi velocity
%, $\lambda_L$ is the London penetration depth, 
and $\ell_{mfp}$ is the normal-state carrier mean free path. For $\beta < \beta_c = (1+\frac{1}{(\omega\tau)^2})^{3/2}$, the sample can be considered to be in the local limit \cite{Hein01b}. We estimate the values of the non-locality parameter $\beta$ for UTe$_2$ in the frequency range of our measurement as follows. The Fermi velocity for UTe$_2$ is estimated to be $v_F = 6\times 10^{3}$ m/s \cite{Metz19}. We take the London penetration depth to be $\lambda_L = 1.5~ \mu$m, which approximates the anisotropic values presented in this paper. This gives values of $\beta = 0.16 \sim 0.06$ for the range of frequencies (4-11 GHz) of our cavity perturbation data. In Fig.~\ref{fig:BetavsOmegaTau}, we plot $\beta$ vs $\omega\tau$, with the local and non-local regions indicated as well as our estimate of $\beta$ values for the UTe$_2$ samples in this experiment. Considering this estimate of the non-locality parameter shown in Fig.~\ref{fig:BetavsOmegaTau} and our four independent estimates $\omega\tau$ shown in Fig.~\ref{fig:omega_tau_comparison}, it is clear that our samples are in the local limit as far as the electrodynamic response is concerned. We discuss the nonlocality parameter further in \S\ref{sec:OmTauAnalysis}.

	%---------------------------------

\subsection{Impedance fitting}
\label{sec:analysis_max_imp_fitting}
In previous work \cite{Seokjin21}, the assumption that  $R_s=X_s=\sqrt{\mu_0\omega\rho_{dc}/2}$ was used in the high temperature limit ($T\gg T_c$) of the normal state to determine $G$ and $X_0$; however, this assumes that $\omega\tau$ vanishes for the fitting regime. Assuming only the local limit and Drude model in the normal state, we can improve the determination of $G$ and $X_0$ by simply fitting to the geometric mean $\sqrt{R_s(T) X_s(T)}=\sqrt{\mu_0\omega\rho_{dc}(T)/2}$, which is valid for the entire normal state, and makes no assumption about the value of $\omega\tau$, as well as the sample size, shape, or demagnetization factor. 
%Additionally, we can then extract $\omega\tau(T)=(X_s^2(T)-R_s^2(T))/2R_s(T) X_s(T)$ in the normal state, which is treated as a composite quantity rather than an anisotropic tensor, as is $\rho_{dc}$.
This method treats $\omega\tau$ and $\rho_{dc}$ as composite quantities rather than anisotropic tensors. We note that in this previous work \cite{Seokjin21}, a single composite $\rho_{dc}$ function was measured for the sample with a four-point contact technique.

However, it is now clear that the electrodynamic properties of UTe$_2$ are anisotropic, and this must be incorporated into our data analysis to enable a more complete understanding of its surface impedance, both above and below T$_c$. The measured surface impedance for each resonant mode arises from a mixture of surface currents induced along each crystallographic axis direction. That is to say that the measured surface impedance is a \textit{composite}. This mixture depends on the electromagnetic structure of each particular resonant mode of the dielectric resonator (DR), as well as the sample location and orientation inside the resonator. We assume that for each resonant mode the measured surface impedance $Z_s$ is a weighted sum of the 3 diagonal surface impedance tensor components $Z_{s,i}$,
\begin{equation}
    Z_s=w_a Z_{s,a} + w_b Z_{s,b} + w_c Z_{s,c},
\label{eq:Zs_comp}
\end{equation}
where the weights $w_i$ satisfy $0\le w_i \le 1$ and $\Sigma w_i = 1$ \cite{Kitano95}. These weights represent the projections of the surface currents along the crystallographic axes. By studying a number of diverse modes with different projection sets, a sampling of $w_i$ is obtained, giving information about the full surface impedance tensor. \rev{Other composite quantities, such as complex conductivity $\sigma$ and penetration depth $\lambda$ would in general have a more complicated dependence on their tensor components. Certain simplifying limits are still possible; such as, at low temperatures ($T\ll T_c$), $\lambda$ would decompose linearly with the same current induced weights $w_i$ as in Eq.~\ref{eq:Zs_comp} since $X_s\approx\mu_0\omega\lambda$ in this limit as discussed in \S\ref{sec:SM_param_summary}. This is demonstrated for the zero-temperature penetration depth $\lambda_0$ in Fig.~\ref{fig:lambda_0_vs_w_sup_mat}.}

There are, correspondingly, three independent scattering times, which means that they cannot all be explicitly eliminated from the normal state behavior without prior knowledge of their functional forms. We can, however, employ a heuristic argument that since $R_i=\sqrt{\mu_0\omega\rho_i/2}$ is the $\omega\tau\rightarrow 0$ limit of the normal state surface resistance for currents flowing in direction $i = (a, b, c)$, these losses should add linearly in the same way as Eq.~(\ref{eq:Zs_comp}). We then match this linear combination to the geometric mean of the composite $R_s$ and $X_s$,
\begin{equation}
    \sqrt{R_s(T) X_s(T)}=w_a R_a(T) + w_b R_b(T) + w_c R_c(T),
    \label{eq:geo_mean_comp}
\end{equation}
which assumes a single composite scattering time $\tau(T)$ in contradiction to our use of an anisotropic resistivity tensor. The approximation underlying the use of Eq. (\ref{eq:geo_mean_comp}) is better satisfied at high temperatures, in samples with higher RRR, and when one of the weights is much larger than the other two. See \S\ref{sec:Synth_data} for a demonstration of the efficacy of this approximation for synthetic data. 

\begin{figure}  %---------------------------------
% \centering
\hspace*{-0.5cm}\includegraphics[scale=0.22]{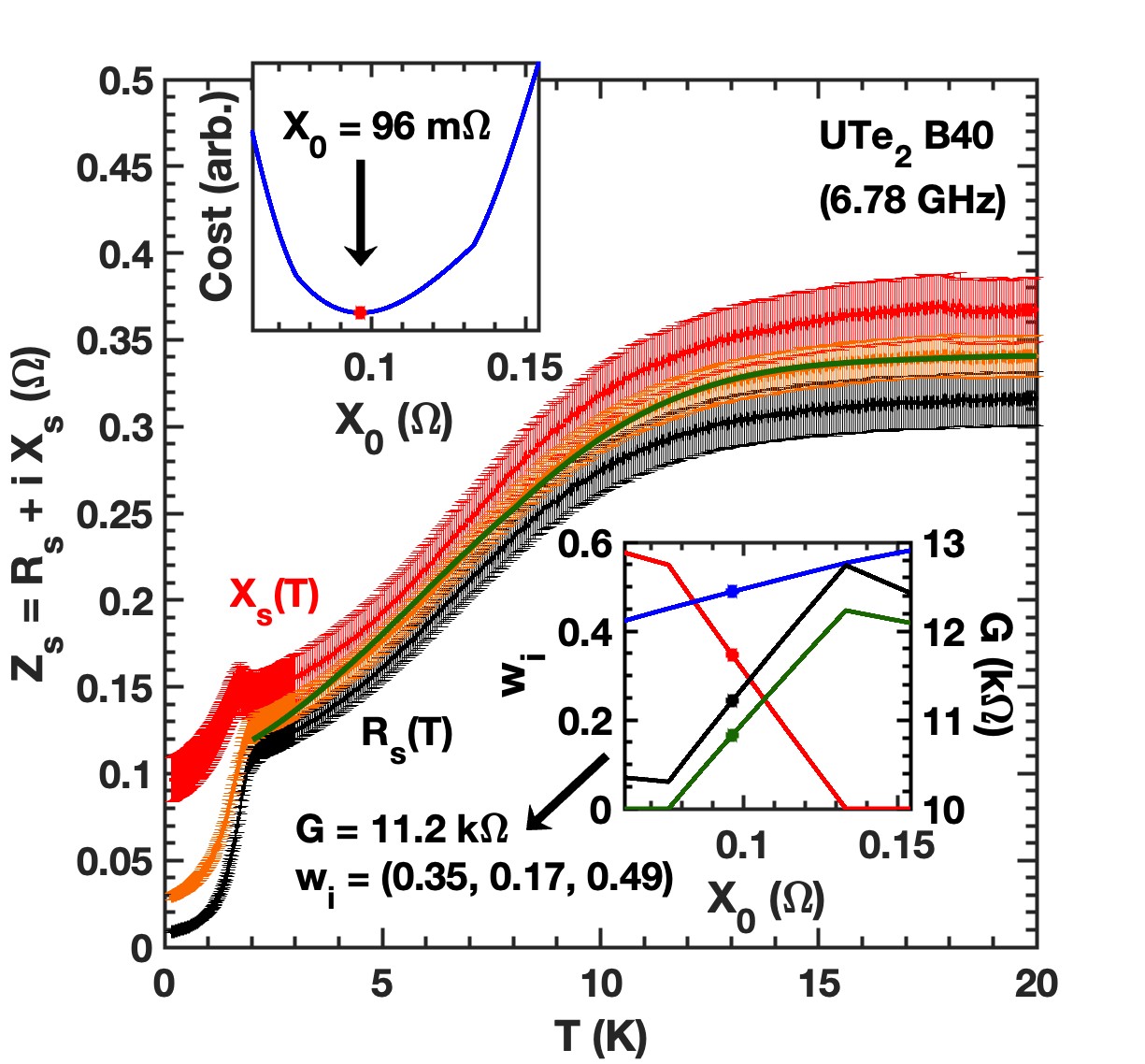}
\caption{Example of the result of the fit for to determine composite surface resistance $R_s(T)$ shown in black and composite surface reactance $X_s(T)$ shown in red for UTe$_2$ sample B40 at 6.78 GHz. The left hand side of Eq.~(\ref{eq:geo_mean_comp}) is shown in orange, which is fitted to the right hand side of Eq.~(\ref{eq:geo_mean_comp}) 
%(assuming the three components of the normal state DC resistivity tensor are known) 
shown in green. 
%This is a different mode (6.78 GHz) of UTe$_2$ B40 from the one shown in Fig.~(\ref{fig:lambda_pow_laws}). 
The upper inset shows the cost minimization with respect to $X_0$ for this mode. The lower inset shows optimal values of $w_a$, $w_b$, $w_c$, and $G$ in red, green, blue, and black, respectively, over the allowed range of $X_0$ with the solution for the optimal $X_0$ indicated as dots.}
\label{fig:Zs_with_fit}
\end{figure}

We use the measured anisotropic DC resistivity tensor of similar UTe$_2$ crystals measured with a Montgomery geometry \cite{Yunsuk22} to do this fitting. See \S\ref{sec:SM_rho_data} for further discussion of our treatment of this resistivity data. The normal-state surface impedance fitting is performed from 2 to 20 K for each mode, individually, which determines the fitting parameters $G$, $X_0$, and the $w_i$ for each mode. The results of this fitting for a representative mode are shown in Fig.~\ref{fig:Zs_with_fit}. This fitting and its uncertainties are discussed in detail in \S\ref{sec:SM_impedance_fitting}\rev{, and summarized in Table~\ref{tab:fitting_uncerts}}. For this mode, we obtain fitting parameters $G=11.2 \pm 0.5\ k\Omega$, $X_0=96 \pm 13\ m\Omega$, $w_a=0.35 \pm 0.12$, $w_b=0.17 \pm 0.10$, and $w_c=0.49 \pm 0.02$. Note that the uncertainty of $w_b$ is comparable to the value of $w_b$ itself, indicating that there may not be a significant response of currents corresponding to the b-axis for this mode.

	%---------------------------------

\subsection{Complex conductivity}
\label{sec:analysis_max_sigma}
We determine the complex conductivity $\sigma=\sigma_1 - i \sigma_2$ from the surface impedance using the expression valid in the local limit $\sigma=i\mu_0\omega/Z_s^2$. The real part $\sigma_1$ is a measure of the loss. Below $T_c$, the imaginary part $\sigma_2$ is a measure of the superfluid response, which is used to determine the penetration depth $\lambda=1/\sqrt{\mu_0\omega\sigma_2}$ and the superfluid density $\rho_s=(\lambda_0/\lambda)^2=\sigma_2/\sigma_{2,0}$, where $\lambda_0$ and $\sigma_{2,0}$ are the zero-temperature values of $\lambda$ and $\sigma_2$ respectively. In conventional superconductors, $\sigma_2\gg \sigma_1$ in the superconducting state at frequencies well below the gap. In UTe$_2$, this limit is not achieved. We find that on average, $\sigma_2(T_0)/\sigma_n$ is $3.2\pm 1.0$ for sample B39 and $3.2\pm 0.6$ for sample B40. Assuming a zero-temperature gap $\Delta_0=0.29$ meV, Mattis-Bardeen theory predicts that for a fully-gapped superconductor at 10 GHz ($\frac{\hbar\omega}{2\Delta_0}=0.07$), that $\sigma_2(0)/\sigma_n=22.7$ \cite{Mattis_Bardeen58}, which is considerably larger than what is observed in the data.
%We find that on average, for most modes, the maximum of $\sigma_2/\sigma_1$ is $19\pm 8$ for sample B39 and $9\pm 8$ for sample B40. For both samples, there is a much larger value of $\sigma_2/\sigma_1$ for one mode, which is 321 and 183 for B39 and B40 respectively. \textcolor{red}{(Are these results statistically significant?)}

\begin{figure}  %---------------------------------
\begin{subfigure}[b]{0.4\textwidth}
\caption{}
\hspace*{-0.5cm}\includegraphics[scale=0.23,width=1.0\textwidth]{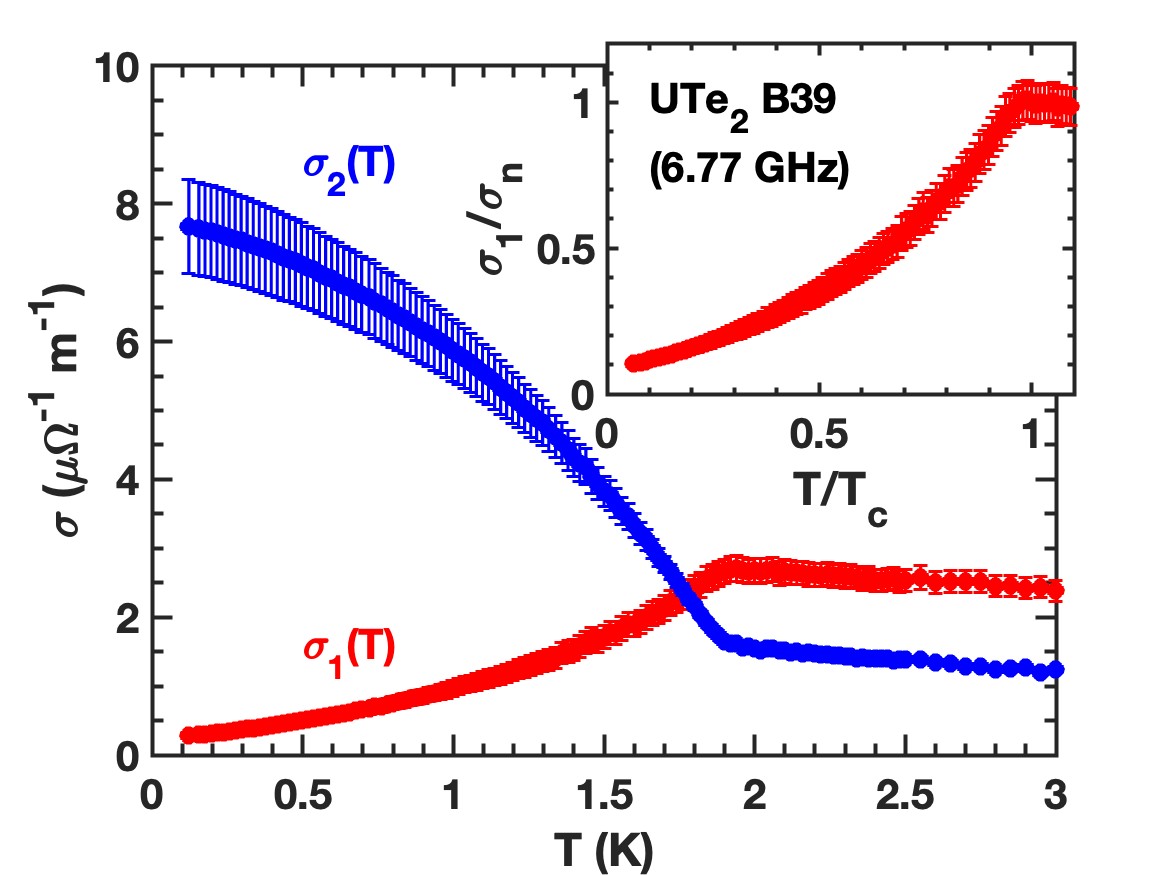}
\label{fig:sigma_vs_T_B39_sup_mat}
\end{subfigure}
\begin{subfigure}[b]{0.4\textwidth}
\caption{}
\hspace*{-0.5cm}\includegraphics[scale=0.23,width=1.0\textwidth]{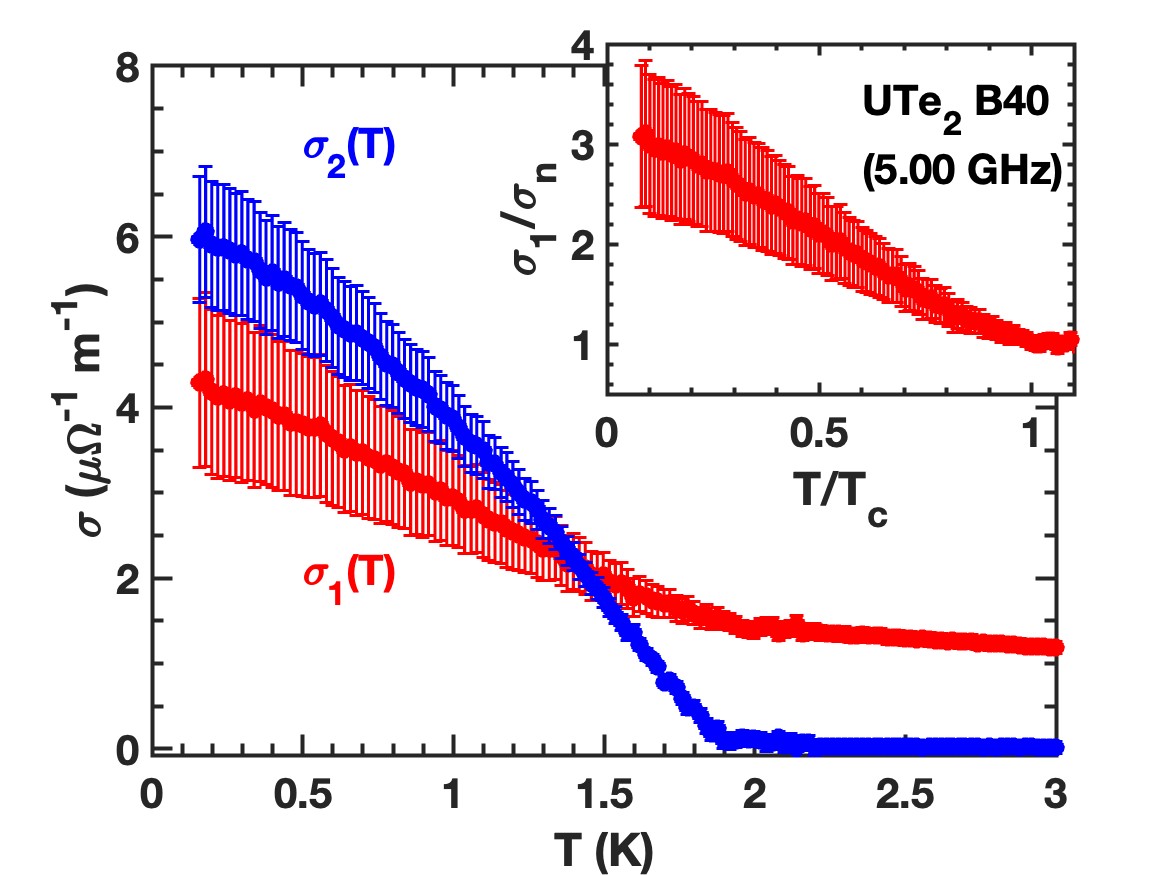}
\label{fig:sigma_vs_T_B40_sup_mat}
\end{subfigure}
\caption{Complex conductivity $\sigma =\sigma_1 -i\sigma_2$ vs temperature for (a) a typical mode and (b) an exceptional mode for UTe$_2$ sample B40 at 5 GHz. Note that $\sigma_2 \gg \sigma_1$ is not satisfied in (b). The insets shows $\sigma_1$ normalized by its value at $T_c$.}
\label{fig:sigma_vs_T_sup_mat}
\end{figure}

Figure \ref{fig:sigma_vs_T_sup_mat} shows two examples of the different temperature dependences observed for $\sigma(T)$. Figure \ref{fig:sigma_vs_T_B39_sup_mat} shows the typical behavior we observe for most modes. Figure \ref{fig:sigma_vs_T_B40_sup_mat} shows the exceptional behavior we observe for just a few modes. For both of these, $\sigma_2$ has roughly the same character of behaving like an order parameter in the superconducting state. In the normal state, the value of $\sigma_2$ is larger for the typical case than that of the exceptional case.
%, which indicates a larger value of $\omega\tau$ just above $T_c$ for the typical case since for a curve of constant $f_s$, increased values of $y=\mu_0\omega\lambda_L^2\sigma_2$ correspond to larger values of $\omega\tau$, as shown in Fig.~\ref{fig:sigma_contours_sup_mat}. 
% this could actually be due to a different $\lambda_L$ rather than $\omega\tau$
For $\sigma_1$, the exceptional case is very much like that of the CVS1 sample studied earlier \cite{Seokjin21} where $\sigma$ reaches its maximum at zero-temperature. The typical case we observe is more consistent with the d-wave result (e.g.  CeCoIn$_5$ in Fig. 2(b) in \cite{Seokjin21}), where $\sigma_1$ decreases below $T_c$ but leaves a finite residual loss at zero-temperature; however, this could also be consistent with some other nodal structure and additionally may depend on the quasi-particle mean free path. The insets of Fig.~\ref{fig:sigma_vs_T_sup_mat} show $\sigma_1$ normalized by its value at $T_c$ for easy comparison between modes and samples. Though the residual loss we find is comparable to that of a d-wave superconductor, our results are not consistent with the d-wave universal residual loss. See \S\ref{sec:SM_res_loss} for further discussion. 

	%---------------------------------

\begin{figure}[!ht]	%---------------------------------
\begin{subfigure}[b]{0.4\textwidth}
\caption{}
\vspace*{-0.6cm}
\hspace*{-2.3cm}
\includegraphics[scale=0.23,width=1.0\textwidth]{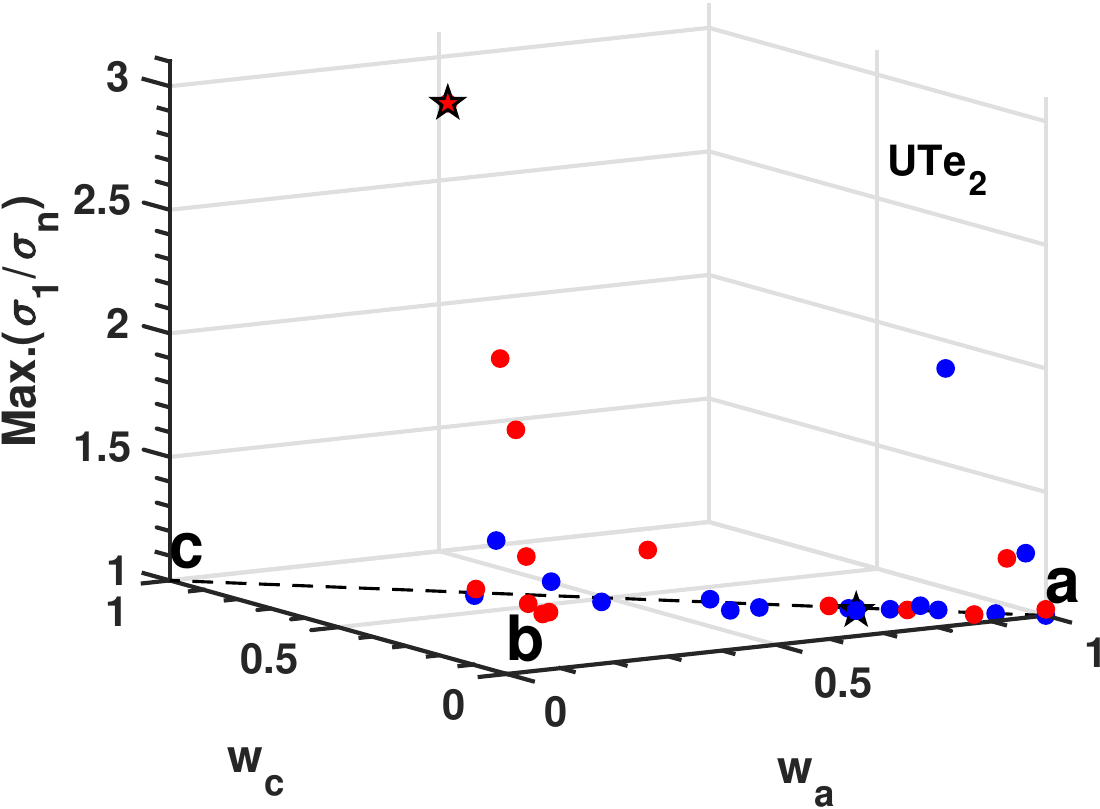}
\label{fig:sigma_1_peak_vs_w_sup_mat}
\end{subfigure}
\begin{subfigure}[b]{0.4\textwidth}
\vspace*{0.6cm}
\caption{}
\vspace*{-1.2cm}
\hspace*{-2.3cm}
\includegraphics[scale=0.23,width=1.0\textwidth]{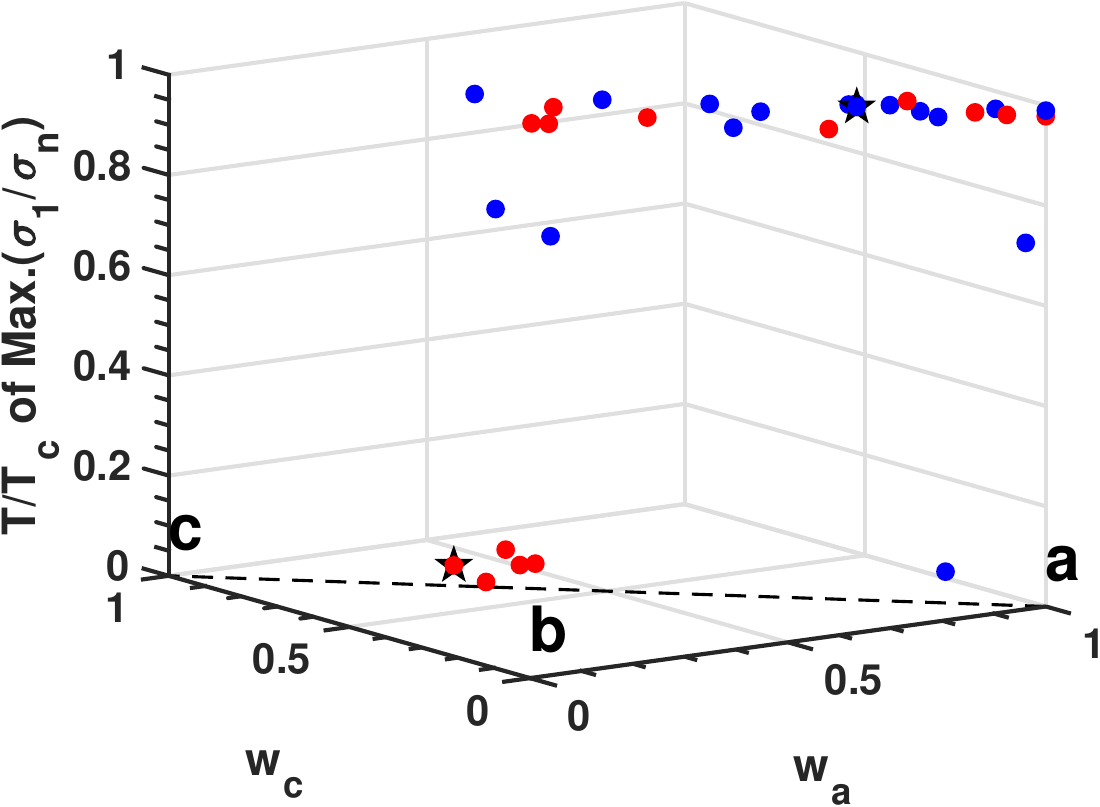}
\label{fig:t_sigma_1_peak_vs_w_sup_mat}
\end{subfigure}
\begin{subfigure}[b]{0.4\textwidth}
\vspace*{0.6cm}
\caption{}
\vspace*{-1.2cm}
\hspace*{-2.3cm}
\includegraphics[scale=0.23,width=1.0\textwidth]{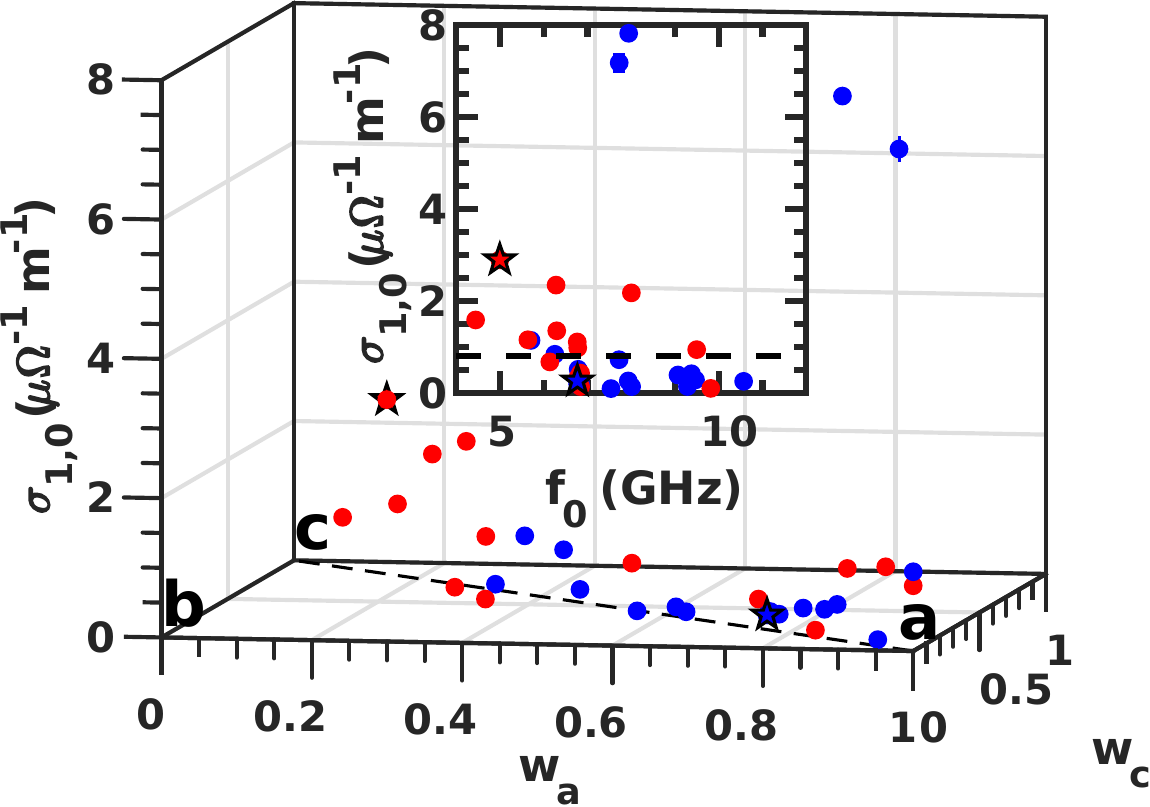}
\label{fig:sigma_1_0_vs_w_sup_mat}
\end{subfigure}
\caption{Plots of (a) the maximum of $\sigma_1$ below $T_c$ normalized by its value at $T_c$ and (b) the temperature at which this occurs normalized by $T_c$, both vs the induced-current weightings. (c) shows the residual value of $\sigma_1$ at zero-temperature, and its inset shows the same quantity vs frequency with our estimate of the universal d-wave loss $\sigma_{00}$ shown as the black dashed line. Samples B39 and B40 are shown in blue and red respectively. The star markers indicate the modes shown in Fig.~\ref{fig:sigma_vs_T_sup_mat}.}
\label{fig:sigma_peak_w_dep_sup_mat}
\end{figure}

We plot the maximum of $\sigma_1(T)/\sigma_n$ in the superconducting state for each mode vs the induced-current weightings in Fig.~\ref{fig:sigma_1_peak_vs_w_sup_mat}. Additionally, we plot the value of $T/T_c$ for which $\sigma_1$ has its maximum in Fig.~\ref{fig:t_sigma_1_peak_vs_w_sup_mat}. These demonstrate that the typical behavior of $\sigma_1$ for most modes is to peak just below $T_c$ and to decrease as temperature decreases, and the exceptional behavior is to increase into a maximum at zero-temperature which is greatly enhanced above the normal state value of $\sigma_1$. These exceptional modes occur for directions of current flow corresponding to the directions of large residual loss, which are indicated in Fig.~\ref{fig:Rs_0_vs_w}. The residual loss can also be interpreted using $\sigma_{1,0}$, the zero-temperature residual value of $\sigma_1$, which we determine from the zero-temperature results $\lambda_0$ and $R_0$ of our power-law fits to $\lambda$ and $R_s$ respectively. We plot $\sigma_{1,0}$ as a function of current-weighting direction in Fig.~\ref{fig:sigma_1_0_vs_w_sup_mat}, which shows peaks in the same locations that Fig.~\ref{fig:Rs_0_vs_w} does but with different relative magnitudes. The inset shows the frequency dependence of $\sigma_{1,0}$, which does not appear to show a systematic frequency dependence for typical (low loss) modes, similarly to the $R_0/\omega^2$ plot in the inset of Fig.~\ref{fig:Rs_0_vs_w}. We note that many of the modes have residual $\sigma_{1,0}$ and $R_0/\omega^2$ values below the universal d-wave (line-nodal) level.  This suggests that the residual losses for these modes may be associated with point nodes.  However, the other modes with large loss compared to the universal line-nodal value imply the existence of another loss mechanism, perhaps that associated with Weyl nodes, and new possible absorption mechanisms, as discussed in \S\ref{sec:discussion} below.

	%---------------------------------

\subsection{Superfluid density}
\label{sec:analysis_max_rho_s}
Related to the $\sigma_2(T)$ and $\lambda(T)$, the superfluid density $\rho_s(T)=\sigma_2(T)/\sigma_2(0)=(\lambda(0)/\lambda(T))^2$ is a measure of excitations out of the ground state, and has theoretical predictions for its temperature dependence \cite{Prozorov06}. For an axial nodal superconductor, a power-law temperature dependence is predicted for $\rho_s(T)$ \cite{Gross86},
\begin{equation}
    \rho_s(T)=1-\eta_{\rho}\left(\frac{T}{T_c}\right)^{\alpha_{\rho}},
    \label{eq:rho_s_pow_law_SM}
\end{equation}
where $\eta_{\rho}$ is related to the impurity scattering rate $\Gamma_{imp}$ and maximum, zero-temperature superconducting gap $\Delta_{\text{max.}}$ only for $\alpha_{\rho}=2$, which would indicate currents parallel to the nodes, or $\alpha_{\rho}=4$, which would indicate currents perpendicular to the nodes. These predictions for $\eta_{\rho}$ are \cite{Gross86},
\begin{equation}
\begin{split}
    \alpha_{\rho}&=2\ (||\ \text{axial point nodes}):\\
    \eta_{\rho}&=\left(\frac{1}{1-3(\frac{\pi}{2} \ln{2} -1)\gamma}\right)\left(\frac{\pi^2}{1-\frac{\pi}{2}\gamma}\right)\left(\frac{k_B T_c}{\Delta_{\text{max.}}}\right)^2 \\
    \alpha_{\rho}&=4\ (\perp\ \text{axial point nodes}):\\
    \eta_{\rho}&=\left(\frac{1}{1-3(1 - \frac{\pi}{8} - \frac{\pi}{4} \ln{2})\gamma}\right)\left(\frac{7\pi^4/15}{(1-\frac{\pi}{2}\gamma)^3}\right)\left(\frac{k_B T_c}{\Delta_{\text{max.}}}\right)^4,
\end{split}
\end{equation}
where $\gamma=\Gamma_{imp}/\Delta_{\text{max.}}$. In these expressions, $\gamma<2/\pi$. Note that given $\Delta_{\text{max.}}$ and $T_c$, there is a minimum $\eta_{\rho}$ necessary to have a solution for $\gamma$. We fit our superfluid density data to Eq.~\ref{eq:rho_s_pow_law_SM} with $\eta_{\rho}$ as a fitting parameter for the cases of $\alpha_{\rho}=2,4$ as well as with $\alpha_{\rho}$ as another fitting parameter. We assume a maximum, zero-temperature gap $\Delta_{\text{max.}}=0.29$ meV to determine $\Gamma_{imp}$. We find that with $\alpha_{\rho}=2$, there is no solution for $\Gamma_{imp}$ for either sample, but there is a solution for $\alpha_{\rho}=4$ for both samples. We do not find any systematic dependence of $\Gamma_{imp}$ on the induced-current weights, so we average among the modes of each sample to obtain $\Gamma_{imp}=0.054\pm0.011$ meV for sample B39 and $\Gamma_{imp}=0.057\pm0.008$ meV for sample B40 for the $\alpha_{\rho}=4$ fits. These fits, however, are very poor compared with the $\alpha_{\rho}=2$ and $\alpha_{\rho}$ as a free parameter fits. For the fits with both $\eta_{\rho}$ and $\alpha_{\rho}$ as free parameters, we find no systematic dependence of $\eta_{\rho}$ or $\alpha_{\rho}$ on the induced-current weights. We find ($\alpha_{\rho}=1.58\pm0.07$, $\eta_{\rho}=0.87\pm0.25$) for sample B39 and ($\alpha_{\rho}=1.67\pm0.20$, $\eta_{\rho}=1.02\pm0.29$) for sample B40.
%Related to the $\sigma_2(T)$, the plot of $(\lambda(0)/\lambda(T))^2$ at low temperatures can be treated as the superfluid density temperature dependence for the lowest energy excitations out of the ground state.  There is a fair bit of theoretical machinery built to describe this behavior, and we can extract information like the impurity scattering rate from comparisons to those predictions.  See the Results section of Seokjin's Nat. Comm. paper.

%\textcolor{red}{This might be a good place to add a plot of superfluid density vs. T and discuss fits to it.  $\rho_s(T)=1-\frac{1}{1-3\frac{\Gamma_{imp}}{\Delta_0(0)}(\frac{\pi}{2}ln2 -1)} \frac{\pi^2}{1-\frac{\pi \Gamma_{imp}}{2\Delta_0(0)}} (\frac{k_B T}{\Delta_0(0)})^2$.  This is for currents in the point-nodal direction of the axial state with impurities.  There is another expression for current flowing perpendicular to the point nodes, with a $(\frac{k_B T}{\Delta_0(0)})^4$ dependence, Eq. (3.21) of Gross (1986) \cite{Gross86}.}  

\begin{figure}	%---------------------------------
\hspace*{-0.5cm}\includegraphics[scale=0.23,width=0.5\textwidth]{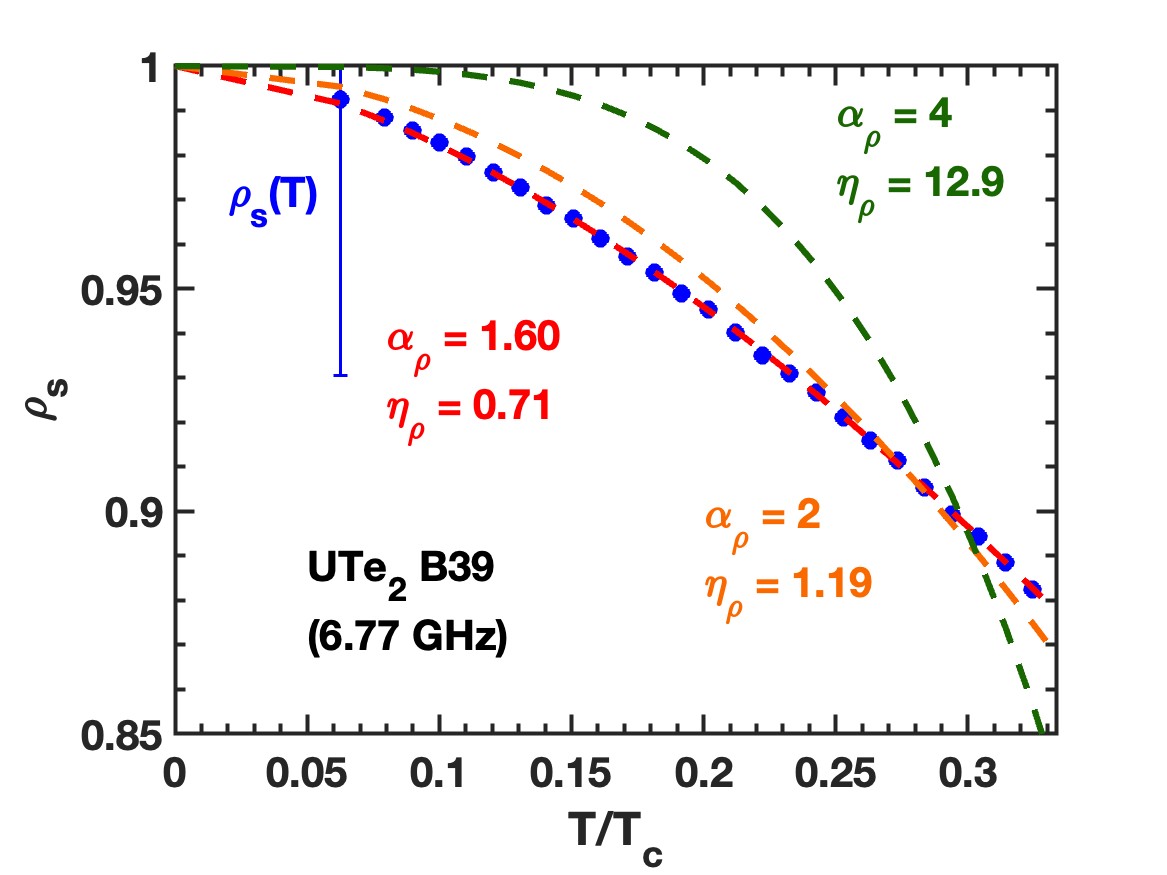}
\caption{Plot of normalized superfluid density  $\rho_s(T)$ vs $T/T_c$ data with error bars for just the first point. Power-law fits for Eq.~\ref{eq:rho_s_pow_law_SM} with only $\eta_{\rho}$ free are shown in orange for $\alpha_{\rho}=2$ and green for $\alpha_{\rho}=4$. A power-law fit with both $\eta_{\rho}$ and $\alpha_{\rho}$ free is shown in red.}
\label{fig:rho_s_fits}
\end{figure}	%---------------------------------

\begin{figure}[!ht]	%---------------------------------
% \centering 
\begin{subfigure}[b]{0.4\textwidth}
% \centering
\caption{}
\vspace*{-0.7cm}
\hspace*{-2cm}
\includegraphics[scale=0.23,width=1.0\textwidth]{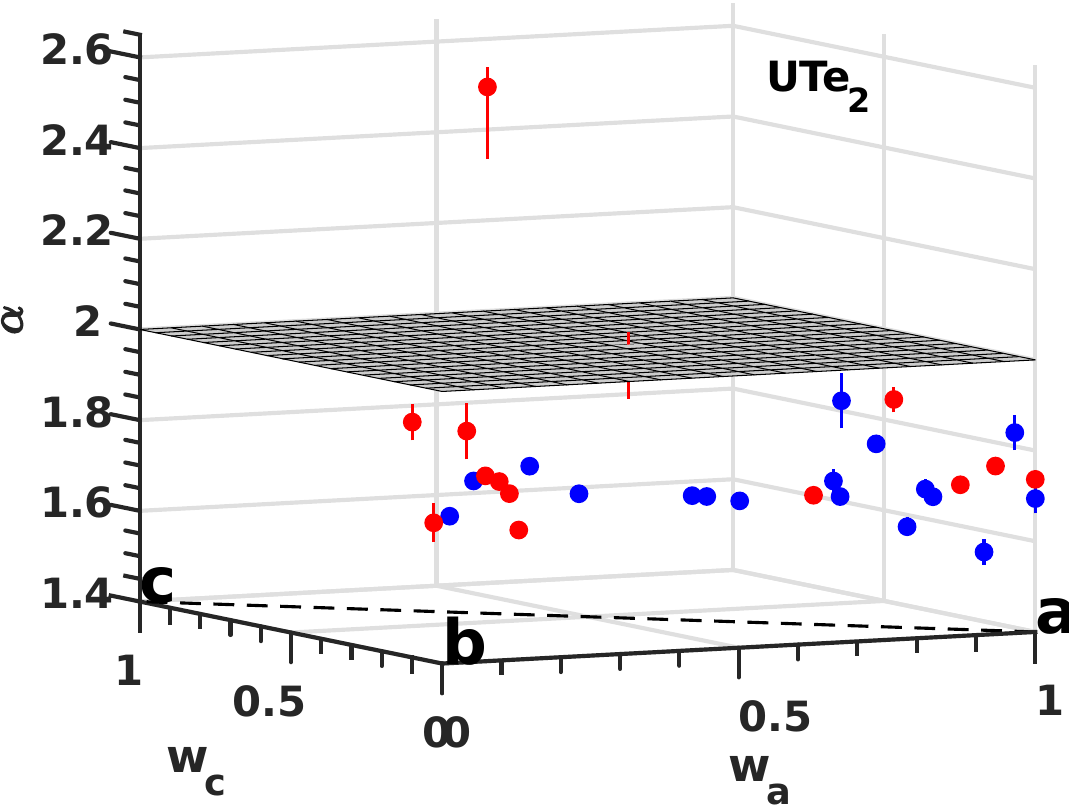}
\label{fig:alpha_vs_w}
\end{subfigure}
\begin{subfigure}[b]{0.4\textwidth}
% \centering
\vspace*{0.6cm}
\caption{}
\vspace*{-1.4cm}
\hspace*{-2cm}
\includegraphics[scale=0.23,width=1.0\textwidth]{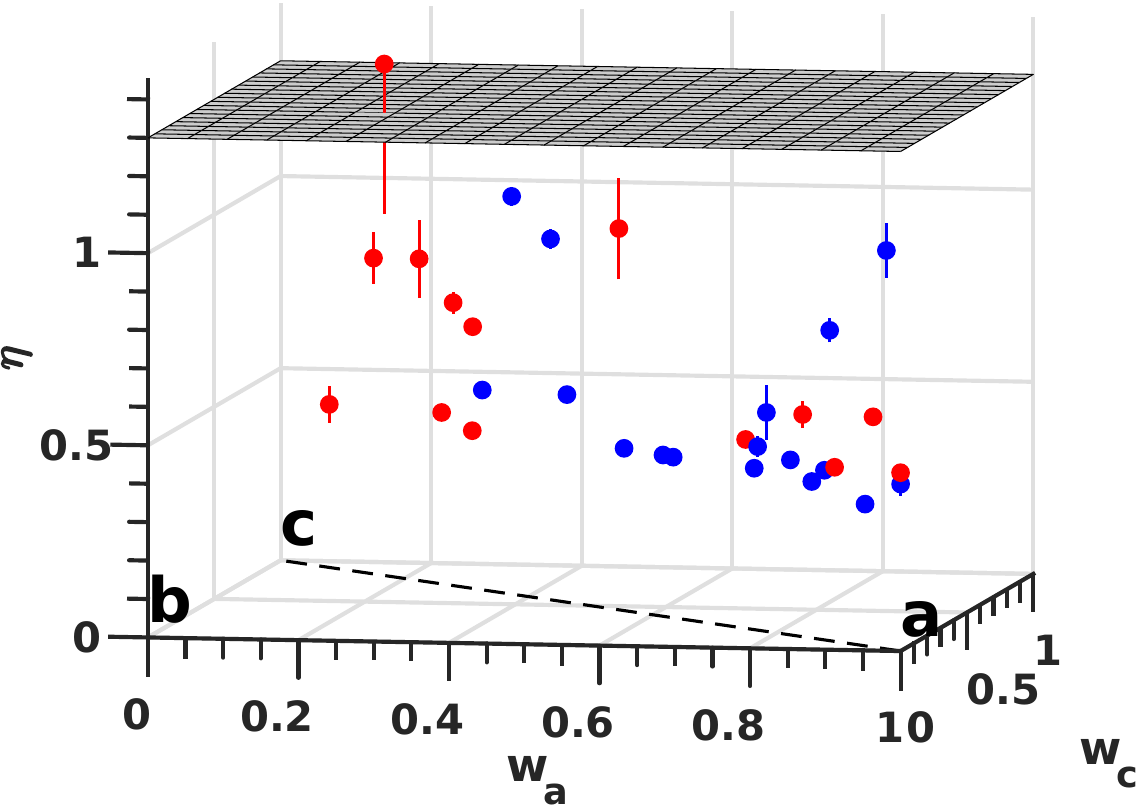}
\label{fig:eta_vs_w}
\end{subfigure}
\begin{subfigure}[b]{0.4\textwidth}
% \centering
\vspace*{-0.3cm}
\caption{}
\vspace*{-0.7cm}
\hspace*{-2.3cm}
\includegraphics[scale=0.23,width=1.0\textwidth]{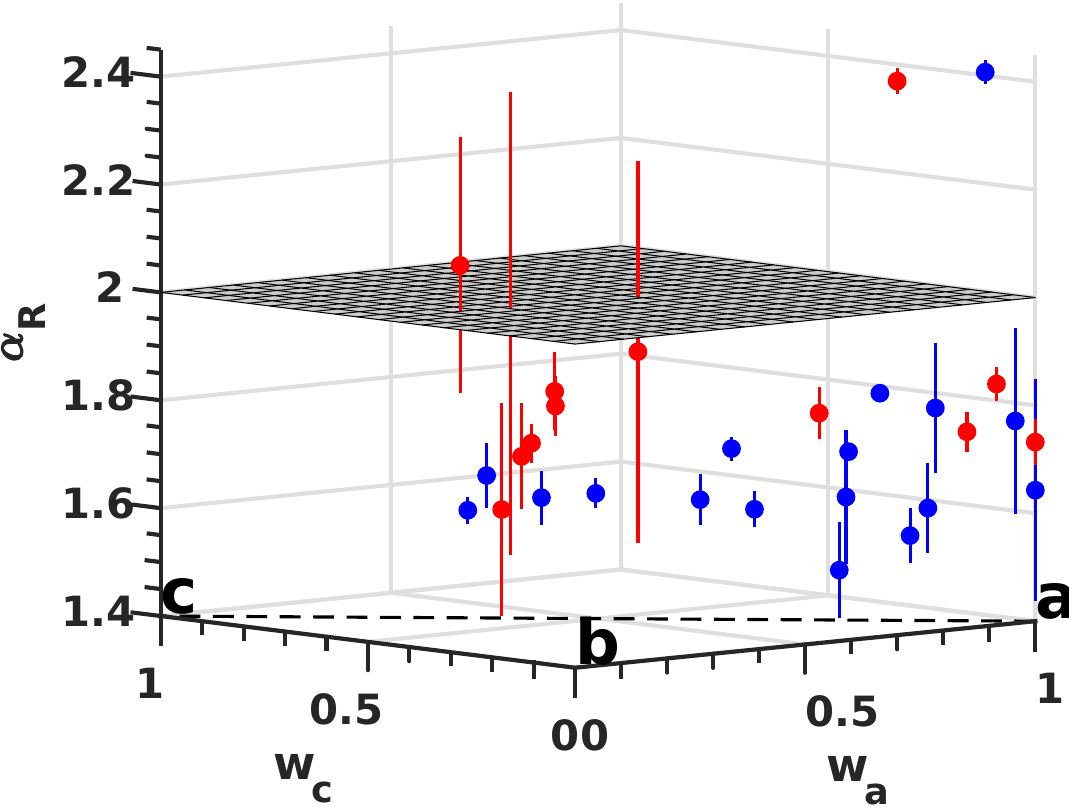}
\label{fig:alpha_R_vs_w}
\end{subfigure}
\caption{Plots of (a) penetration depth power-law exponent $\alpha$ and (b) coefficient $\eta$, as well as (c) surface resistance power-law exponent $\alpha_R$ vs current-flow contribution of each crystallographic axis. UTe$_2$ B39 and B40 are shown in blue and red, respectively. The simple p-wave predictions for penetration depth with currents parallel to the point nodes are indicated.}
\label{fig:w_dep}
\end{figure}	%---------------------------------

\begin{figure}[!ht]	%---------------------------------
\hspace*{-0.5cm}\includegraphics[scale=0.23,width=0.5\textwidth]{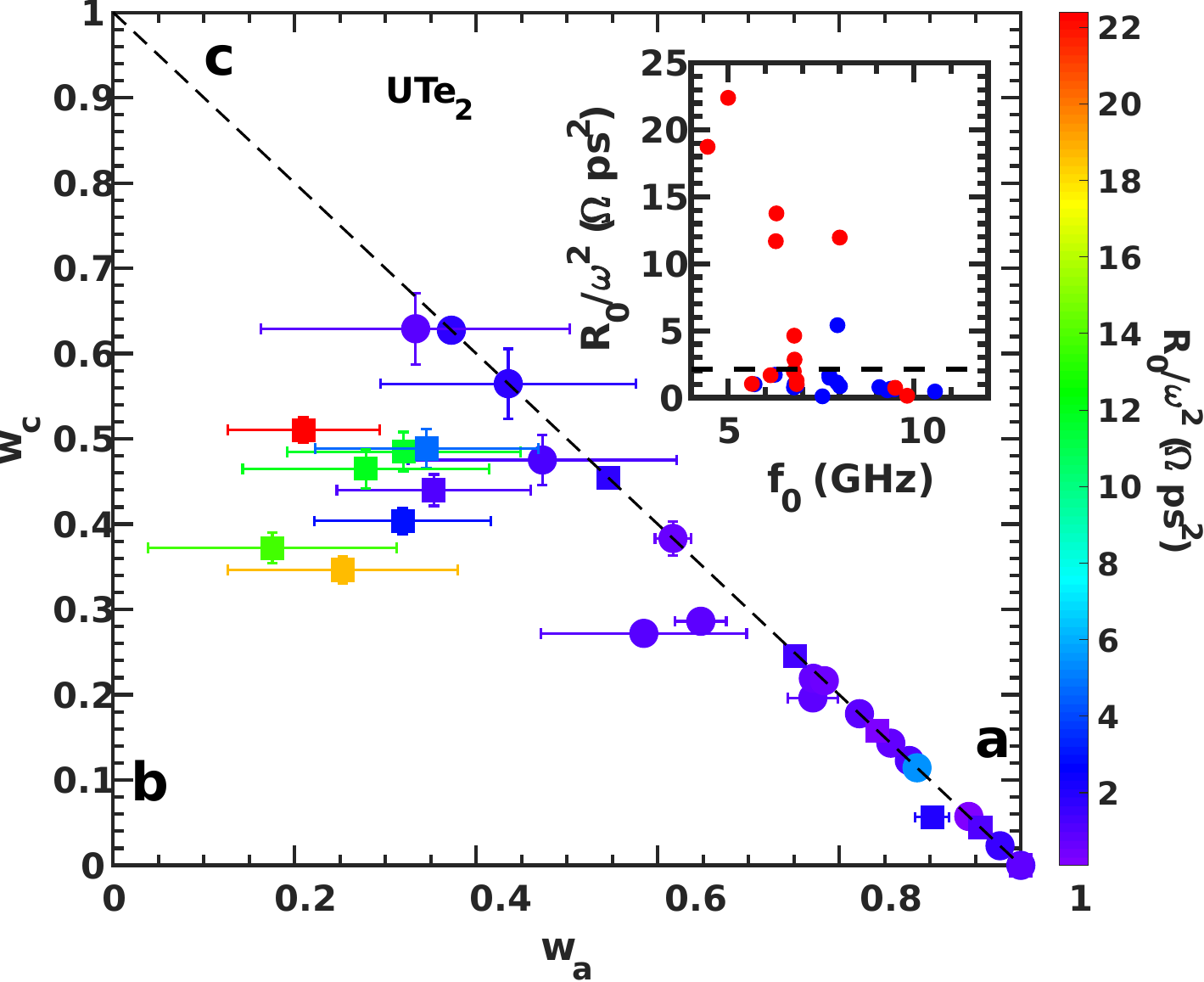}
\caption{Plot of nominally frequency independent zero-temperature loss $R_0/\omega^2$ vs current-flow contribution of each crystallographic axis. UTe$_2$ B39 and B40 are shown as circles and squares, respectively.  The colors indicate the value of $R_0/\omega^2$ in units of $\Omega$ ps$^2$. The inset shows $R_0/\omega^2$ vs frequency with UTe$_2$ B39 and B40 shown in blue and red respectively. The estimate of the universal d-wave loss is also included in the inset as the dashed line.}
\label{fig:Rs_0_vs_w}
\end{figure}	%---------------------------------

\section{Discussion/ Interpretation}
\label{sec:discussion}
%\paragraph{Discussion}
We plot the power-law exponent $\alpha$ from Eq.~(\ref{eq:lambda_pow_law}) for composite $\lambda(T)$ data after fitting for two UTe$_2$ samples in Fig.~\ref{fig:alpha_vs_w}. The power-law coefficient $\eta$ is similarly presented in Fig.~\ref{fig:eta_vs_w}. Finally, the power-law exponent $\alpha_R$ from Eq.~(\ref{eq:Rs_pow_law}) for composite $R_s(T)$ data is shown in Fig.~\ref{fig:alpha_R_vs_w}. The error bars shown in Fig.~\ref{fig:w_dep} represent the uncertainty in the determination of the fit parameters, which is discussed in \S\ref{sec:SM_impedance_fitting},\ref{sec:lambda_fits}. Typical values of these uncertainties are also summarized in Table~\ref{tab:fitting_uncerts} in \S\ref{sec:SM_impedance_fitting}. In addition to the $\alpha=2, \eta=1.3$ prediction for currents parallel to point nodes shown as planes in Fig.~\ref{fig:alpha_vs_w}, \ref{fig:eta_vs_w}, Figs.~\ref{fig:alpha_vs_w_sup_mat},\ref{fig:eta_vs_w_sup_mat} in \S\ref{sec:SM_param_summary} show the perpendicular-to-nodes prediction of $\alpha=4, \eta=2.1$ \cite{Klemm88}. 

Our $\alpha$ and $\alpha_R$ power-law exponent data shows no systematic dependence on the crystallographic contributions, with average values less than 2 for both samples. Likewise, $\eta$ does not show a systematic dependence on the induced-current weights; however, $\lambda_0$ does show a systematic dependence, which can be seen in Fig.~\ref{fig:lambda_0_vs_w_sup_mat} in \S\ref{sec:SM_param_summary}. Note that there is one mode with much larger $\alpha$ and $\eta$ than the rest of the modes, which we discuss further in \S\ref{sec:SM_param_summary}. Since our composite values of $\alpha$, $\eta$, and $\alpha_R$ show no systematic dependence on our estimates of the axis weights, we present average values of $\alpha$, $\eta$, and $\alpha_R$ for each sample as well as the standard deviation of the modes as a measure of the uncertainty of these averages in Table \ref{tab:lambda_and_loss}. Considering these uncertainties, $\alpha$ and $\alpha_R$ are practically equivalent.
%Since these fluctuations of $\alpha$ between modes for both UTe$_2$ samples shown in Table~\ref{tab:lambda_and_loss} are comparable to the typical fitting uncertainty for $\alpha$ in Table~\ref{tab:fitting_uncerts} of \S\ref{sec:SM_impedance_fitting}, these variations are certainly not significant. While the variations in $\eta$ are much larger than the uncertainty in its determination, it does not appear to show a systematic dependence on the induced-current weights. 
The other quantities show a systematic dependence on the crystallographic responses, so we present the range of values observed for all modes measured. We compare our results with that of other researchers in Table~\ref{tab:lambda_and_loss}. In the previous measurements with this resonator, $\alpha\approx 2$ was found for composite $\lambda$ for the $f_0=11.26$ GHz mode with a CVT1 sample \cite{Seokjin21}. Axis-resolved $\alpha$ values have been determined using tunnel-diode oscillator screening measurements \cite{Ishihara23}, which find $\alpha\lesssim 2$ for CVT1 samples and a MSF1 grown sample. These measurements from \cite{Ishihara23} determine $\alpha$ for each of the three crystallographic axes but are not sensitive to the surface resistance or $\sigma_1$.

The results for $\alpha$ and $\eta$ shown in Figs.~\ref{fig:alpha_vs_w}, \ref{fig:eta_vs_w} are inconsistent with a single pair of point nodes (simple p-wave) in the weak scattering limit. We would expect the data to be somewhere between the parallel ($\alpha =2$) and perpendicular ($\alpha=4$) predictions with only a few modes close to the direction of the nodes matching the parallel-to-nodes results, while many more modes would produce the perpendicular-to-nodes results since the directions perpendicular to the nodes form a line rather than a point in phase space. These results are also inconsistent with a single line node in the weak scattering limit, for which $\alpha=1$ and $\alpha=3$ are expected parallel and perpendicular to the line node, respectively \cite{Klemm88}. However, our data could be consistent with either a line node or a pair of point nodes in the case of unitary scattering. For both these cases, $\alpha\approx 2$ is expected. Further, $\eta\approx 0.5-0.7$ is expected for point nodes with unitary scattering, and $\eta\approx 0.8$ is expected for a line node with unitary scattering, \cite{Klemm88}, both of which are consistent with our data. The presence of multiple pairs of point nodes slightly off of one or more of the high symmetry axes could also be consistent with these results \rev{\cite{Chris25}}, which is used to explain the results of \cite{Ishihara23}.  We note that a nearly isotropic $\Delta \lambda \sim T^{1.9-2}$ was observed in PrOs$_4$Sb$_{12}$ and attributed to multi-domain orientations of a point nodal order parameter \cite{Chia03}.

Note that we have also fit the penetration depth to the `dirty d-wave' and nonlocal models (see \S\ref{sec:lambda_alt_fits}), however the quality of these fits are markedly inferior to the power-law fits (see Table  \ref{tab:fitting_uncerts}). A similar situation was encountered in CeCoIn$_5$, where $\Delta \lambda(T) \sim T^{\alpha}$ with $\alpha=1.5$ was observed \cite{Ozcan03}.  In that case it was proposed that Fermi liquid renormalization took place below T$_c$, and the unusual power-law was attributed to an increase in quasiparticle effective mass with decreasing temperature, and this conclusion was strengthened by further experiments \cite{Truncik13}.  In the case of UTe$_2$ there is no evidence for non-Fermi liquid behavior, hence this explanation for the power-law exponents seems unlikely.

Another possible explanation for the behavior seen in Fig.~\ref{fig:alpha_vs_w} is the presence of Weyl nodes in a topological superconducting state. Although broken time-reversal invariance is not observed in recent generation UTe$_2$ samples, theory suggests that different irreducible representations (e.g. B$_{2u}$ and B$_{3u}$) may be nearly degenerate, leading to a mixed order parameter \cite{Choi24}.  Alternatively, recent theory suggests that a combined Kondo/spin-liquid model can show a single-step-transition into a chiral topological superconducting state \cite{Cole24}. The theory for a topological Weyl superconductor predicts the existence of a gapless 2D Majorana fermion surface fluid with unique electrodynamic properties.  First, a Weyl node superconductor is expected to show power-law exponent $\alpha=2$ for the penetration depth for currents in all directions \cite{Foster21}, which is qualitatively consistent with our data and that of \cite{Seokjin21,Ishihara23}, though some measured values of $\alpha$ are considerably less than 2. Secondly, a new absorption mechanism is predicted to exist involving absorption of a photon from a surface state causing a transition to an empty bulk state. This mechanism is predicted to be active at zero temperature and to show a characteristic anisotropy, being larger for electric fields polarized parallel to the Weyl nodes compared to the perpendicular orientation. This result is predicted to be robust to disorder as well \cite{Foster21}. % The direction of the Weyl nodes is discussed below.
\rev{Predictions of surface states in UTe$_2$ due to Dirac Majorana modes have also been put forward \cite{Roising24,Chris25}.  Finally, recent STM measurements on UTe$_2$ show evidence of zero-energy Andreev conductance on the $(0\bar{1}1)$ surface termination \cite{Gu25}.}

We use $R_0/\omega^2$ as a measure of the residual loss because we expect it to be roughly frequency independent (as observed in CeCoIn$_5$ \cite{Ormeno02}, for example), and its determination requires minimal assumptions and data processing. See \S\ref{sec:SM_res_loss} for more details. For multiple modes, the observed losses are larger than the intrinsic residual loss predicted for a line-nodal d-wave superconductor with the same maximum energy gap ($R_0^{line-nodal}/\omega^2\approx 2.1\ \Omega\ ps^2$), but for most modes, the residual loss is less than the d-wave prediction, indicating that our residual loss data is not consistent with line nodes. See \S\ref{sec:SM_res_loss} for further analysis. We also plot $R_0/\omega^2$ (in color) in Fig.~\ref{fig:Rs_0_vs_w}. 
%Error bars for the induced-current weights are also included here to estimate the uncertainty in determining these fitting parameters, as discussed above and summarized in Table~\ref{tab:fitting_uncerts} in \S\ref{sec:SM_impedance_fitting}. These error bars are omitted from Figs.~\ref{fig:alpha_vs_w},\ref{fig:eta_vs_w}. 
This plot shows a systematic dependence of residual loss on the induced-current weights which peaks for current flow in between the b- and c-axes. The maximum frequency-independent loss we measured occurred for $w_i=(0.21,0.28,0.51)$ with UTe$_2$ B40. This could indicate the presence of Weyl nodes near this direction, considering the uncertainties. We also note that for UTe$_2$ B39, the maximum frequency-independent loss occurred at $w_i=(0.89,0.00,0.11)$, but this peak is much smaller than that of sample B40 and only consists of one mode. These results are also included in the summary in Table~\ref{tab:lambda_and_loss}.  
%\textcolor{red}{Independent thermal conductivity measurements in a field of similar UTe$_2$ crystals (JP group) suggest that \textit{line} \textcolor{blue}{(are you sure?)} nodes exist in the b-direction.[reference]}

\section{Conclusions}
%\paragraph{Conclusions}
In conclusion, we have used resonant cavity perturbation techniques to study the electrodynamic properties of two CVT grown UTe$_2$ single crystals with $T_c$ near 2 K. We are able to determine the composite surface impedance and magnetic penetration depth for a variety of microwave-frequency modes using minimal analysis and assumptions. We find a robust sub-$T^2$ power-law in the low temperature ($T\le T_c/3$) magnetic penetration depth and surface resistance, which is mostly independent of mode and fitting parameters used to determine the impedance. These results do not agree with a single pair of point nodes under weak scattering. Our results are more consistent topological Weyl superconductivity, especially considering the large resisdual loss we observe which peaks between the b- and c-directions with a relatively small a-axis contribution.

\section{Acknowledgments}
The work of AC-J and SMA was supported by NSF-DMR/2004386, and ARO/FSDL under grant W911NF-24-1-0153.  The work of AS, YSE, IMH, SRS, JP and NPB was supported by the U.S. Department of Energy Award No. DE-SC-0019154 (sample characterization), the Air Force Office of Scientific Research under Grant No. FA9950-22-1-0023 (materials synthesis), and the Gordon and Betty Moore Foundation’s EPiQS Initiative through Grant No. GBMF9071, the National Institute of Standards and Technology, and the Maryland Quantum Materials Center. SRS acknowledges support from the National Institute of Standards and Technology Cooperative Agreement 70NANB17H301.

\appendix

\section{Samples and Resonator Measurements}
\label{sec:App_general}
Here we discuss sample preparation methods and the selection of resonant modes of the cavity that reveal the anisotropic electrodynamic properties of the samples.
\subsection{Sample preparation}
\label{sec:SM_sample_prep}
The UTe$_2$ samples used in this experiment are first shaped using coarse polishing paper while exposed to air. For the B40 sample, this was grit 1000, P2500, and for B39, it was a similar grit. The goal of this procedure is to remove jagged edges and craters, as well as to produce a roughly rectangular shape, but with no specific crystallographic orientation. The 6 surfaces are then polished again using fine polishing paper in a nitrogen-filled glove bag with $\le 0.1\%$ $\text{O}_2$ by volume. For the B40 sample, this was 3 $\mu$m, and for B39, it was a similar grit. The purpose of this procedure is to remove any surface oxides on the crystal.  However, all uranium oxides are known to be insulating \cite{Seokjin21}, although it is not known if UTe$_2$ oxides share this property. Before removing it from the glove bag, the sample is encased in N-grease, an electrically inert, viscous material at room temperature, which solidifies at cryogenic temperatures. This protects the polished surfaces from oxidization. The prepared UTe$_2$ samples have side lengths of roughly $0.5-1$ mm.

We also study a NbSe$_2$ sample provided by 2D Semiconductors as a comparison with UTe$_2$. This NbSe$_2$ sample is micaceous, allowing us to shape it by exfoliating layers with a scalpel to expose smooth ab-planes. The c-direction is evident from the mostly flat ab-planes of the as-received samples. A slight misalignment of our cut surfaces with the actual ab-plane could still be possible, though a precise alignment of the crystal is not necessary for our analysis. The crystal is also soft enough to cut planes in the perpendicular direction. The orientation of these cuts is arbitrary since NbSe$_2$ is isotropic in the ab-plane. We do not perform any polishing for this sample. The prepared NbSe$_2$ sample has side lengths of roughly $0.5$ mm in the a- and b-directions and $0.25$ mm in the c-direction.

Finally, we also compare with a Nb sample cut from a larger piece of a heat-treated Nb cavity resonator \cite{Turneaure68} using a hack-saw. This larger piece contains many grain boundaries, which we try to avoid in the sample we cut from it. This sample is first coarse polished with P1000 and P220 polishing paper to shape it and to remove the rough features from cutting. We also remove any grain boundaries from the single crystal portion we are trying to keep using this coarse polishing. We then perform fine polishing on the 6 sides with 3 $\mu m$ polishing paper. All of this polishing is done in air. The prepared Nb sample has side lengths of roughly $1$ mm.

\subsection{Selection of perturbed resonant cavity modes}
\label{sec:SM_mode_selection}
The sample is introduced into a hollow cylindrical, rutile dielectric resonator \cite{Seokjin21} to perturb its resonances. The sample is attached to the end of a sapphire-rod `hot finger' \cite{Sridhar88,Rubin88,Mao95,Trunin98} using Apiezon N-grease, which holds it on the symmetry axis in the middle of the resonator and allows heat to be conducted from a heater outside the resonator cavity directly to the sample over a range from 100 mK to 20 K, while isolating it from the copper walls of the cavity and the rutile, which are nominally held at 100 mK. We inject a microwave signal between 2 and 12 GHz with a small loop on one side of the cavity which inductively couples to the modes of the cavity. A hollow rutile dielectric cylinder is present in the center of the cavity, which closely surrounds the sample. This concentrates the electromagnetic fields near the sample for some resonant modes, enhancing the sample filling fraction. The transmitted signal is similarly received by a second loop on the opposite side of the cavity. A Low Noise Factory cryogenic low noise microwave amplifier LNF-LNC6\_20B s/n 413B at 4 K is used to amplify the output of the transmitted signal. At 5 K, this amplifier has a reported gain of roughly 35 dB and noise temperature of $5-12$ K from $4-20$ GHz. This 2-port system can then be described by a $2\cross2$ scattering matrix. The transmission component $S_{21}$ of the scattering matrix is measured, and the resonant frequency $f_0$ and quality factor $Q$ are determined by fitting \cite{Petersan98}. The resonant properties are measured both with the sample present in the cavity $(f_{0,\text{tot.}},Q_{\text{tot.}})$ and with only residual N-grease on the sapphire hotfinger after the sample has been removed $(f_{0,r},Q_r)$, which allows us to isolate an effective $\Delta f_{0\rev{,\text{sample}}}$ and $Q_{\rev{\text{sample}}}$ arising from the sample alone \cite{Mao95,Seokjin21}, 
%\textcolor{red}{[Cite these equations from Seokjin - see page 112 of his thesis]},
\begin{equation}
\begin{split}
    \Delta f_{0\rev{,\text{sample}}}(T) &=\Delta f_{0,\text{tot.}}(T) - \Delta f_{0,r}(T) \\
    1/Q_{\rev{\text{sample}}}(T) &= 1/Q_{\text{tot.}}(T) - 1/Q_r(T),
\end{split}
\label{eq:background}
\end{equation}
where $\Delta$ denotes the change in a quantity from its value at the lowest measurement temperature $T_0\sim 150$ mK.

The geometry factor $G$ has dimensions of resistance and relates the field structure at the sample location to that of the rest of the resonator for each mode. Specifically, 
\begin{equation}
    G=\mu_0\omega\int_{\text{cavity}}|\textbf{H}(\textbf{r})|^2 d^3\textbf{r}/\oint_{\text{sample}}|\textbf{H}(\textbf{r})|^2 d^2\textbf{r},
    \label{eq:Gdefinition}
\end{equation}
where $\textbf{H}$ is the magnetic field, the integral in the numerator is over the cavity volume, and the integral in the denominator is over the sample surface. Note that the value of $G$ is unique to each resonant cavity mode and each sample.  Here we assume that the sample creates a predominantly magnetic perturbation to the resonant modes considered.

Our resonator is similar to a closed cylindrical cavity with perfect electric conducting (PEC) boundaries, which has a known, infinite sequence of resonant modes. Our cavity is not empty, but also contains a concentric, dielectric rutile cylinder, sapphire `hot finger', and sapphire substrate that supports the rutile, which add to the electromagnetic volume of the cavity. In the absence of a sample, the observed resonances are perturbations of these due to copper walls rather than PEC, the inductive coupling loops, and the loss in the dielectrics. We additionally apply heat directly to the sapphire `hot finger' to vary its temperature, though this heat radiates and diffuses into other components of the system as well. These perturbations can shift the resonant frequencies and add loss to the system, changing the quality factors of the modes. These effects can vary from mode to mode. The introduction of an inductive coupling with the loops additionally selects a subset of these eigen-modes which can be induced by this particular choice of coupling. We observe $\sim 300$ modes of the cavity in the $2-12$ GHz range. The measured transmission spectrum tends to break into clusters of several modes, which are typically separated by $50-100$ MHz. Within these clusters, modes are typically separated by $5-20$ MHz.

\begin{figure}  %---------------------------------
% \centering
\hspace*{-0.5cm}\includegraphics[scale=0.23]{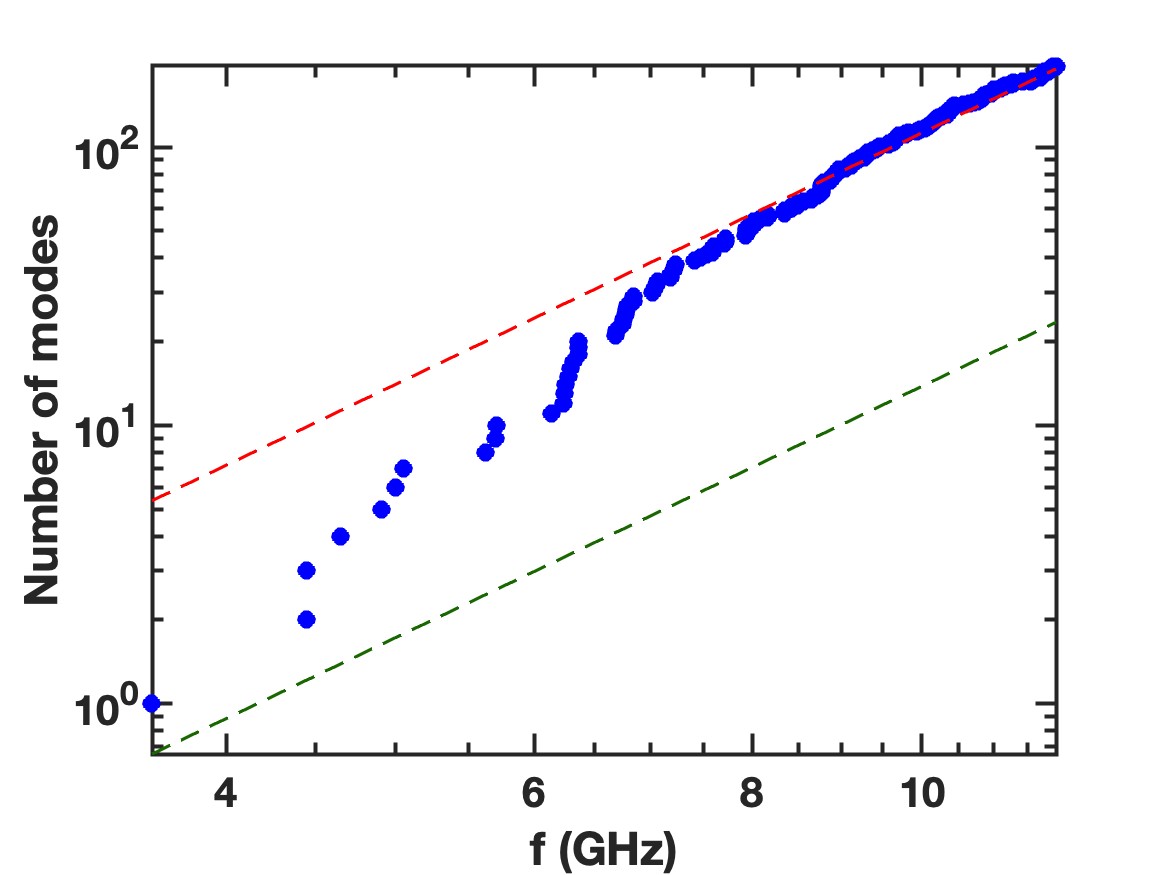}
\caption{Plot of the number of measured resonant modes in the cavity vs frequency in blue. The green dashed line is the naive Weyl estimate of $N(f)$ due to the cavity and dielectrics. The red dashed line is a fit to a $N(f)\sim f^3$ power-law for high frequencies.}
\label{fig:num_modes_vs_freq_sup_mat}
\end{figure}	%---------------------------------

The Weyl estimate of the number of resonant modes up to frequency $f$ in a three-dimensional cavity with effective volume $V$ is given by $N(f)=\frac{8\pi}{3}V(f/c)^3$, where $c$ is the speed of light in vacuum \cite{Hill09}. We show the number of modes vs frequency for our resonator in the absence of a sample in Fig.~\ref{fig:num_modes_vs_freq_sup_mat}. This estimate of about 300 modes at 12 GHz is an underestimate as some modes are too weak or close together in frequency to be recognized or distinguished in the $S_{21}$ measurement.  The internal volume of the resonant cavity in the absence of any dielectrics is $3.375\times 10^{-5}\ m^3$. The dielectrics inside the cavity increase the effective volume of the cavity, given in the asymptotic limit as $V_{EM} = \sum_{i=1}^{N_d} \sqrt{\epsilon_{r,i}}\ V_i$, where each of the $N_d$ dielectrics of permittivity $\epsilon_{r,i}$ occupies volume $V_i$. We take the dielectrics to be ($\epsilon_r=10$ and $V_{sapphire\ substrate}=1.134\times 10^{-6}\ m^3$ and $V_{sapphire\ hot\ finger}=1.178\times 10^{-8}\ m^3$ for sapphire), ($\epsilon_{r,eff}=157$ \cite{Tobar98} and $V_{rutile}=7.147\times 10^{-7}\ m^3$ for rutile), and ($\epsilon_r=1$ and $V_{vacuum}=3.189\times 10^{-5}\ m^3$ for vacuum).  This results in an effective electromagnetic volume of $V_{EM}=4.446\times 10^{-5}\ m^3$. We plot the Weyl estimate for the number of modes up to a given frequency corresponding to this volume as the green dashed line in Fig.~\ref{fig:num_modes_vs_freq_sup_mat}.

We estimate the actual effective volume of the cavity to be $3.622\times 10^{-4}\ m^3$ by fitting the observed number of modes to the Weyl estimate for frequencies above 7 GHz. We also plot this estimate as the red dashed line in Fig.~\ref{fig:num_modes_vs_freq_sup_mat}.  The observed asymptotic scaling of the number of modes, $N(f) \sim f^3$, assures us that we are coupling to a significant fraction of the modes of the three-dimensional cavity. 

The inclusion of a conductive sample will additionally perturb these resonances due the exclusion of electromagnetic fields from its bulk. The purpose of the dielectric rutile cylinder, which surrounds the sample and part of the sapphire `hot finger', is to concentrate fields near the sample to increase the intensity of the fields which are excluded relative to those of the rest of the cavity, known as its filling fraction. The perturbed frequency $\omega$ due to this field-exclusion effect is given by \cite{Slater46},
\begin{equation}
    \omega^2=\omega_a^2 \left( 1 + \int_{V_p} (H_a^2-E_a^2)\ dV \right),
\end{equation}
where $\omega_a$, $E_a$, and $H_a$ are the unperturbed frequency, electric field, and magnetic field, and $V_p$ is the volume of the perturbation which excludes the electromagnetic fields. From this, it can be seen that modes with a dominant magnetic field at the sample location should experience a positive frequency shift due to sample insertion, while a dominant electric field would result in a negative frequency shift. If the electric and magnetic fields are comparable, then the perturbation to the resonant frequency could be very small. For each of the samples we have measured, we observe cases of both left (lower) and right (higher) frequency shifts due to sample insertion, including some comparatively, very small shifts. We observe shifts due to sample insertion which are typically between 1 MHz to lower frequency and 2 MHz to higher frequency. This indicates that the samples tend to have larger magnetic perturbations than electric perturbations.

\begin{figure}	%---------------------------------
\includegraphics[scale=0.23]{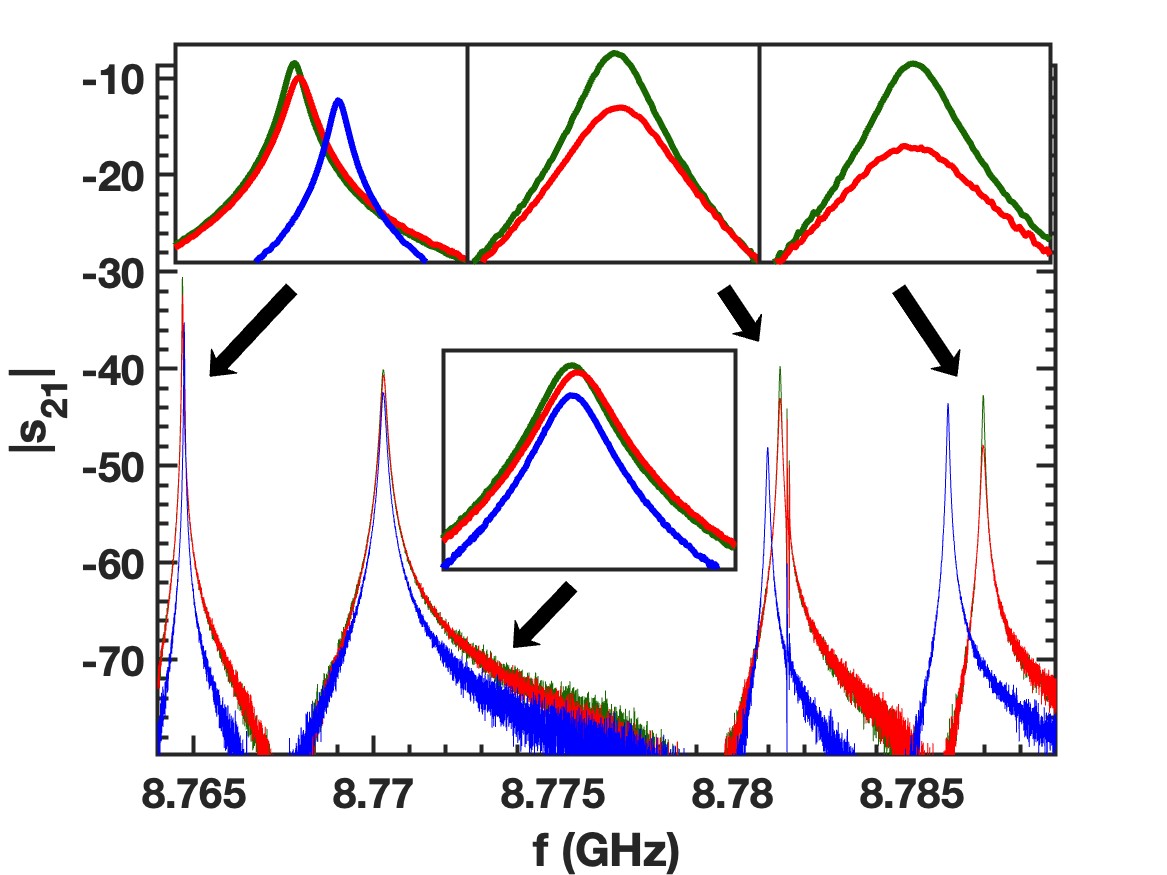}
\caption{Plot of the different types of perturbations to the transmission magnitude $|S_{21}|$ through the resonator. The blue curve shows the resonator with no sample. The green and red curves shows the resonator with the UTe$_2$ B39 sample in the superconducting and normal states respectively. The upper left inset shows the first peak at 8.765 GHz which has a conventional electric perturbation. The lower inset shows the second peak at 8.770 GHz which is not significantly perturbed by the sample. The upper middle inset shows the third peak at 8.781 GHz which has an anomalous magnetic perturbation. The upper right inset shows the last peak at 8.787 GHz which has a conventional magnetic perturbation.}
\label{fig:pert_examples}
\end{figure}	%---------------------------------

As the sample is heated, more of the fields should be admitted back into the volume of the sample as the magnetic penetration depth increases. This should be especially evident for the magnetic perturbations as a superconductor is heated above $T_c$, though electric fields also tend to be screened more effectively at lower temperatures. For both the electric and magnetic cases, the perturbation due to heating is then expected to oppose that of sample insertion. We observe shifts due to sample heating from the superconducting state at base temperature to the normal state at 4 K typically from 20 kHz down to 40 kHz up in frequency, though shifts $\sim 100$ kHz in both directions have been observed. The increase in frequency heating shifts (electric) tend to be larger than the down in frequency heating shifts (magnetic), which is contrary to the larger magnetic perturbations we observed due to sample insertion. For most modes, the direction of the perturbation due to heating opposes that of sample insertion; however, some modes show the opposite behavior. This likely indicates a weak coupling of the mode to the sample and the dominance of the temperature dependence of the other materials in the cavity, such as the sapphire `hot finger'. We classify electric and magnetic modes by a down or up shift due to sample insertion, respectively. If the shift due to heating the sample is in the same direction, we refer to this behavior as anomalous. 

Fig.~\ref{fig:pert_examples} shows some examples of the various resonant mode perturbation types for UTe$_2$ sample B39. The blue curve shows the modes of the resonator when no sample is present. The green and red curves shows the modes of the resonator with the UTe$_2$ B39 sample in the superconducting and normal states, respectively. The first mode in this section is electrically perturbed, which can be seen in the upper left inset. The second mode is shown in the bottom inset, which has almost no dependence on the sample insertion. Note that this shift due to sample insertion is much smaller than that of the electric perturbation compared with their corresponding temperature dependences. The third mode has an anomalous magnetic perturbation. The upper middle inset shows the two sample cases, for which heating causes an upward shift in frequency. There are additionally two very fine resonances just off the right side of the third peak which show no significant perturbation due to the sample. The fourth mode of this selection has a conventional magnetic perturbation. Note the substantial drop in quality factor for the two magnetic perturbation cases, indicating a strong coupling to the sample, as well as the clearly visible perturbations due to sample insertion for these modes, which are much larger than that of the electric perturbation. 

We study a selection of modes in two parts. First we measure the modes with the sample present (sample run), then remove the sample and measure again with just residual N-grease on the sapphire 'hot finger' (background run). Each run constitutes a complete cycling of the cryostat to base temperature and back to room temperature. Since the sample run is first, we can only use the effect of sample heating to inform us of each mode's coupling strength to the sample. Only once we do the background run, do we know the actual strength of the sample perturbation and whether these are electrically or magnetically dominated. To select the modes to measure, we look for the largest downward frequency shifts, and largest quality factor drops, with increasing temperature. This method is predisposed to select magnetically perturbed modes. Once we perform the background run, we find that all modes we are able to use for surface impedance studies do have magnetic perturbations.
%Once we measure the resonator without the sample for the purpose of background subtraction, we find that all modes we are able to use for surface impedance studies do have magnetic perturbations.

After measuring a background run with residual N-grease, we performed two subsequent background runs without N-grease on the sapphire `hot-finger' to see the effect of the N-grease and the reproducibility of resonant frequency and quality factor measurements after cycling the system. Between two runs without N-grease, we removed an re-inserted the sapphire `hot finger' as we would need to do when changing samples. This case represents the minimum possible disruption to the system between actual measurements. We found a distribution of frequency shifts of the modes from the first to the second clean sapphire `hot finger' cases, with a mean of 3 MHz among all modes. We also found a corresponding mean inverse quality factor shift of $-7\times 10^{-7}$. We average the frequency and inverse quality factor between the two clean sapphire 'hot finger' cases for each mode to get an average behavior for the resonator without N-grease. The residual N-grease caused a mean frequency shift of $-2$ MHz from this average background without N-grease and a corresponding mean inverse quality factor shift of $6\times 10^{-7}$. This indicates that cycling the system can cause frequency shifts comparable to those of sample insertion, and that the N-grease acts like a lossy electric perturbation on average.

\section{Determination of Surface Impedance by fitting}
\label{sec:App_imp_fitting}
Here we discuss the fitting used to determine the surface impedance from resonance data and independently measured DC resistivity data.  The goal is to determine the values of the parameters $G$ and $X_0$, unique to each resonant mode of each sample.  Another objective is to determine the current-weighting directions for each unique mode and sample.  These parameters then determine the magnetic penetration depth tensor for each sample. 

\subsection{UTe$_2$ Anisotropic Resistivity data}
\label{sec:SM_rho_data}
Our objective is to create reliable fitting functions for the normal-state resistivity tensor of orthorhombic UTe$_2$ crystals. 
This tensor will be used to determine the current-weighting factors from measurements of the sample surface impedance in the normal state between 2 K and 20 K, where the three components of the resistivity tensor have distinctly different temperature dependencies.

Montgomery geometry samples can be very accurately oriented by cleaving, but the signal is very weak at low temperatures, resulting in an unreliable determination of the resistivity below a certain temperature, which is about 6 K for this data. On the other hand, Hall (bar) geometry samples have the potential for a small amount more misalignment but can produce better signal at low temperatures, which allowed Yun-Suk Eo to confirm a $T^2$ power-law for the resistivity in each crystallographic direction at low temperatures down to $T_c$ \cite{Yunsuk22}.

We fit each resistivity function $\rho_i(T)$ of the Montgomery geometry data to a quadratic polynomial $\rho_{i,1}(T) = A^{(i)} + B^{(i)} T^2$ for $2\ K \lesssim T < T_0$ and to a fifth order polynomial $\rho_{i,2}(T) = \sum_{n=0}^5 C_n^{(i)}T^n$ for $T_0 < T < 20\ K$ for some cross-over temperature $T_0$. We model the resistivities as,
\begin{equation}
\rho_i(T)=
    \begin{cases}
        \rho_{i,1}(T), & T \le T_0 \\
        \theta(T)\rho_{i,1}(T) + [1-\theta(T)]\rho_{i,2}(T), & T_0 < T \le T_1 \\
        \rho_{i,2}(T), & T > T_1, \\
    \end{cases}
\end{equation}
where $\theta(T)=\frac{T_1-T}{T_1-T_0}$ is a weighting function to merge the two different polynomial temperature dependences between the temperatures $T_0$ and $T_1$. A summary of the parameters used in this model are given in Table~\ref{tab:rho_poly}.

\begin{table}
    \begin{center}
    \caption{Polynomial fit parameters used to model Montgomery geometry resistivity data $\rho(T)$ from Ref.~\cite{Yunsuk22} for UTe$_2$ samples.}
    \begin{tabular}{ | c | c | c | c | }
    $T_0\ (K)$ & \multicolumn{3}{c}{$6$} \\
    $T_1\ (K)$ & \multicolumn{3}{c}{$8$} \\
    \hline
     & $\rho_a$ & $\rho_b$ & $\rho_c$ \\ 
    \hline
    $A\ (\mu\Omega\ m)$ & $0.126$ & $0.550$ & $0.621$ \\
    $B\ (\mu\Omega\ m/K^2)$ & $7.60\times 10^{-3}$ & $2.36\times 10^{-2}$ & $6.49\times 10^{-2}$ \\
    $C_0\ (\mu\Omega\ m)$ & $1.10$ & $4.06$ & $16.2$ \\
    $C_1\ (\mu\Omega\ m/K)$ & $-0.495$ & $-1.63$ & $-7.65$ \\
    $C_2\ (\mu\Omega\ m/K^2)$ & $0.100$ & $0.290$ & $1.48$ \\
    $C_3\ (\mu\Omega\ m/K^3)$ & $-7.76\times 10^{-3}$ & $-1.89\times 10^{-2}$ & $-0.120$ \\
    $C_4\ (\mu\Omega\ m/K^4)$ & $2.89\times 10^{-4}$ & $5.61\times 10^{-4}$ & $4.41\times 10^{-3}$ \\
    $C_5\ (\mu\Omega\ m/K^5)$ & $-4.18\times 10^{-6}$ & $-6.37\times 10^{-6}$ & $-6.13\times 10^{-5}$
    \label{tab:rho_poly}
    \end{tabular}
    \end{center}
\end{table}

\subsection{Surface Reactance constraints}
\label{sec:SM_Xs_constraints}
In general, for a metal $X_s(T) \ge R_s(T)$, assuming that the surface impedance is determined by the complex conductivity alone \cite{Trunin00}. (One exceptional case to the expectation that $X_s>R_s$ in the superconducting state has been noted for odd-frequency pairing in a chiral p-wave superconductor \cite{Bakur18}.) In the treatment of cavity perturbation data of superconducting samples, it is often assumed that $R_s(T) = X_s(T)$ in the normal state \cite{Seokjin21,Trunin00}; however, this assumption can result in $X_s(T)<R_s(T)$ or even $X_s(T)<0$ for some temperatures in the superconducting state. This occurs
when $X_s(T)$ changes faster than $R_s(T)$ in the superconducting state. To avoid this, $X_s(T)>R_s(T)$ must instead be allowed in the normal state, requiring a larger value of $X_0$. We interpret this as being due to a finite non-zero value of $\omega\tau$ in the normal state. It is additionally possible to have $X_s(T) > R_s(T)$ if the normal state screening length is greater than the relevant dimension of the crystal \cite{Gough94}. Alternatively, the excessive change in the reactance in the superconducting state could be attributed to effects beyond the complex conductivity, such as surface roughness or thermal expansion \cite{Trunin00}. In the cuprate superconductors, surface roughness of the crystal on length scales larger than $\lambda(T)$ at some temperatures between 0 and $T_c$ would cause a decrease in the reactance \cite{Trunin00}.  Thermal expansion is known to have a significant effect on $X_s(T)$ (but not $R_s(T)$) in the cuprates at $T \gtrsim 60$ K requiring corrections \cite{Trunin00}.

\begin{figure}  %---------------------------------
\begin{subfigure}[b]{0.4\textwidth}
% \centering
\caption{}
\hspace*{-0.5cm}\includegraphics[scale=0.3,width=1.0\textwidth]{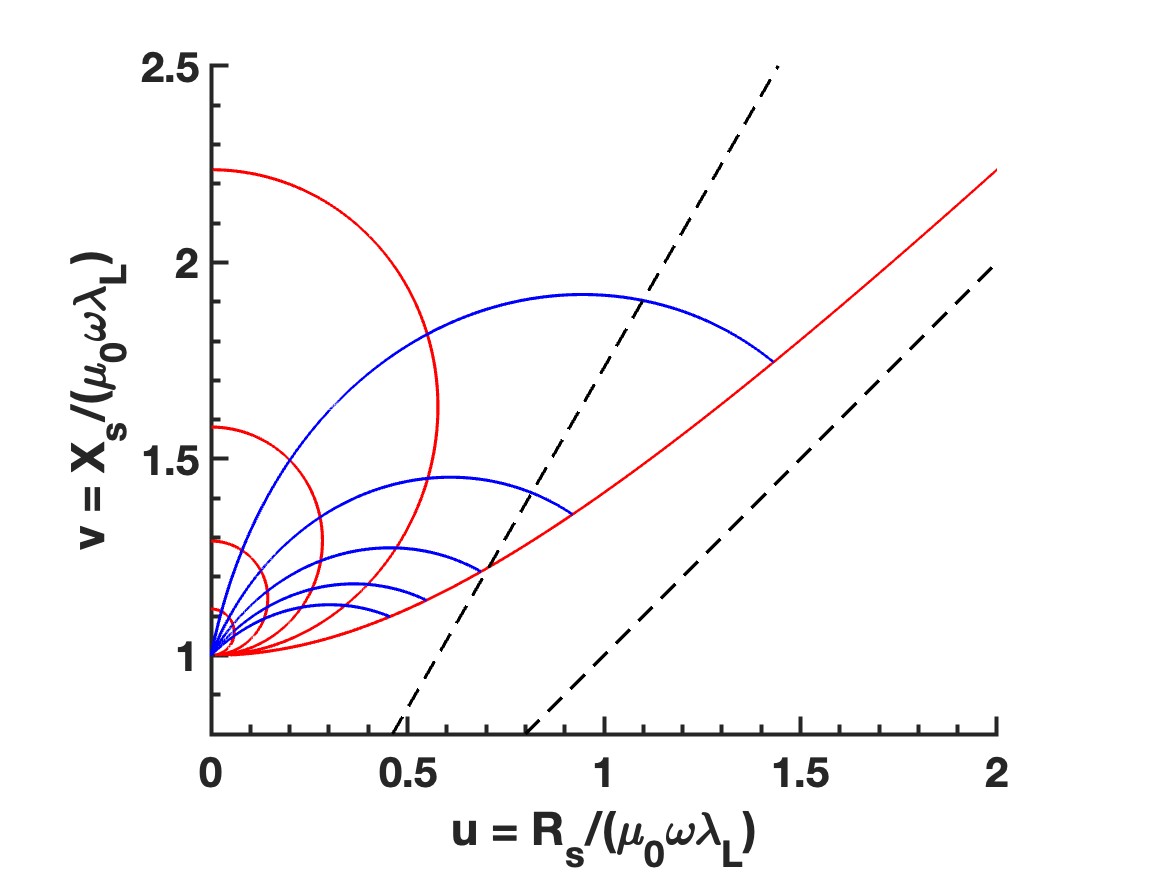}
\label{fig:Zs_contours_sup_mat}
\end{subfigure}
\begin{subfigure}[b]{0.4\textwidth}
% \centering
\caption{}
\hspace*{-0.5cm}\includegraphics[scale=0.3,width=1.0\textwidth]{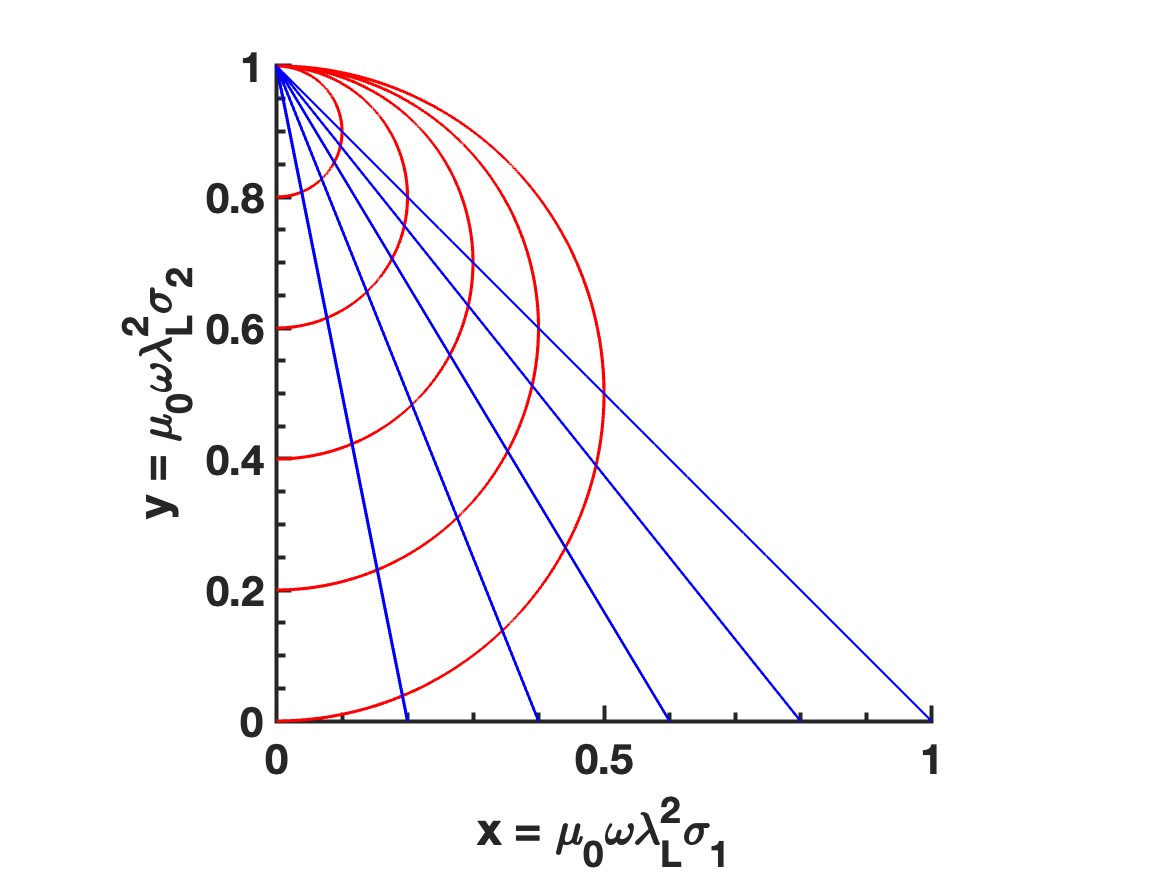}
\label{fig:sigma_contours_sup_mat}
\end{subfigure}
\caption{Complex geometry of the (a) surface impedance $Z_s=R_s+iX_s$ and (b) complex conductivity $\sigma=\sigma_1-i\sigma_2$, where $\lambda_L$ is the London penetration depth and $\omega$ is angular frequency. The axes are dimensionless versions of (a) $R_s, X_s$, and (b) $\sigma_1, \sigma_2$.  Curves of constant superfluid fraction $f_s = 0,0.2,0.4,0.6,\text{and}\ 0.8$ are shown in red and constant scattering time $\omega\tau=0.2,0.4,0.6,0.8,\text{and}\ 1.0$ are shown in blue. The lower dashed line in (a) shows the condition \rev{$X_s=R_s$}, and the upper dashed line shows \rev{$X_s=\sqrt{3}R_s$}. To find the direction of increasing $f_s$, follow any curve of constant $\omega\tau$ towards $(0,1)$ and vice versa.}
\label{fig:Zs_and_sigma_geometry_sup_mat}
\end{figure}

Assuming a single Drude conductivity peak, it can be seen from the expression for the normal state scattering time in an isotropic superconductor, $\omega\tau=(X_s^2-R_s^2)/(2R_s X_s)$, that $X_s \ge R_s$ is required to maintain $\omega\tau \ge 0$. Assuming an isotropic two-fluid model, in the normal state (with superfluid fraction $f_s=0$), $Z_s$ is constrained by the hyperbola $X_s^2-R_s^2 = (\mu_0\omega\lambda_L)^2$ as shown in the right-most red curve in Fig.~\ref{fig:Zs_contours_sup_mat}. Values of $f_s>0$ and the $\sigma$ complex plane will be discussed in \S\ref{sec:2f_model} below. Note that the London penetration depth $\lambda_L$ can be temperature dependent, in which case, the constraint still holds, but it is not a true hyperbola. For example, it was observed that the effective mass of quasiparticles in CeCoIn$_5$ is temperature dependent in the superconducting state \cite{Ozcan03,Truncik13}. Other temperature-dependent changes in carrier density and/or effective mass include the development of Kondo hybridization of bands in heavy-Fermion superconductors. In the superconducting state, $Z_s$ lies above this hyperbola in the $R_s + iX_s$ complex plane. This analysis can be extended to the anisotropic case using Eq.~\ref{eq:Zs_comp} to obtain a composite of three hyperbolas which lies above the $R_s=X_s$ line since each component of the impedance tensor individually does. Likewise, the composite superconducting impedance lies above this composite of three hyperbolas. 

In the limit of non-local electrodynamics, it is predicted that $X_s(T) = \sqrt{3}R_s(T)$ in the normal state \cite{Reuter48}. Since we assume our data is in the local limit, we constrain the value of $X_0$ is such that $X_s(T)\le \sqrt{3}R_s(T)$ in the normal state. For some modes, $X_0$ reaches the upper limit of this constraint, which could indicate the onset of non-local electrodynamics.

In the context of Fig.~\ref{fig:Zs_contours_sup_mat}, the $X_0/G$ degree of freedom corresponds to vertically shifting the data in this complex plane. For all temperatures, the data must always lie above the lower dashed line, but more strongly, it must always lie on or above the outer-most red curve (hyperbola). The upper limit of $X_0$ is shown as the upper dashed line. In the normal state, the data would also need to lie below this slope $\sqrt{3}$ line.

	%---------------------------------

\subsubsection{A note on $T_c$}
%We determine $T_c$ by eye from the point in the insets of Fig.~\ref{fig:lambda_pow_laws} where $R_s(T)/G$ drops suddenly with decreasing temperature.
We determine $T_c$ by eye from the temperature where $R_s(T)/G$ drops suddenly with decreasing temperature. At this temperature, $X_s(T)/G$ can begin to increase into a reactance peak, as discussed in \S\ref{sec:SM_reactance_peak} below. We have observed the same critical temperature across different resonant modes measured for a given sample, and we do not have a conclusive indication of multiple superconducting transitions/phases for the samples used in these experiments, so we will use a single value for $T_c$. Alternatively, $T_c$ could be determined from Fig.~\ref{fig:Zs_and_sigma_geometry_sup_mat} by finding the temperature (parametric variable of these plots) at which the data departs from the outer-most red curve ($f_s=0$) into the region of $f_s>0$. This is also demonstrated for data in Fig.~\ref{fig:sigma_plane_fit}.

\subsection{Impedance fitting}
\label{sec:SM_impedance_fitting}
To determine $G$ and $X_0$ and estimate the response of surface currents corresponding to each crystallographic axis for each mode, we fit to,
\begin{equation}
    \sqrt{R_s(T) X_s(T)}=w_a R_a(T) + w_b R_b(T) + w_c R_c(T),
    \label{eq:geo_mean_comp_SM}
\end{equation}
for temperatures $T$ above $T_c$. This analysis relies on knowledge of the temperature dependences of the resistivity tensor, as discussed in \S\ref{sec:SM_rho_data}, to determine $R_i(T)=\sqrt{\mu_0\omega\rho_i(T)/2}$. Note that the left-hand-side of Eq.~\ref{eq:geo_mean_comp_SM} assumes a single composite scattering time $\tau(T)$ in contradistinction to our use of an anisotropic resistivity tensor. Eq.~\ref{eq:geo_mean_comp_SM} has five parameters: $X_0$ and $G$ (used to determine $R_s(T)$ and $X_s(T)$) and $w_a$, $w_b$, and $w_c$; however, there are only four free parameters due to the normalization constraint on $w_i$ ($\Sigma w_i = 1$ \cite{Kitano95}). This system of equations is nonlinear in $X_0$; however, for a fixed value of $X_0$, it can be made linear in the remaining three free parameters. To do this, we factor out $G$ and introduce un-normalized weight parameters $v_i=w_i/G$ and the dimensionless minimum temperature reactance $x=X_0/G$.  With this, the fit function becomes,
\begin{equation}
    \sqrt{\frac{1}{Q_{\rev{\text{sample}}}(T)}\left(\frac{-2\Delta f_{0\rev{,\text{sample}}}(T)}{f_{0,\text{tot.}}(T_0)} + x \right)} = \sum_{i=a,b,c} v_i R_i(T).
    \label{eq:geo_mean_lin}
\end{equation}
Once the three $v_i$ are determined, we can use the normalization condition on $w_i$ to determine $G=\left(\sum_i v_i \right)^{-1}$, which allows us to convert back to the original parameters. Each measured temperature $T_i$ gives an instance of Eq.~(\ref{eq:geo_mean_lin}) with the same four parameters $x$, $v_a$, $v_b$, and $v_c$ to be determined. Since the number of measured temperatures in the normal state $N_{T,N}\sim 200$, this system is overdetermined. In matrix form, this system becomes $\bm{M}\Vec{v}=\Vec{F}(x)$, where $M_{ij}=R_j(T_i)$ and $F_i(x)=\sqrt{\frac{1}{Q_{\rev{\text{sample}}}(T_i)}\left(\frac{-2\Delta f_{0\rev{,\text{sample}}}(T_i)}{f_{0,\text{tot.}}(T_0)} + x \right)}$. We make a grid of $x$ values within the constraints discussed in \S\ref{sec:SM_Xs_constraints} and uniquely determine $v_i$ which best approximates Eq.~(\ref{eq:geo_mean_lin}) in the least-squares sense for each individual value of $x$. This can result in one or more of $v_i$ having a negative value, which is not allowed since they are proportional to current weighting factors $w_i$. In such a case, we must consider induced-current weights due to a subset of the resistivity tensor components. For the case of only one resistivity tensor component, the corresponding $w_i=1$, so a valid solution is guaranteed. For each value of $x$, we consider the results of all seven possible combinations of one, two, and three allowed resistivity tensor components. We then choose the valid solution which results in the minimum least-squares difference between the left- and right-hand sides of Eq.~(\ref{eq:geo_mean_lin}). We finally minimize the least-squares cost function with respect to $x$ to determine the full solution. The cost function and the parameters of the fit are still continuous with respect to $x$ even when the number of allowed resistivity tensor components changes, though their derivatives are discontinuous. See the insets of Fig.~(\ref{fig:Zs_with_fit}) for an example of this optimization.

As the non-linear parameter $x=X_0/G$ is varied above and below the optimal value, the cost function increases as shown in the upper inset of Fig.~(\ref{fig:Zs_with_fit}). To estimate the uncertainty in $X_0$, we use the maximum (of left and right) distance from its optimal value for which the cost function increases by $3\%$. In this range of variation in $X_0$ for which the cost function increases by less than $3\%$, we take the maximum departure of $G$ and $w_i$ from their optimal values as their respective uncertainties. The lower inset of Fig.~(\ref{fig:Zs_with_fit}) shows an example of how $G$ and $w_i$ vary with the choice of $X_0$. The range of values of these fitting parameters between modes and samples as well as their uncertainty estimates are given in Table~\ref{tab:fitting_uncerts}. The fractional uncertainty for $G$ is relatively small compared with that of $X_0$ indicating that the quality of the fit is less sensitive to $G$ than to $X_0$. The uncertainties for $w_a$ and $w_b$ are greater than that of $w_c$, which is evident in the slopes of these quantities vs $X_0$ in the lower inset of Fig.~\ref{fig:Zs_with_fit}. This occurs because the distinctive temperature dependence of the surface impedance corresponding to the c-axis resistivity is not as strongly affected by $X_0$.

To reiterate, this fitting to determine the surface impedance from resonance data assumes the local limit, the Drude model, a composite scattering time, and the linear addition of impedances for each axis to form a composite microwave surface impedance. The Drude model with a composite scattering time also implies a composite resistivity $\rho_{dc}$, though we use the anisotropic and temperature-dependent resistivity tensor to determine the normal state surface resistance for each axis.  This analysis is performed on many resonant modes (a total of 17 for B39 and 14 for B40), giving a very complete picture of the electrodynamics of UTe$_2$.

\begin{table*}[t]
    \begin{center}
    \caption{A comparison of the range of values between different modes of several parameters and their uncertainties due to fitting for two UTe$_2$ samples.}
    \begin{tabular}{ | c | c | c | c | c | }
      & Min. & Max. & Uncertainty & Fractional uncertainty \\ 
      & (B39/B40) & (B39/B40) & (B39/B40) & (B39/B40) \\ 
     \hline
     $X_0\ (m\Omega)$ & 39/47 & 128/118 & $10/9$ & $0.1/0.1$ \\
     $G\ (k\Omega)$ & 3.7/1.4 & 26.7/27.6 & $0.6/0.5$ & $0.03/0.03$ \\
     $w_a$ & 0.33/0.18 & 1.00/1.00 & $0.07/0.09$ & $0.17/0.35$ \\
     $w_b$ & 0.00/0.00 & 0.14/0.45 & $0.06/0.08$ & $1.74/0.50$ \\
     $w_c$ & 0.00/0.00 & 0.63/0.51 & $0.02/0.01$ & $0.04/0.27$ \\
     $\alpha$ & 1.57/1.61 & 1.90/2.61 & 0.02/0.05 & 0.01/0.02 \\
     $\eta$ & 0.37/0.46 & 1.04/1.43 & 0.03/0.12 & 0.05/0.10 \\
     $\lambda_0\ (nm)$ & 638/998 & 2012/2418 & 1/4 & 0.001/0.002 \\
     $\alpha_R$ & 1.49/1.64 & 2.42/2.40 & 0.09/0.18 & 0.05/0.10 \\
     $\eta_R$ & 1.61/1.01 & 143.78/123.21 & 0.44/1.16 & 0.07/0.20 \\
     $R_0\ (m\Omega)$ & 0.23/0.54 & 13.5/30.2 & 0.07/0.18 & 0.03/0.04 \\
     $T_I/T_c$ & 0.435/0.535 & 3.76/4.02 & 1.75/4.03 & 0.71/1.91 \\
     $\beta$ & 0.44/0.60 & 2.60/4.01 & 1.42/3.98 & 0.59/1.60
     \label{tab:fitting_uncerts}
    \end{tabular}
    \end{center}
\end{table*}

\section{Determination of Scattering time $\tau$}
\label{sec:App_scatt_time}
Here we discuss multiple independent methods to estimate the scattering time below and above T$_c$.
\subsection{Analysis of Reactance peak $X_s(T)$ below $T_c$}
\label{sec:SM_reactance_peak}
The peak in surface reactance as a function of temperature just below $T_c$ is a consequence of the competition between the loss of normal screening, and the increase in superfluid screening, which act in quadrature, and lead to a net minimum in effective screening \cite{Hein01,Ormeno02,Ormeno06,Baker09,Baker10,Ghigo18}. 

Quantitative analysis of the reactance peak can be carried out, as suggested by Hein, Ormeno, and Gough \cite{Hein01b}.  By plotting the frequency shift of the microwave resonator against the inverse quality factor, with temperature $T<T_c$ as a parameter, one obtains a parameter-free plot of reactance change vs. resistance change.  In other words one plots $\xi(T) = (X_s(T)-X_s(T=0))/R_s(T_c)$ vs. $R_s(T)/R_s(T_c)$, with temperature as a parameter.  This quantity has a peak value above zero at temperature $T_*<T_c$.  Such a plot utilizes only raw data, and makes no assumption about the geometry factor $G$, minimum temperature reactance $X_0$, weights $w_i$, etc.  Its interpretation only assumes validity of a two-fluid behavior with a single effective quasiparticle momentum-relaxation time $\tau_{eff}$.  Ref.\ \cite{Hein01b} argues that this effective scattering time accounts for nonlocality and for the fact that it represents the quasiparticle relaxation time, rather than the Drude relaxation time $\tau$.  It is argued that $\tau_{eff}>\tau$. 

\begin{figure}  %---------------------------------
% \centering
%\hspace*{-0.5cm}
\hspace*{-0.2cm}\includegraphics[scale=0.23]{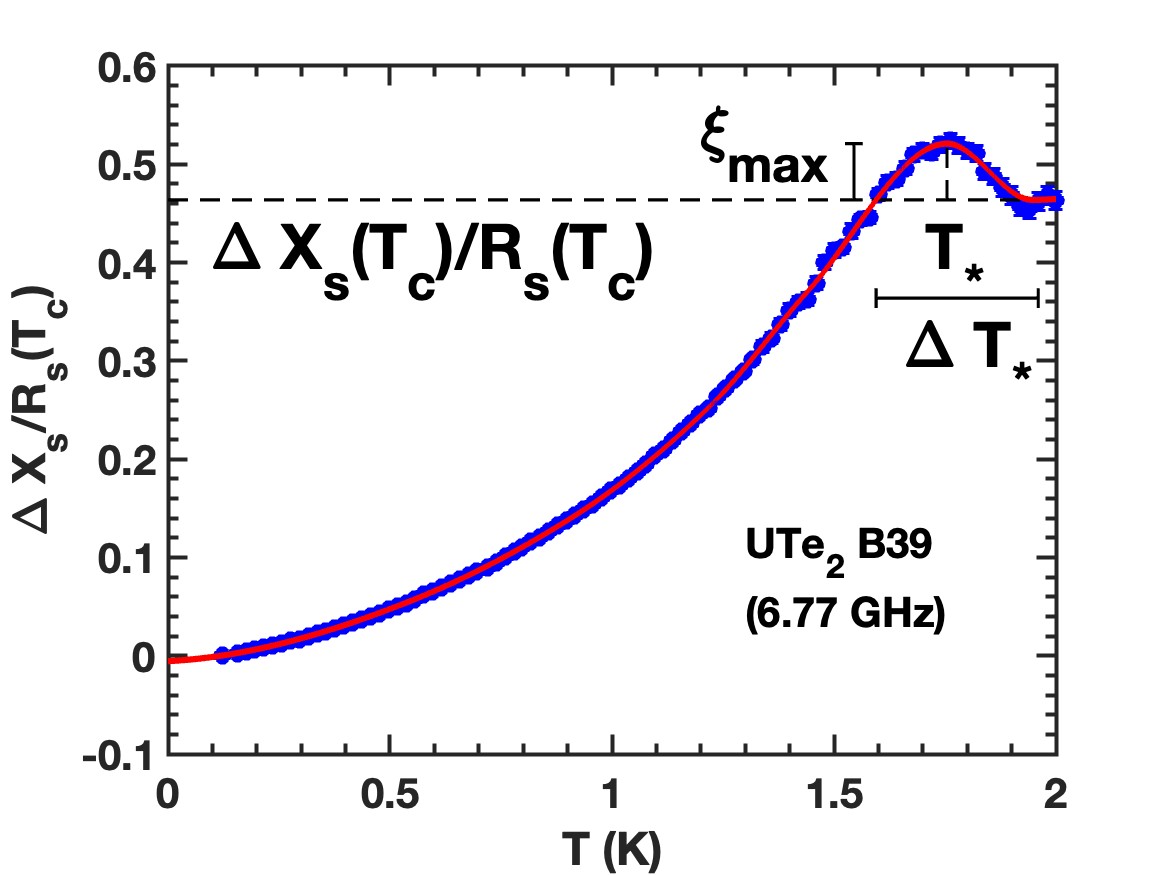}
\caption{Reactance peak data plotted as $\Delta X_s(T)/R_s(T_c)$ (utilizing only raw data) vs temperature is shown in blue. We define $\Delta X_s=0$ at the base temperature. The red curve shows a smoothing of this data, which is used to determine the reactance peak parameters $\xi_{max}$, $T_*$, and $\Delta T_*$, defined as shown based on the value of $\Delta X_s(T_c)/R_s(T_c)$.}
\label{fig:Xs_peak_vs_T_sup_mat}
\end{figure}

\begin{figure*}[t!]  %---------------------------------
\begin{subfigure}[b]{0.475\textwidth}
\caption{}
\hspace*{-0.5cm}\includegraphics[scale=0.23,width=1.0\textwidth]{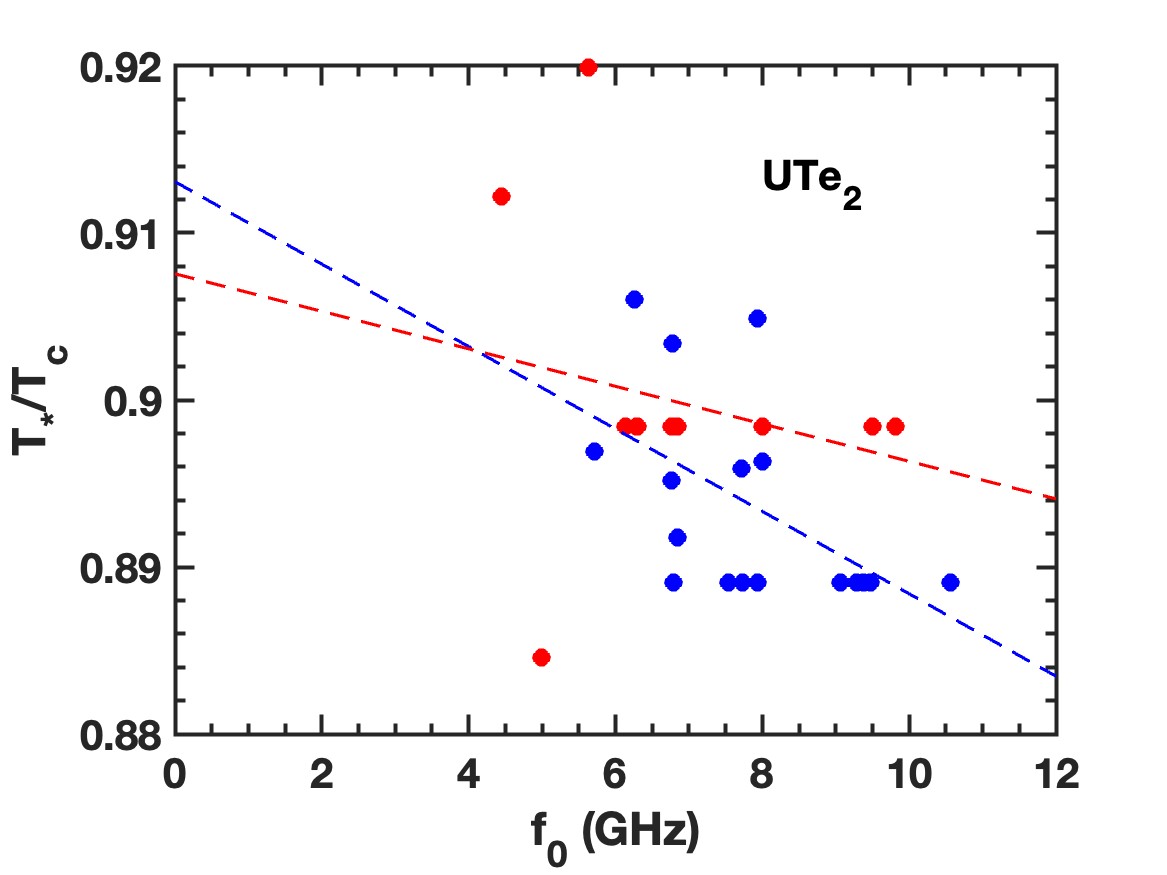}
\label{fig:t_peak_vs_f_sup_mat}
\end{subfigure}
\hfill
\begin{subfigure}[b]{0.475\textwidth}
\caption{}
\hspace*{-0.5cm}\includegraphics[scale=0.23,width=1.0\textwidth]{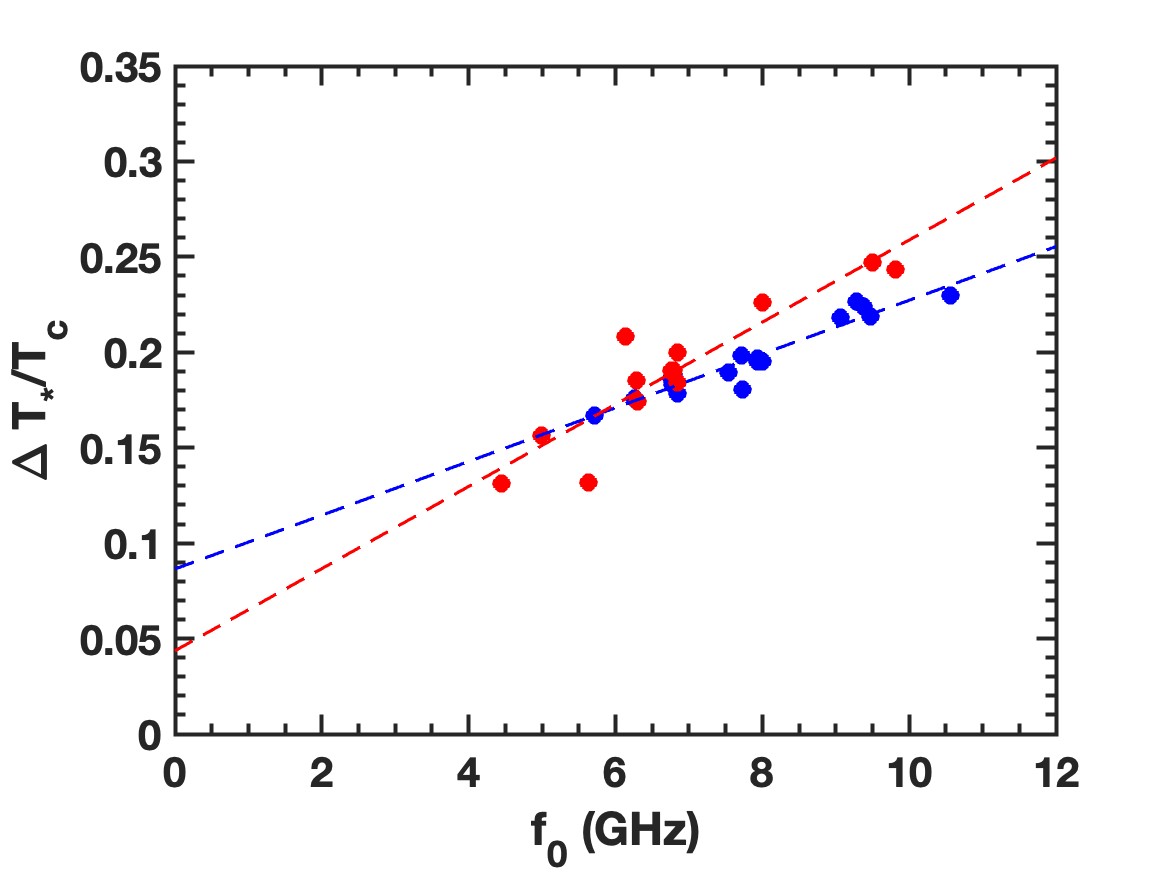}
\label{fig:t_width_vs_f_sup_mat}
\end{subfigure}
\vskip\baselineskip
\begin{subfigure}[b]{0.475\textwidth}
\caption{}
\hspace*{-0.5cm}\includegraphics[scale=0.23,width=1.0\textwidth]{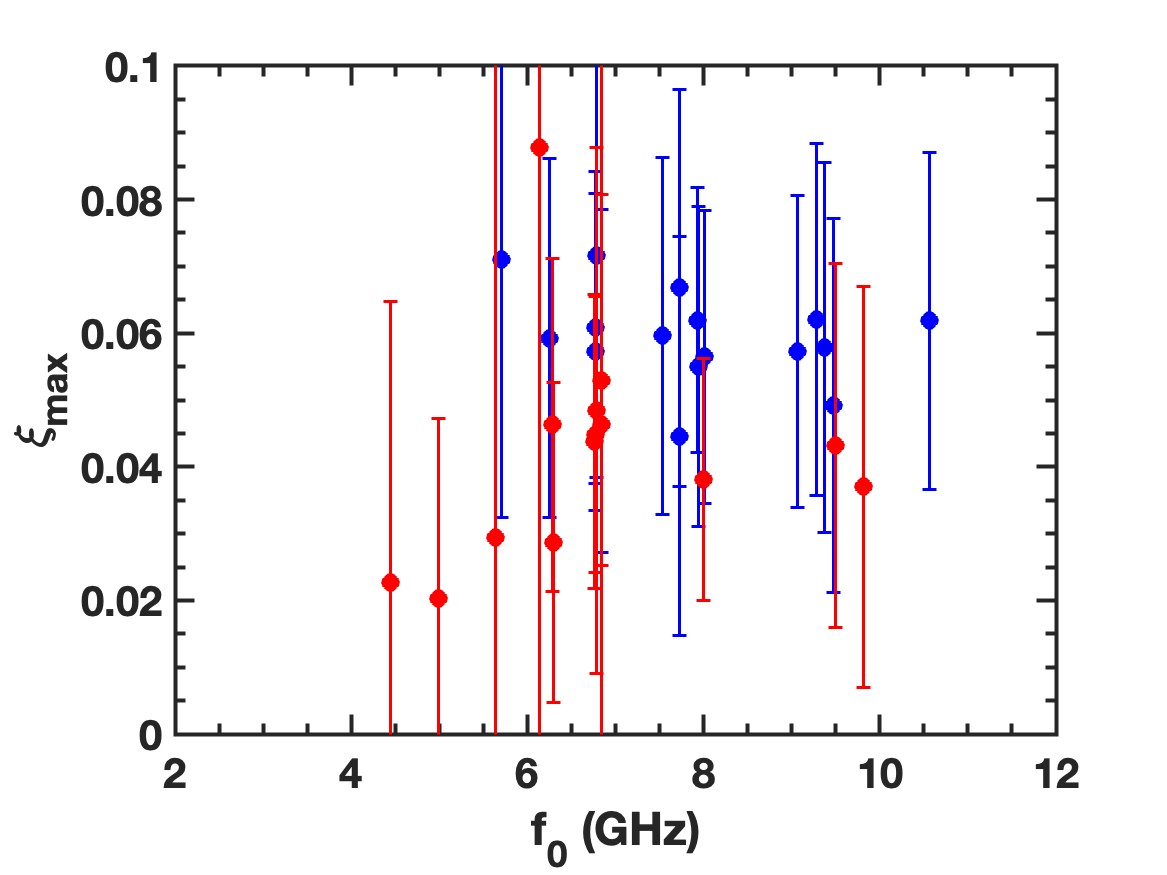}
\label{fig:xi_max_vs_f_sup_mat}
\end{subfigure}
\hfill
\begin{subfigure}[b]{0.475\textwidth}
\caption{}
\hspace*{-0.5cm}\includegraphics[scale=0.23,width=1.0\textwidth]{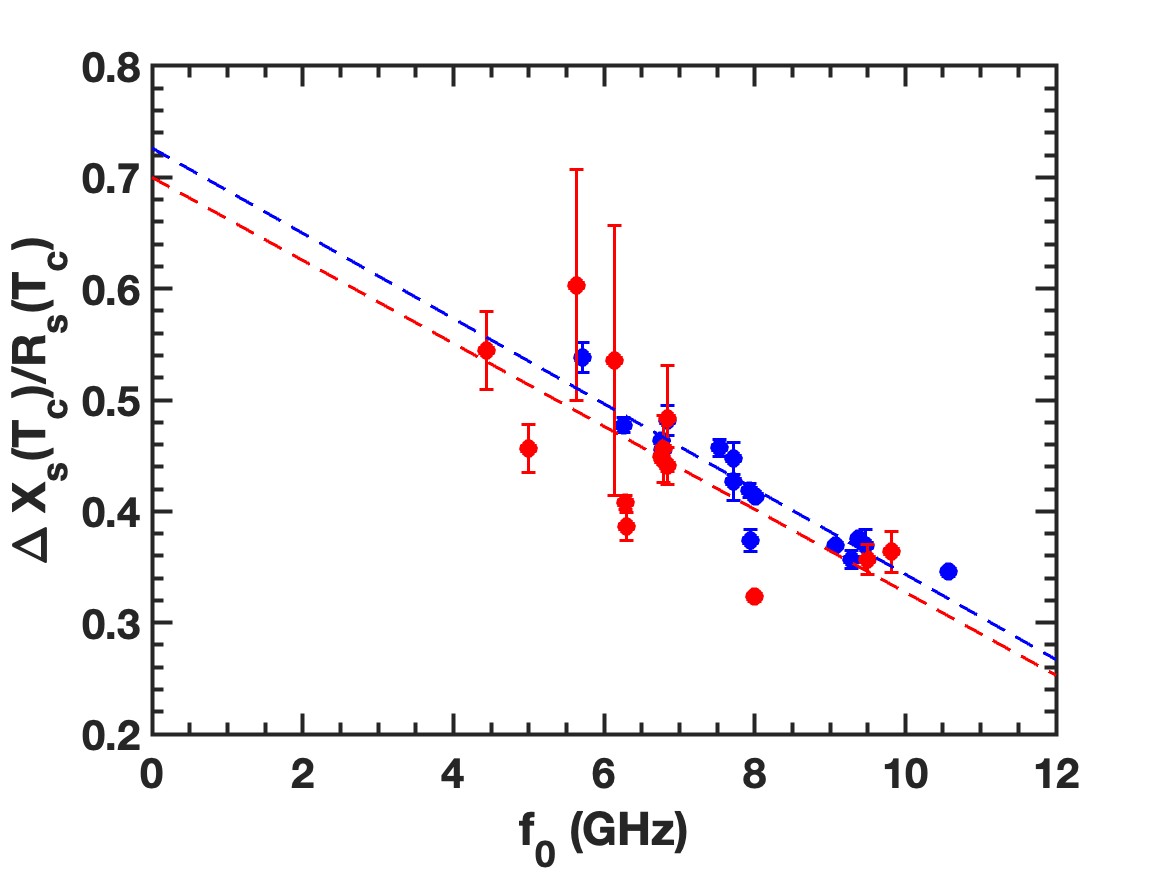}
\label{fig:xs_N_vs_f_sup_mat}
\end{subfigure}
\caption{Plots of the frequency dependence of the parameters describing the reactance peak in $X_s(T)$ for all modes studied for both UTe$_2$ samples. UTe$_2$ B39 and B40 are shown in blue and red, respectively. Linear fits to these data sets are also included in the corresponding colors where trends are observed. (a) shows the temperature $T_* < T_c$ of the reactance peak in $X_s(T)$, normalized by $T_c$. (b) shows the temperature width $\Delta T_*$ of the reactance peak in $X_s(T)$, normalized by $T_c$. (c) shows $\xi_{max}$, which is the height of the $X_s(T)$ reactance peak normalized by $R_s(T_c)$. (d) shows the total change in $X_s$ from base temperature to $T_c$ normalized by $R_s(T_c)$. The determination of all of these quantities involve only the use of raw experimental data.}
\label{fig:Xs_peak_params}
\end{figure*}

A peak in $X_s(T)$ below T$_c$ is observed in \textit{all} measured modes of \textit{both} UTe$_2$ samples. An example of the reactance peak vs temperature is shown in Fig.~\ref{fig:Xs_peak_vs_T_sup_mat}. The $\Delta X_s(T)$ data is smoothed to determine the temperature of the local maximum $T_*$, width of the peak $\Delta T_*$, height of the peak $\xi_{max}$, and the value of $\Delta X_s(T_c)/R_s(T_c)$.
Figure \ref{fig:t_peak_vs_f_sup_mat} shows the temperature $T_* < T_c$ at which the temperature-dependent reactance $X_s(T)$ reaches its maximum vs the mode frequency for all modes of UTe$_2$ B39 and B40, shown in blue and red, respectively. In the two-fluid picture, the peak in $X_s(T)$ occurs at the point where the superfluid fraction $f_s \sim \omega\tau/2$ \cite{Ormeno06}.

	%---------------------------------

Assuming that the scattering time increases with decreasing temperature in the superconducting state, the peak temperature $T_*$ is expected to migrate to lower temperatures as the measurement frequency is increased \cite{Ormeno02,Ormeno06}, consistent with the data for $T_*(f)$ for UTe$_2$ in Fig. \ref{fig:t_peak_vs_f_sup_mat}. As this happens, the peak also becomes broader in temperature (see the definition of $\Delta T_*$ in Fig.\ \ref{fig:Xs_peak_vs_T_sup_mat}), as shown in Fig.~\ref{fig:t_width_vs_f_sup_mat}. We also observe that $\Delta X_s(T_c)/R_s(T_c)$, the change in the dimensionless surface reactance from base temperature to $T_c$, also generally decreases as frequency increases, as shown in Fig. \ref{fig:xs_N_vs_f_sup_mat}. We include linear fits to these frequency dependences in Fig. \ref{fig:Xs_peak_params} for the two UTe$_2$ samples. Interestingly, we do not find a systematic frequency dependence for the dimensionless height, $\xi_{\text{max}}$, of the reactance peak.
Note that the reactance peak can be very close to $T_c$ for superconductors with small $\omega\tau$ \cite{Ormeno02,Ormeno06}. 
%The frequency dependence of the reactance peak is discussed further in \S\ref{sec:SM_reactance_peak}. 
Reactance peaks are not observed for our measurements of Nb and NbSe$_2$. The universal observation of a reactance peak for UTe$_2$ may be indicative of a larger value of $\omega\tau$ than that of Nb and NbSe$_2$.

	%---------------------------------

By examining the height of the reactance peak $\xi_{max}$, as shown in Fig.~\ref{fig:xi_max_vs_f_sup_mat}, one can determine the value of $\omega\tau_{eff} > \omega \tau$ at the temperature of the peak \cite{Hein01b}.  It is found that $\xi_{max}=0.059 \pm 0.007$ for sample B39 and $\xi_{max}=0.042 \pm 0.017$ for sample B40.  Based on the two-fluid model analysis proposed in Ref. \cite{Hein01b}, one finds that the corresponding effective scattering time in the superconducting state is $\omega \tau_{eff} =1.2 \pm 0.2$ for B39 and $\omega \tau_{eff} =1.9^{+1.2}_{-0.7}$ for B40 at $T_*$, placing upper limits on $\omega \tau$ below, but near, $T_c$. These results suggest that $\omega\tau \lesssim 1$ in the normal state of UTe$_2$.

%We next compare multiple methods to determine $\omega\tau$ for UTe$_2$. In the normal state, the Drude model gives $\omega\tau(T)=(X_s^2(T)-R_s^2(T))/2R_s(T) X_s(T)$ for an isotropic sample. The two-fluid result for $\omega\tau$ is given in Eq.~\ref{eq:omega_tau_2f}, which is valid in both the normal and superconducting states. The two-fluid model and its determination of $\omega\tau$ is discussed further below in \S\ref{sec:2f_model}. Near $T_c$, we find good agreement between the Drude and the two-fluid models, as shown in Fig.~\ref{fig:omega_tau_comparison}. 

\subsection{Assumption-Free Determination of $\omega \tau$ in the Normal State}
\label{sec:OmTauAnalysis}

\begin{figure}  %---------------------------------
% \centering
\hspace*{-0.5cm}\includegraphics[scale=0.23]{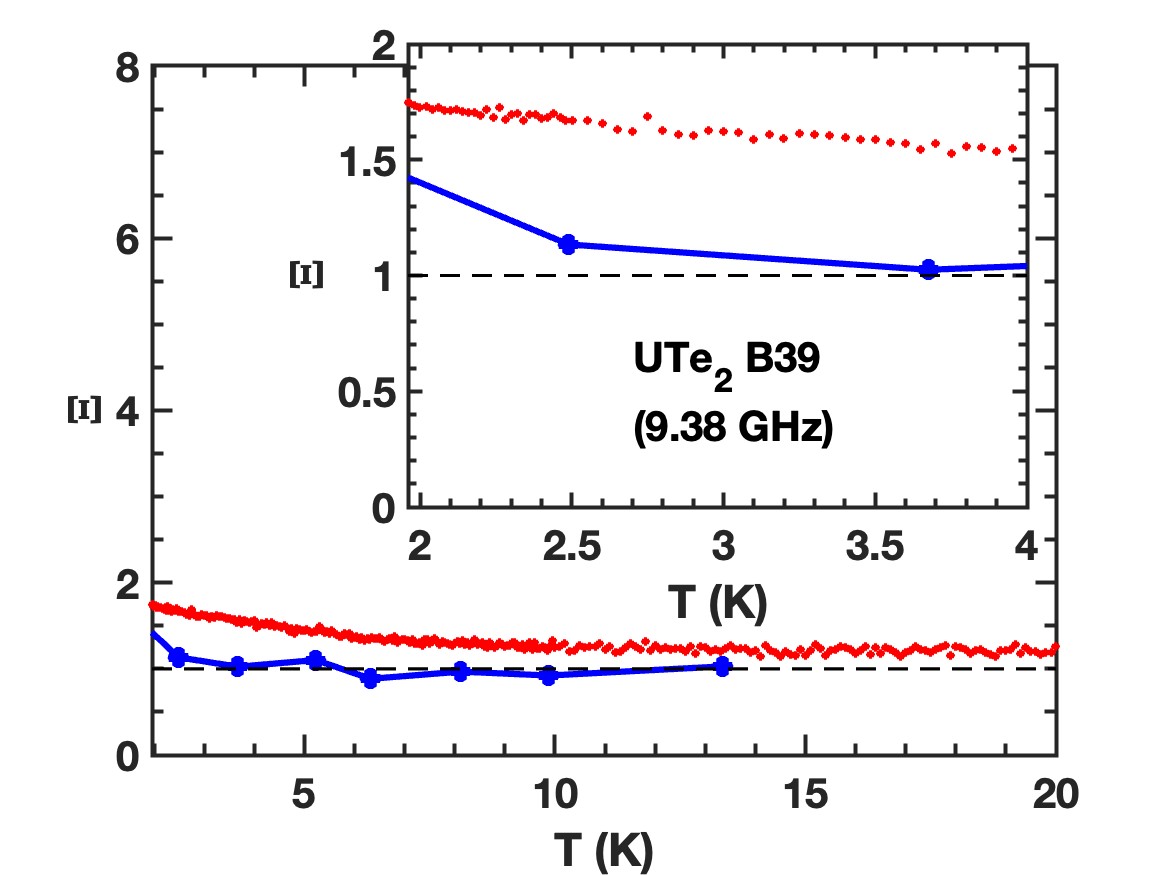}
\caption{Plot of different determinations of $\Xi(T)$. 
%The blue curve shows $\Xi(T)$ obtained from the ratio of numerical derivatives of temperature-dependent bandwidth and frequency shift raw data. 
The blue curve shows $\Xi(T)$ obtained from the numerical derivative of temperature-dependent bandwidth ($\Delta f_{BW}$) vs frequency shift ($\Delta f_0$) raw data. The red curve shows $\Xi(T)$ obtained from the ratio of $R_s$ and $X_s$, which is only valid in the local-limit. The curves should approach unity, shown by the black dashed line, in the limit of small $\omega\tau$. The inset shows the region near $T_c$.}
\label{fig:Xi_vs_T_sup_mat}
\end{figure}	%---------------------------------

Here we outline a simple analysis of the normal state raw data ($\rev{\Delta}f_{\rev{0,\text{sample}}}(T)$, $Q_{\rev{\text{sample}}}(T)$ for $T>T_c$) to determine $\omega\tau(T)$ using only the assumption that an isotropic single-Drude model describes the normal state electrodynamics.  This analysis does \uline{not} assume anything about the geometric factor $G$, the zero-temperature reactance $X_0$, the weighting functions $w_i$, etc. 
 Reference \cite{Hein01b} notes that a plot of raw data in the form of 
\begin{equation} \label{eq:BigXi}
    \Xi(T) \equiv - [ \partial(\Delta f_{BW})/\partial T)]/[\partial(2\Delta f_0)/\partial T],
\end{equation}
where $f_{BW}(T)$ is the 3-dB bandwidth ($Q_{\rev{\text{sample}}}=f_{0\rev{,\text{tot.}}}/f_{BW}$) \rev{of the sample} and $\rev{\Delta}f_{0\rev{,\text{sample}}}$ is the \rev{change in the} resonant frequency \rev{of the sample}, both as function\rev{s} of temperature.  Reference \cite{Hein01b} shows that $\Xi(T)$ has a unique dependence on $\omega\tau(T)$ depending only on the frequency-dependent dimensionless nonlocality parameter $\beta = \frac{v_F}{\omega \lambda_L} = \frac{\ell_{mfp}/\lambda_L}{\omega\tau}$, where $v_F$ is the Fermi velocity, $\lambda_L$ is the London penetration depth, and $\ell_{mfp}$ is the normal-state carrier mean free path.  In other words, once $\beta$ is known for a given mode, we can then convert the raw complex frequency shift data into an estimate of $\omega\tau(T)$ in the normal state.  In addition, if $\beta < \beta_c = (1+\frac{1}{(\omega\tau)^2})^{3/2}$, then the sample can be considered to be in the local limit \cite{Hein01b}. For an isotropic sample in the local limit, $\Xi(T)=X_s(T)/R_s(T)$, and $\omega\tau(T)=(\Xi^2(T)-1)/(2\Xi(T))$. In Fig.~\ref{fig:Xi_vs_T_sup_mat}, we compare $\Xi(T)$ calculated from raw data using Eq.~\ref{eq:BigXi} with the ratio of $X_s(T)$ and $R_s(T)$ after the impedance fitting. 

%We can estimate the values of the non-locality parameter $\beta$ for UTe$_2$ in the frequency range of our measurement as follows.  The Fermi velocity for UTe$_2$ is estimated to be $v_F = 6\times 10^{3}$ m/s \cite{Metz19}.  We take the London penetration depth to be $\lambda_L = 1.5~ \mu$m, which approximates the anisotropic values presented in this paper.  This gives values of $\beta = 0.16 \sim 0.06$ for the range of frequencies (4-11 GHz) of our cavity perturbation data.  In Fig. \ref{fig:BetavsOmegaTau} we show a plot of $\beta$ vs $\omega\tau$, with the local and non-local regions delineated by the blue line ($\beta_c$).  The estimated $\beta$ values for the UTe$_2$ samples are shown in the red box. Considering this estimate of the non-locality parameter shown in Fig.~\ref{fig:BetavsOmegaTau} and our four independent estimates $\omega\tau$ shown in Fig.~\ref{fig:omega_tau_comparison}, it is clear that our samples are in the local limit as far as electrodynamic response is concerned.

%\textcolor{red}{Add a plot of slope $\Xi$ vs $\omega\tau$ with $\beta$ as a parameter.}

The conversion of $\Xi(T)$ to $\omega \tau(T)$ is dependent upon the value of the nonlocality parameter $\beta$ \cite{Hein01b}.  In the local limit ($\beta \ll 1$, well-satisfied in our case), the relation is known to be $\Xi_{local} = 1/(\sqrt{1+(\omega\tau)^2}-\omega\tau)$. For larger values of $\beta$, the relationship is non-monotonic. We present this local limit estimate of $\omega\tau$ in Fig.~\ref{fig:omega_tau_comparison}.

\subsection{Two-fluid model determintion of $\omega \tau$}
\label{sec:2f_model}
\begin{figure}	%---------------------------------
\hspace*{-0.5cm}\includegraphics[scale=0.23,width=0.5\textwidth]{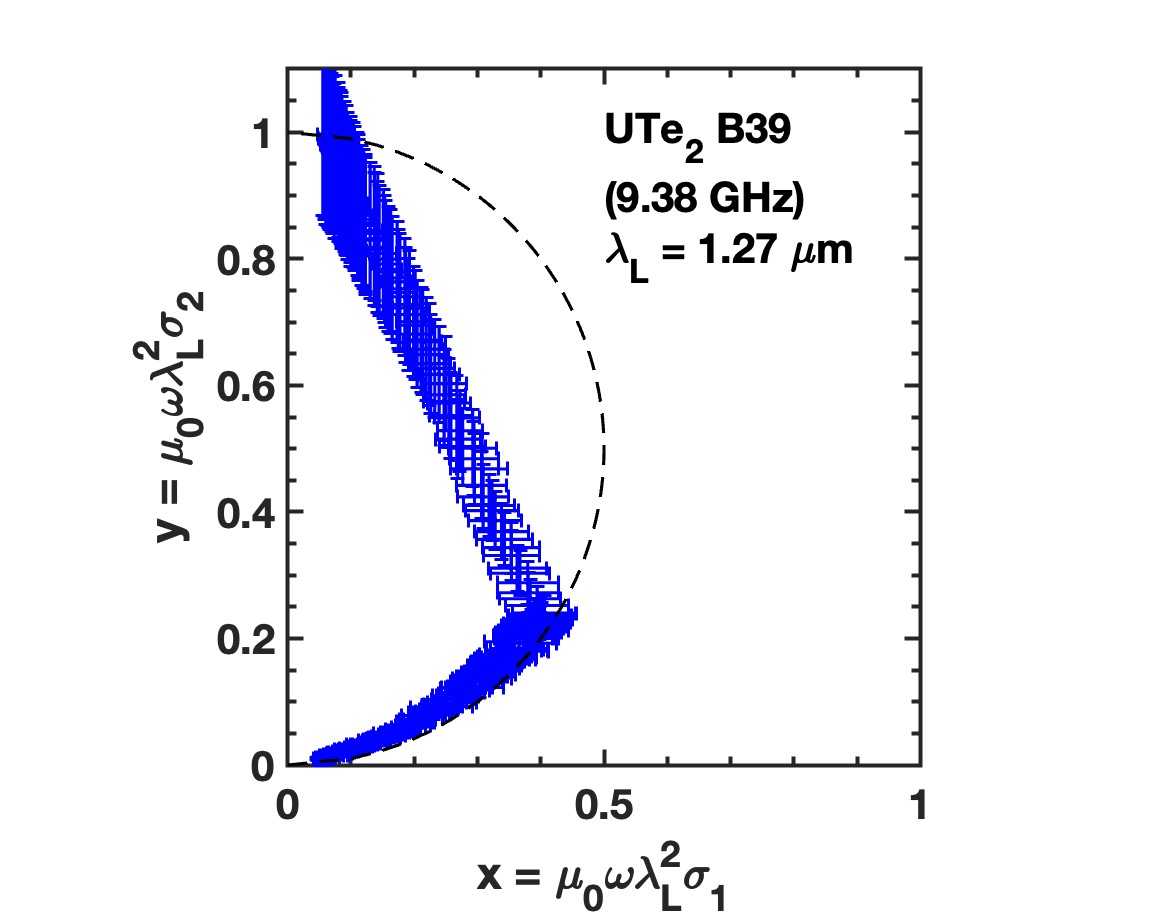}
\caption{Plot of complex $\sigma$ data in the geometry of Fig.~\ref{fig:sigma_contours_sup_mat}. The curve of $f_s=0$ shown as a black dashed line is used to used to fit the normal state portion of the data to obtain a composite London penetration depth $\lambda_L$.}
\label{fig:sigma_plane_fit}
\end{figure}

The two-fluid model extends the Drude model and its scattering time $\tau$ into the superconducting state by assuming that the electrodynamics is a mixture of normal metal and superconductor properties. In terms of the complex conductivity $\sigma$, this can be expressed as,
\begin{equation}
    \sigma=\frac{1}{\mu_0\lambda_L^2}\left( \frac{f_s}{i\omega}+\frac{f_n\tau}{1+i\omega\tau} \right),
    \label{eq:sigma_2f}
\end{equation}
where $f_n$ and $f_s$ are the normal- and super-fluid fractions, for which $f_n+f_s=1$. For $f_s=0$, Eq.~\ref{eq:sigma_2f} reduces to the Drude model. The parameters of the two-fluid model can be determined from $\sigma$ if $\lambda_L$ is known,
\begin{equation}
    \begin{split}
        \omega\tau&=\frac{x}{1-y}\\
        f_n&=\frac{(1-y)^2+x^2}{1-y},
    \end{split}
    \label{eq:omega_tau_2f}
\end{equation}
where $x-iy=\mu_0\omega\lambda_L^2\sigma=\mu_0\omega\lambda_L^2(\sigma_1-i\sigma_2)$. Eq.~\ref{eq:omega_tau_2f} informs the geometry of Fig.~\ref{fig:Zs_and_sigma_geometry_sup_mat}, specifically that the curve of $f_s=0$ is a circle (the outer-most circle) in the complex sigma plane shown in Fig.~\ref{fig:sigma_contours_sup_mat} and a hyperbola in the complex impedance plane shown in Fig.~\ref{fig:Zs_contours_sup_mat}. We can fit our $\sigma$ (or alternatively $Z_s$) data to this predicted behavior to determine $\lambda_L$. An example of this process applied to experimental data is shown in Fig.~\ref{fig:sigma_plane_fit} which gives $\lambda_L=1.27\pm0.07\ \mu m$. %Note that $\sigma$ goes above its maximum bound ($f_s=1$, $y=1$)
To ensure that all points are on or within the dashed line indicating $f_s=0$ in Fig.~\ref{fig:sigma_plane_fit}, it is necessary that $\mu_0\omega\lambda_L^2 \le \frac{\sigma_2}{\sigma_1^2 + \sigma_2^2}$. While this ensures that the interpretation of $\omega\tau$ and $f_n$ with Eq.~\ref{eq:omega_tau_2f} are physical, the quality of the fit of the normal state data to the circle suffers. This isotropic two-fluid model is still a reasonable analysis tool for our data which allows for an estimate of the scattering time, shown in Fig.~\ref{fig:omega_tau_comparison}, as well as the super- and normal-fluid fractions. The systematic departure of the normal state data from the dashed circle likely indicates that the treatment of the composite $\sigma$ data using a single  temperature-independent composite $\lambda_L$ is inappropriate. A London penetration depth corresponding to each axis may be necessary to describe the normal state data; additionally, these lengths could be temperature dependent if there is some renormalization of carrier density and/or effective mass. 
%Note that $\sigma$ goes outside its allowed bound, shown as the dashed line in Fig.~\ref{fig:sigma_plane_fit}, which should not be possible. As indicated in Fig.~\ref{fig:sigma_contours_sup_mat}, all values of $f_s>0$ lie inside the circle. This causes the divergence in the two-fluid model calculation of $\omega\tau$ shown in red in Fig.~\ref{fig:omega_tau_comparison}. The departure of the data from the dashed circle indicates that this estimate of $\lambda_L$ is too large, or that the treatment of the composite $\sigma$ data using a single  temperature-independent composite $\lambda_L$ is inappropriate. A London penetration depth corresponding to each axis may be necessary to describe the normal state data; additionally, these lengths could be temperature dependent if there is some renormalization of carrier density and/or effective mass. 

%Despite these inconsistencies, this treatment of the data is well behaved for most of the temperature dependence. In this well behaved region of temperatures, we are able to estimate the scattering time, shown in Fig.~\ref{fig:omega_tau_comparison}, as well as the super- and normal-fluid fractions.

	%---------------------------------

\section{Low-temperature $\Delta \lambda(T)$ and $R_s(T)$ fits}
\label{sec:App_sigma_and_low_temp_fits}
Here we examine the low-temperature behaviors of the magnetic penetration depth and surface resistance, discussing various fits and residual values.

\subsection{Penetration depth low-temperature fits}
\label{sec:lambda_fits}
We present a generalized form which will be useful to do multiple fittings in the sections below,
\begin{equation}
    \frac{x(T)-x_0}{x_0} = a y(T;c),
    \label{eq:general_low_temp_form_SM}
\end{equation}
where $x(T)$ is any observed temperature dependent data, $y(T,c)$ is any family of temperature dependent functions which depends non-linearly on a single parameter $c$ and for which $y(0,c)=0$, $x_0$ is the non-zero value of $x(T)$ at zero-temperature implied by $y(T,c)$, and $a$ is a positive pre-factor for $y(T,c)$. By measuring $x(T)$ at $N_T\ge 3$ discrete temperatures, we obtain a linear system for a fixed value of $c$ described by the matrix form, $\bm{M}\Vec{b}=\Vec{y}(c)$, where $\bm{M}=\left( 1, \Vec{x} \right)$, $x_i = x(T_i)$, $\Vec{b}=\begin{pmatrix}
    -1/a\\
    1/(a x_0)
\end{pmatrix}$, and $y_i=y(T_i,c)$. The solution of this system for a fixed value of $c$ in the least-squares sense is $\Vec{b}(c)=\left(M^T M\right)^{-1}\left(M^T \Vec{y}(c)\right)$ The non-linear parameter $c$ can then be varied to achieve an optimal fit, such as by using $1-R(c)^2$ as a cost function, where $R(c)$ is the linear correlation coefficient between $\Vec{y}(c)$ and $\Vec{x}$.

Similarly to \S\ref{sec:SM_impedance_fitting}, the uncertainty in the non-linear parameter $c$ can be estimated by observing how much it must vary to produce a $3\%$ increase in the cost function. The uncertainties in the two linear parameters can then be estimated by the maximum variation from their optimal values in the range of $c$ for which the increase in the cost function is still less than $3\%$. The range of parameters and their uncertainties for the following fits are included in Table~\ref{tab:fitting_uncerts}.

\subsubsection{Power-law penetration depth fits}
\label{sec:lambda_pow_law_fits}
The form,
\begin{equation}
    \frac{\lambda(T)-\lambda_0}{\lambda_0} = \eta \left(\frac{T}{T_c}\right)^{\alpha},
    \label{eq:lambda_pow_law_SM}
\end{equation}
is ubiquitous for nodal superconductivity with predictions for $\alpha$ and $\eta$ depending on the nature of the nodes and the direction of the current relative to the nodes \cite{Gross86,Klemm88}. Eq.~\ref{eq:lambda_pow_law_SM} is of the form of Eq.~\ref{eq:general_low_temp_form_SM} for $x(T)=\lambda(T)$, $x_0=\lambda_0$, $a=\eta$, $c=\alpha$, and $y(T,c)=(T/T_c)^c$. We fit our penetration depth data to this power-law form in Fig.~\ref{fig:lambda_pow_laws} for various choices of $X_0/G$. In this case, we specifically have $x(T)=\lambda(T)/G$, $x_0=\lambda_0/G$, $a=\eta$, $c=\alpha$, and $y(T,c)=(T/T_c)^c$. Though $\lambda_0$ cannot be determine without $G$, $\alpha$ and $\eta$ are explicitly independent of $G$. For the $f_0=6.77$ GHz mode with UTe$_2$ B39, the range of $X_0$ used in Fig.~\ref{fig:lambda_pow_laws} gives $\alpha=1.69-1.71$, $\eta=0.432-0.991$, and $\lambda_0/G=0.194-0.417\ \text{nm}/\Omega$. For the $f_0=5.64$ GHz mode with UTe$_2$ B40, the range of $X_0$ used in Fig.~\ref{fig:lambda_pow_laws} gives $\alpha=1.76-1.78$, $\eta=0.636-1.15$, and $\lambda_0/G=0.411-0.712\ \text{nm}/\Omega$. Note that $G$ is different for these different modes and samples. For both these samples, the power-law exponent $\alpha$ is nearly independent of $X_0$ for the modes shown, as well as the other modes we studied (a total of 17 for UTe$_2$ B39 and 14 for UTe$_2$ B40). In contrast, $\eta$ does vary with $X_0$. Since $\lambda_0$ is proportional to $X_0$, $\eta$ must decrease with increasing $X_0$ to keep $\Delta\lambda$ independent of $X_0$. 
%We additionally find no systematic variation of $\alpha$ between modes, so we take the average and standard deviation among the modes to get $\alpha=1.70\pm0.08$ for sample B39 and $\alpha=1.82\pm0.25$ for sample B40. These averages use a fit to fix $G$ and $X_0$ for each mode, which is discussed below. The variation of $\alpha$ due to changing the value of $X_0/G$ is then much smaller than its variation between modes. The behavior of $\alpha$, $\eta$, and $\lambda_0$ between modes and samples is further discussed below, and summarized in Table~(\ref{tab:lambda_and_loss}) in \S\ref{sec:SM_param_summary}.

\rev{Additionally, we find that, overall, there does not appear to be any correlation of $\alpha$ with $Q_{\text{sample}}$; however, the large outlier for $\alpha$ corresponds to a relatively low $Q_{\text{sample}}$, while the largest $Q_{\text{sample}}$ mode has the smallest $\alpha$. Furthermore, no apparent systematic dependence (either overall or of the outlier) of $\alpha$ on $Q_r$ was observed.}

\subsubsection{'Dirty d-wave' penetration depth fits}
\label{sec:lambda_alt_fits}
In addition to the 3-parameter power-law fit in Eq.~\ref{eq:lambda_pow_law_SM}, we also tried fitting our low-temperature ($T<T_c/3$) penetration depth data to the 3-parameter `dirty d-wave' temperature dependence,\cite{Hirsch93b}
\begin{equation}
    \frac{\lambda(T)-\lambda_0}{\lambda_0} = \beta \frac{(T/T_c)^2}{(T/T_c)+(T_I/T_c)},
    \label{eq:lambda_pow_law_alt}
\end{equation}
where $T_I$ is an impurity-dependent temperature scale for the crossover between $\Delta \lambda(T) \sim T$ ($T \gg T_I$, clean d-wave) to $\Delta \lambda(T) \sim T^2$ ($T \ll T_I$, dirty d-wave) behavior.  If one finds that $T_I > T_c$, then it is unlikely that the `dirty d-wave' scenario is correct (\cite{BaeThesis20}, p. 149). 

\begin{figure}  %---------------------------------
\begin{subfigure}[b]{0.4\textwidth}
% \centering
\caption{}
\hspace*{-0.5cm}\includegraphics[scale=0.23,width=1.0\textwidth]{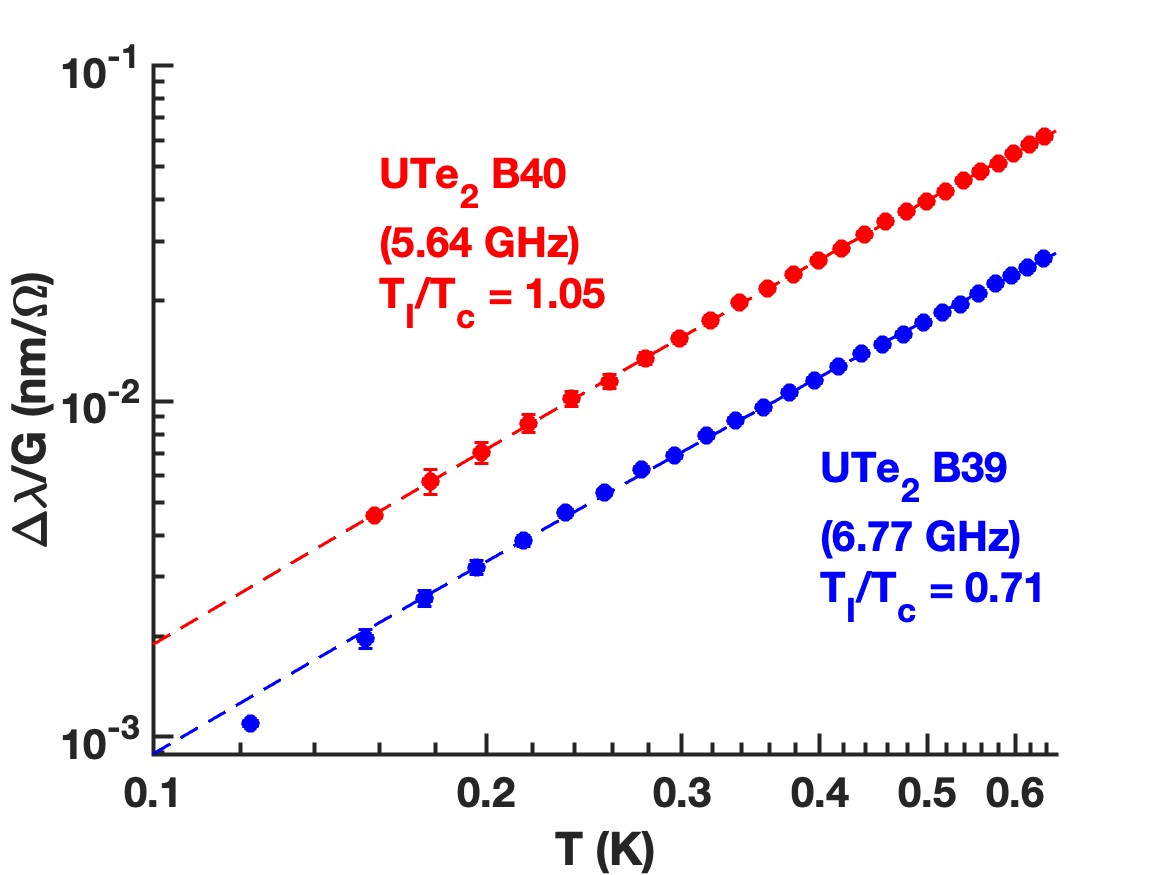}
\label{fig:dirty_d_fit_sup_mat}
\end{subfigure}
\begin{subfigure}[b]{0.4\textwidth}
% \centering
\caption{}
\hspace*{-0.5cm}\includegraphics[scale=0.23,width=1.0\textwidth]{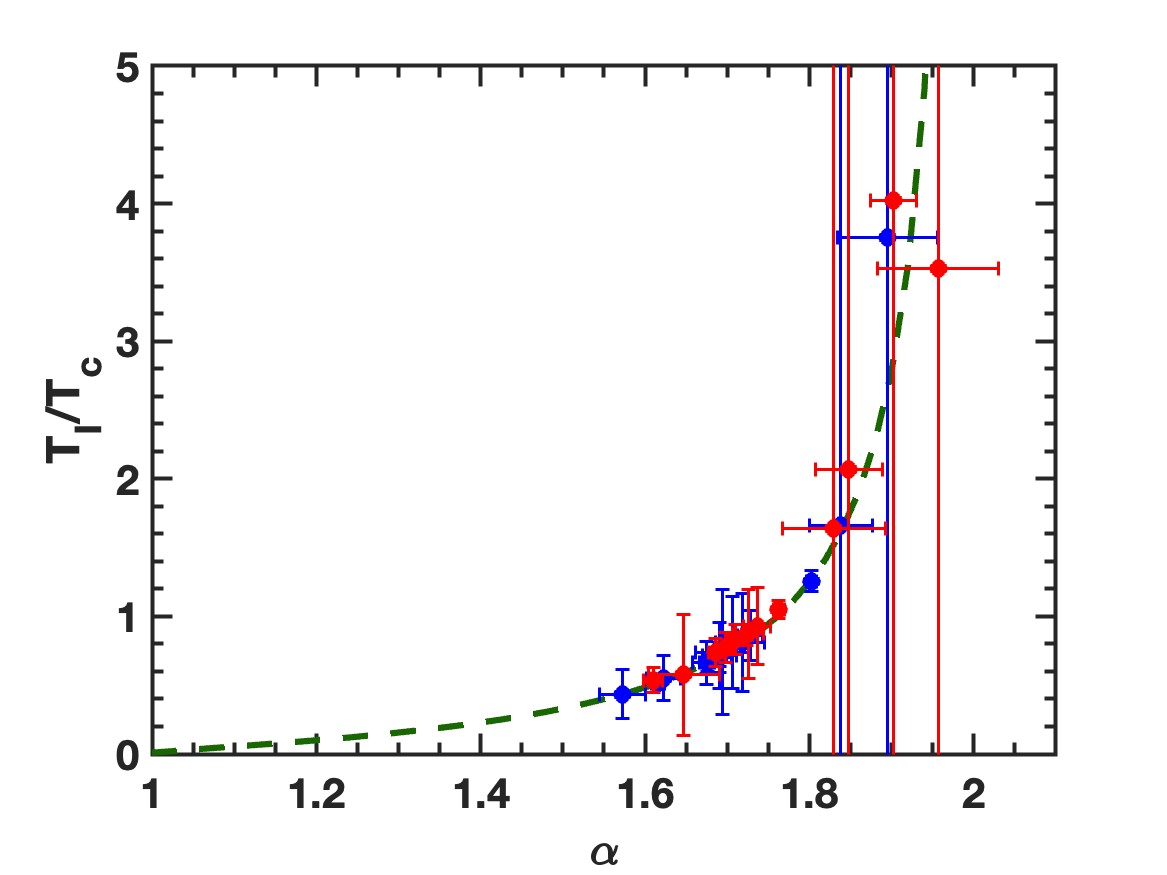}
\label{fig:t_I_vs_alpha_sup_mat}
\end{subfigure}
\caption{(a) Example of `dirty d-wave' fits to penetration depth data (b) Results of the impurity temperature $T_I$ from dirty d-wave fits (Eq.~\ref{eq:lambda_pow_law_alt}) normalized by $T_c$ for two UTe$_2$ samples vs $\alpha$ from power-law fits (Eq.~\ref{eq:lambda_pow_law_SM}) for the same modes and samples. UTe$_2$ B39 is shown in blue and UTe$_2$ B40 is shown in red. The results of the dirty d-wave fit to perfect power-law data is shown as a green dashed line.}
\label{fig:dirty_d_sup_mat}
\end{figure}	%---------------------------------

An alternative physical origin for the penetration depth temperature dependence in Eq.\ \ref{eq:lambda_pow_law_alt} has been put forward.  Kosztin
and Leggett have proposed that nonlocal electrodynamic response will arise in d-wave superconductors due to the presence of the node in the superconducting gap \cite{KL97}.  They predict a nonlocality crossover temperature of $T_{NL}=\frac{\xi_0}{\lambda_0}\Delta_{\text{max.}}/k_B$ (in place of $T_I$ in Eq.\ \ref{eq:lambda_pow_law_alt}), well below which the penetration depth should be quadratic in temperature.  However, using estimates for the coherence length ($\xi_0 = 3-6\ nm$, based on $H_{c2}$ data \cite{Aoki24b}), penetration depth ($\lambda_0 = 1.5\ \mu m$ from the present work) and maximum gap ($\Delta_{\text{max.}} = 0.29$ meV)\cite{Metz19} gives an estimate that $T_{NL} \sim 10^{-2} T_c$.  This is roughly two orders of magnitude lower than the observed values of $T_I$ shown in Fig.\ \ref{fig:t_I_vs_alpha_sup_mat}, ruling out nonlocality in a d-wave superconductor as the microscopic origin of the observed power-law behavior of the penetration depth at low temperatures.

Eq.~\ref{eq:lambda_pow_law_alt} is of the form of Eq.~\ref{eq:general_low_temp_form_SM} for $x(T)=\lambda(T)$, $x_0=\lambda_0$, $a=\beta$, $c=T_I/T_c$, and $y(T,c)=\frac{(T/T_c)^2}{(T/T_c) + c}$. These fits appear to be roughly as viable in reproducing $\lambda(T)$ as Eq.~\ref{eq:lambda_pow_law_SM} is, as shown in Fig.~\ref{fig:dirty_d_fit_sup_mat}. The range of impurity temperatures, $T_I$, across modes and samples is plotted in Fig.~\ref{fig:t_I_vs_alpha_sup_mat} vs the corresponding power-law fit value, $\alpha$. Most values of $T_I$ range from $T_c/2$ to $T_c$, which is reasonable for a `dirty d-wave' interpretation. To understand whether or not Eq.~\ref{eq:lambda_pow_law_SM} and Eq.~\ref{eq:lambda_pow_law_alt} are equivalent interpretations of our data, we attempt to fit Eq.~\ref{eq:lambda_pow_law_SM} to Eq.~\ref{eq:lambda_pow_law_alt} and vice versa. With $T_c=1.96$ K, these forms are very similar for our range of measured temperatures. For temperatures below roughly 100 mK, however, the two models differ dramatically. We find that for Eq.~\ref{eq:lambda_pow_law_SM} calculated with the grid of temperatures used in our measurements, the value of $\alpha$ implies a value of $T_I$ in Eq.~\ref{eq:lambda_pow_law_alt}. In a similar manner, $T_I$ would imply a value for $\alpha$ for Eq.~\ref{eq:lambda_pow_law_alt}. We plot this relationship between $\alpha$ and $T_I$ as the green dashed line in Fig.~\ref{fig:t_I_vs_alpha_sup_mat}, which agrees well with our fittings to data. For $\alpha=1$, $T_I=0$, and for $\alpha=2$, $T_I$ diverges. Eq.~\ref{eq:lambda_pow_law_alt} is incompatible with $\alpha>2$. We find one instance of a mode with $\alpha>2$ in our data. This result for this particular mode is then incompatible with the `dirty d-wave' model; however, the uncertainty in $\alpha$ for that mode is large enough that it is reasonable that it could actually be less than two. 

The power-law and `dirty d-wave' interpretations are evidently equivalent and indistinguishable in our measurement regime.  However, we see from the column labeled `Fractional uncertainty' in Table  \ref{tab:fitting_uncerts} that the determination of the parameters in the power-law fits ($\alpha$, $\eta$) is significantly better than the parameters for the `dirty d-wave' fits ($T_I/T_c$, $\beta$).  In fact there are systematic deviations observed in the 'dirty d-wave' fits that are not present in the power-law fits.  From this we conclude that the power-law fits for the low-temperature penetration depth are statistically favored over those of the `dirty d-wave' temperature dependence. 

\subsubsection{Comparison with s-wave fits}
\label{sec:SM_lambda_exp_fit}
\begin{figure}	%---------------------------------
\hspace*{-0.5cm}\includegraphics[scale=0.23,width=0.5\textwidth]{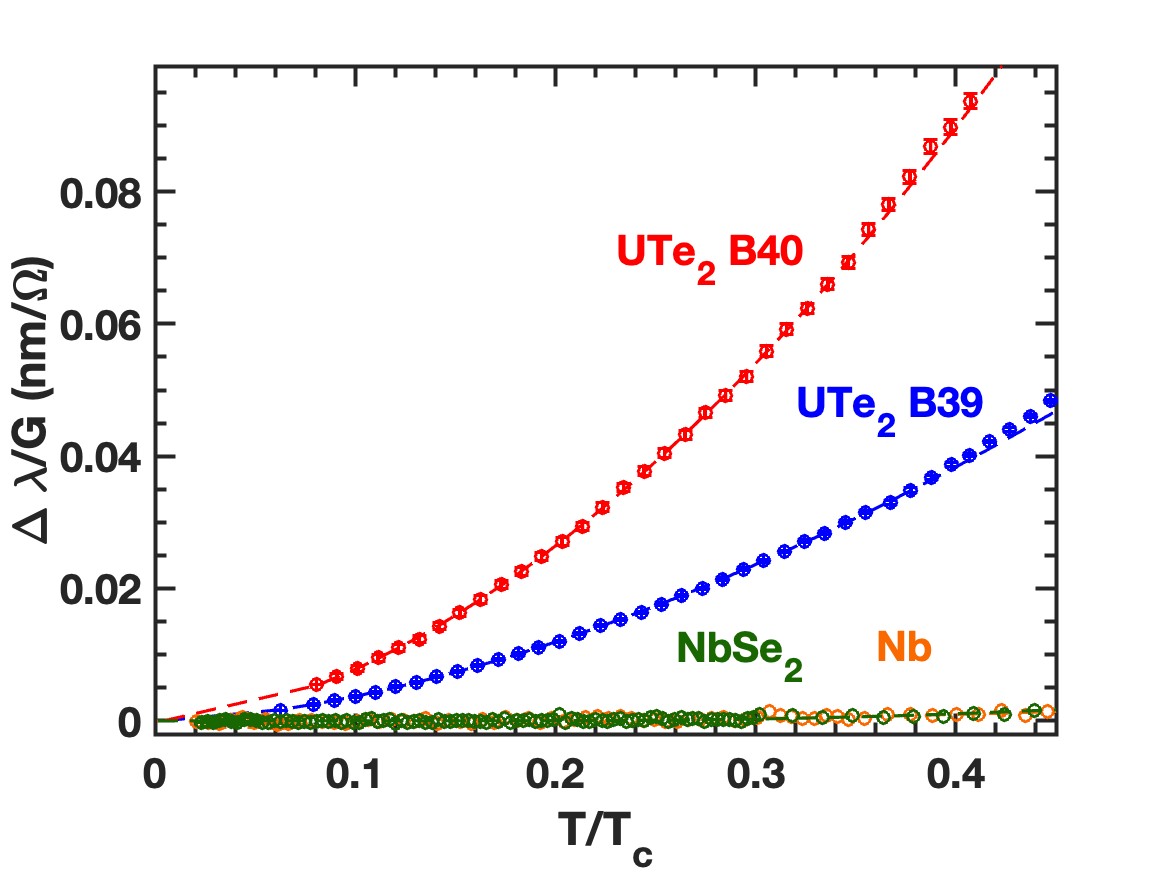}
\caption{Plots of $\Delta\lambda/G$ vs. $T/T_c$ for two UTe$_2$ samples and two fully-gapped superconductors. Power-law fits, given by Eq.~(\ref{eq:lambda_pow_law_SM}), to the UTe$_2$ data are shown as dashed lines of the corresponding colors. Exponential fits, given by the expression for $\Delta\lambda$ from \cite{Halbritter71,Pam94}, to the Nb and NbSe$_2$ data are shown as dashed lines of the corresponding colors.}
\label{fig:delta_lambda_comparison}
\end{figure}

The temperature dependence of the penetration depth for the UTe$_2$ samples is qualitatively very different from that of fully-gapped superconductors, as demonstrated in Fig.~\ref{fig:delta_lambda_comparison}. These data are from four different resonant modes with different $G$, $X_0$, and for the UTe$_2$ samples, different contributions of current flow along it's crystallographic axes. 
%We have chosen a particular value for $X_0/G$ for each mode; though, as demonstrated in Fig.~\ref{fig:lambda_pow_laws}, $\Delta\lambda(T)/G$ is independent of $G$ and $X_0$ in this temperature range. The fitting method to determine $\Delta \lambda/G$ for UTe$_2$ is discussed in \S\ref{sec:SM_impedance_fitting} above. 
For an s-wave superconductor, the low-temperature penetration depth is given by $\frac{\lambda(T)-\lambda_0}{\lambda_0} = \sqrt{\frac{\pi\Delta_{\text{min.}}}{2k_B T}}\ e^{-\Delta_{\text{min.}}/(k_B T)}$ \cite{Halbritter71,Pam94},
where $\Delta_{\text{min.}}$ is the zero-temperature, minimum gap energy on the Fermi surface. We determine $\lambda_0$ and $\Delta_{\text{min.}}$ by fitting our s-wave penetration depth data to this form. Since $\Delta_{\text{min.}}$ also appears in the pre-factor of this expression, $\Delta_{\text{min.}}$ depends on the selected value of $X_0/G$. Rather than the fitting method discussed in \S\ref{sec:SM_impedance_fitting}, we select $X_0/G$ so that the value of $\Delta_{\text{min.}}$ from this fit reproduces estimates of $\Delta_{\text{min.}}$ for Nb and NbSe$_2$ from literature \cite{Bonnet67,Clayman71}. This exponential suppression of excitations out of the ground state means that the penetration depth will be nearly constant for these low temperatures, which can be seen in Fig.~\ref{fig:delta_lambda_comparison}; whereas, for the UTe$_2$ samples, the penetration depth rises much more quickly out of the ground state, consistent with nodal behavior.  
%Note that we have also fit the penetration depth to the `dirty d-wave' and nonlocal models (see Supp, Mat. Sec. \ref{sec:lambda_fits}), however the quality of these fits are markedly inferior to the power-law fits (see Table  \ref{tab:fitting_uncerts}).

	%---------------------------------

\subsection{Microwave Surface Resistance Power-Law Temperature Dependence}
\label{sec:SM_resistance_pow_law}
\begin{figure}	%---------------------------------
\begin{subfigure}[b]{0.4\textwidth}
% \centering
\caption{}
\includegraphics[scale=0.23,width=1.0\textwidth]{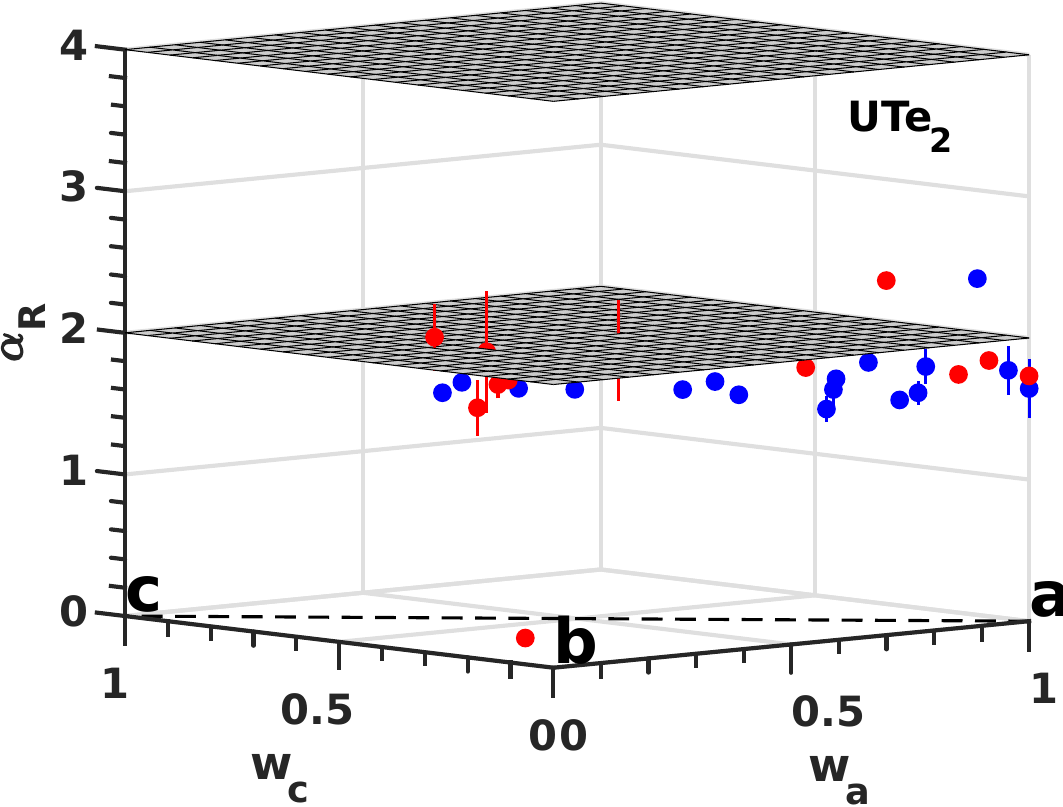}
\label{fig:alpha_R_vs_w_sup_mat}
\end{subfigure}
\begin{subfigure}[b]{0.4\textwidth}
% \centering
\caption{}
\includegraphics[scale=0.23,width=1.0\textwidth]{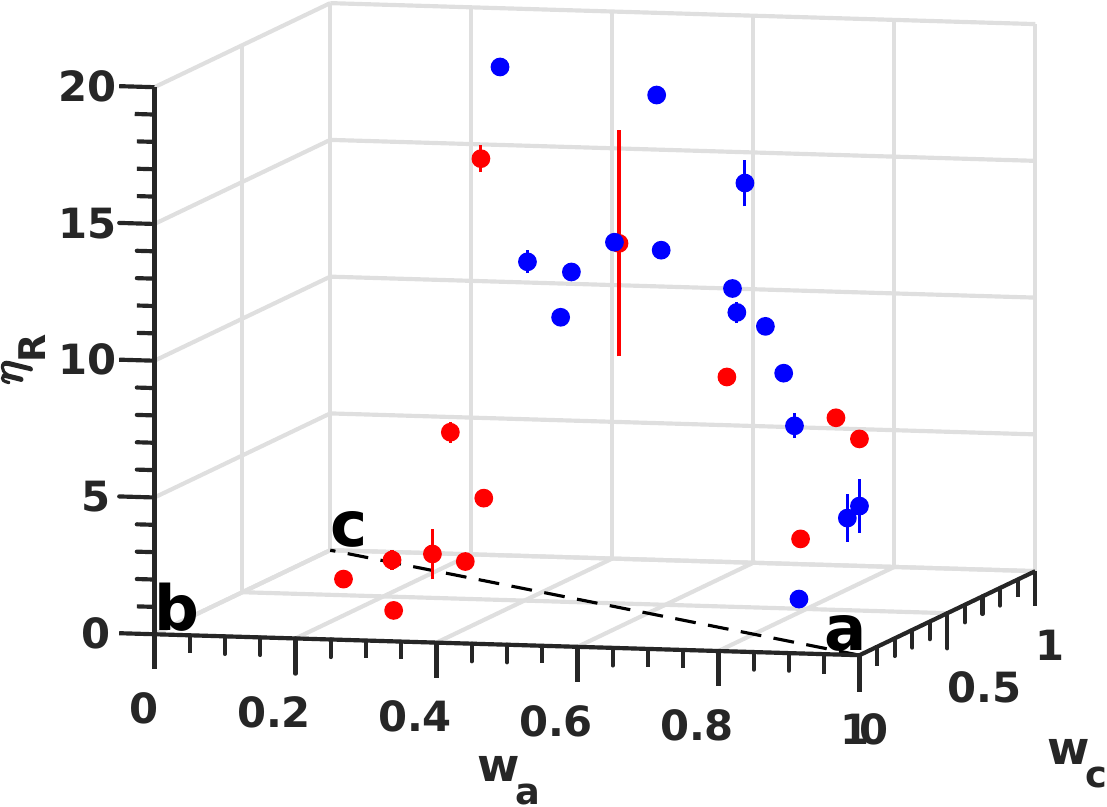}
\label{fig:eta_R_vs_w_sup_mat}
\end{subfigure}
\caption{Plots of power-law parameters for surface resistance for two UTe$_2$ samples vs contribution of each crystallographic axis. UTe$_2$ B39 is shown in blue and UTe$_2$ B40 is shown in red. (a) shows the surface resistance power-law exponent $\alpha_R$ and the p-wave predictions of $\alpha=2$ parallel and $\alpha=4$ perpendicular to the point nodes for the penetration depth as a reference. (b) shows the surface resistance power-law coefficient $\eta_R$. There are additionally $\eta_R$ values of 144 and 123 (not shown) for UTe$_2$ B39 and B40 respectively, which are near the a-direction ($w_a=1$).}
\label{fig:w_dep_R_sup_mat}
\end{figure}

We fit the surface resistance to the form of Eq.~\ref{eq:Rs_pow_law},
%\begin{equation}
%    \frac{R_s(T)-R_0}{R_0} = \eta_R \left(\frac{T}{T_c}\right)^{\alpha_R},
%    \label{eq:R_pow_law}
%\end{equation}
for $T<T_c/3$, where $R_0$ is the zero-temperature surface resistance, and $\alpha_R$ and $\eta_R$ are the dimensionless surface resistance power-law exponent and coefficient, respectively. Eq.~\ref{eq:Rs_pow_law} is of the form of Eq.~\ref{eq:general_low_temp_form_SM} for $x(T)=R_s(T)$, $x_0=R_0$, $a=\eta_R$, $c=\alpha_R$, and $y(T,c)=(T/T_c)^c$. See Fig.~\ref{fig:Rs_pow_law_sup_mat} for examples of this power-law fitting. In general, we find that this expression fits the data very well for \textit{all} measured modes. Fig.~\ref{fig:alpha_R_vs_w_sup_mat} shows the dependence of the surface resistance power-law exponent $\alpha_R$ vs the induced-current-direction weights. As with the penetration depth power-law exponents, these do not show a dependence on the induced-current weights. The average results are $\alpha_R=1.65\pm 0.23$ for UTe$_2$ B39 and $\alpha_R=1.57\pm 0.55$ for UTe$_2$ B40. Fig.~\ref{fig:eta_R_vs_w_sup_mat} shows the dependence of the surface resistance power-law coefficient $\eta_R$ vs the induced-current-direction weights. These do show a systematic dependence on the induced-current weights in contrast with the penetration depth power-law coefficients. There is a peak near $w_b=0, w_a=w_c=0.5$ and a very large spike near $w_a=1$ to 144 and 123 for UTe$_2$ B39 and B40 respectively. In these plots, UTe$_2$ B39 is shown in blue and UTe$_2$ B40 is shown in red.

\rev{Similarly to $\alpha$ in \S\ref{sec:lambda_pow_law_fits}, $\alpha_R$ shows no apparent overall correlation with $Q_{\text{sample}}$. The two large outliers for $\alpha_R$ correspond to the largest values of $Q_{\text{sample}}$. Furthermore, no apparent systematic dependence (either overall or of the outlier) of $\alpha_R$ on $Q_r$ was observed.}

	%---------------------------------

The interpretation of surface resistance data of superconductors is much more complicated than that of penetration depth data \cite{Halb74,Hirsch93}.  Nevertheless, the relatively robust and consistent behavior seen in the low temperature $R_s(T)$ suggests that a simple mechanism may be responsible for this behavior.

\subsection{Residual Microwave Loss}
\label{sec:SM_res_loss}
$R_0$ and $\sigma_{1,0}$ are measures of the residual loss at zero-temperature. While $\sigma_{1,0}$ is theoretically independent of mode frequency, it is highly dependent on the value of $X_0$. $R_0$ requires much less data processing to calculate, but it is dependent on the mode frequency; however, at a finite base temperature $T_0$, $R_s(T_0)/\omega^2$ is expected to be frequency independent for a BCS superconductor in the absence of a residual loss mechanism, so we expect $R_0/\omega^2$ will be less frequency dependent than $R_0$. We observe that $\sigma_{1,0}$ and $R_0/\omega^2$ do not demonstrate a systematic frequency dependence for typical (low loss) modes as shown in the insets of Figs.~\ref{fig:sigma_1_0_vs_w_sup_mat},\ref{fig:Rs_0_vs_w_sup_mat}.

We note that there are several possible origins of residual loss in nodal superconductors.  For p$_x \pm i$p$_y$ superconductors it has been shown that bound states will be present at and near exposed surfaces, at all energies inside the energy gap when averaged over all directions on the Fermi surface \cite{Mats99}. Andreev bound states contributing to residual loss are also predicted on rough surfaces of chiral p-wave superconductors \cite{Bakur18}.  

% R_0/\omega^2 = 10*10^(-3)/(2 \pi 11*10^9)^2 = 2.1 \Omega (ps)^2

The temperature dependence of the surface resistance $R_s$ and real part of the complex conductivity $\sigma_1$ depend on a number of properties of the superconductor.  First is the question of what types of nodes exist in the superconducting order parameter on the Fermi surface.  For the case of line-nodal superconductors, there is a predicted intrinsic value for the residual loss $\sigma_{00}$, as calculated by Hirshfeld, Putikka, and Scalapino \cite{Hirsch93}.  These authors also show that the low temperature behavior of the losses do not simply reflect the number of quasiparticles excited, but depend also on the quasiparticle scattering rate, the scattering phase shift, and the magnetic penetration depth.  The case of point nodes in the superconducting order parameter is less well studied, but it is known that a universal value of residual $\sigma$ cannot be established.

Assuming line nodes on a cylindrical Fermi surface results in a residual loss of $R_0^{line-nodal} = \frac{1}{2} \omega^2 \mu_0^2 \lambda_L^3 \sigma_{00}$, where the intrinsic universal residual conductivity is $\sigma_{00} = \frac{ne^2\hbar}{m\Delta_{\text{max.}}} = \frac{\hbar}{\mu_0\lambda_L^2\Delta_{\text{max.}}}$ \cite{Hirsch93}.  Assuming $\Delta_{\text{max.}}=0.29$ meV and $\lambda_L = 1.5$ $\mu$m, we find that $\sigma_{00}=0.80\ \mu\Omega^{-1}\ m^{-1}$, and $R_0^{line-nodal}\approx 8.4$ m$\Omega$ at 10 GHz (corresponding to $R_0^{line-nodal}/\omega^2 = 2.1\ \Omega\  ps^2$), a considerable amount of residual loss.  We use this value as a known standard for comparison with the $R_0/\omega^2$ values for UTe$_2$.

\subsection{Summary and comparison of $\lambda(T)$ fit parameters}
\label{sec:SM_param_summary}
\begin{figure}	%---------------------------------
\begin{subfigure}[b]{0.4\textwidth}
\caption{}
\hspace*{-1cm}\includegraphics[scale=0.23,width=1.0\textwidth]{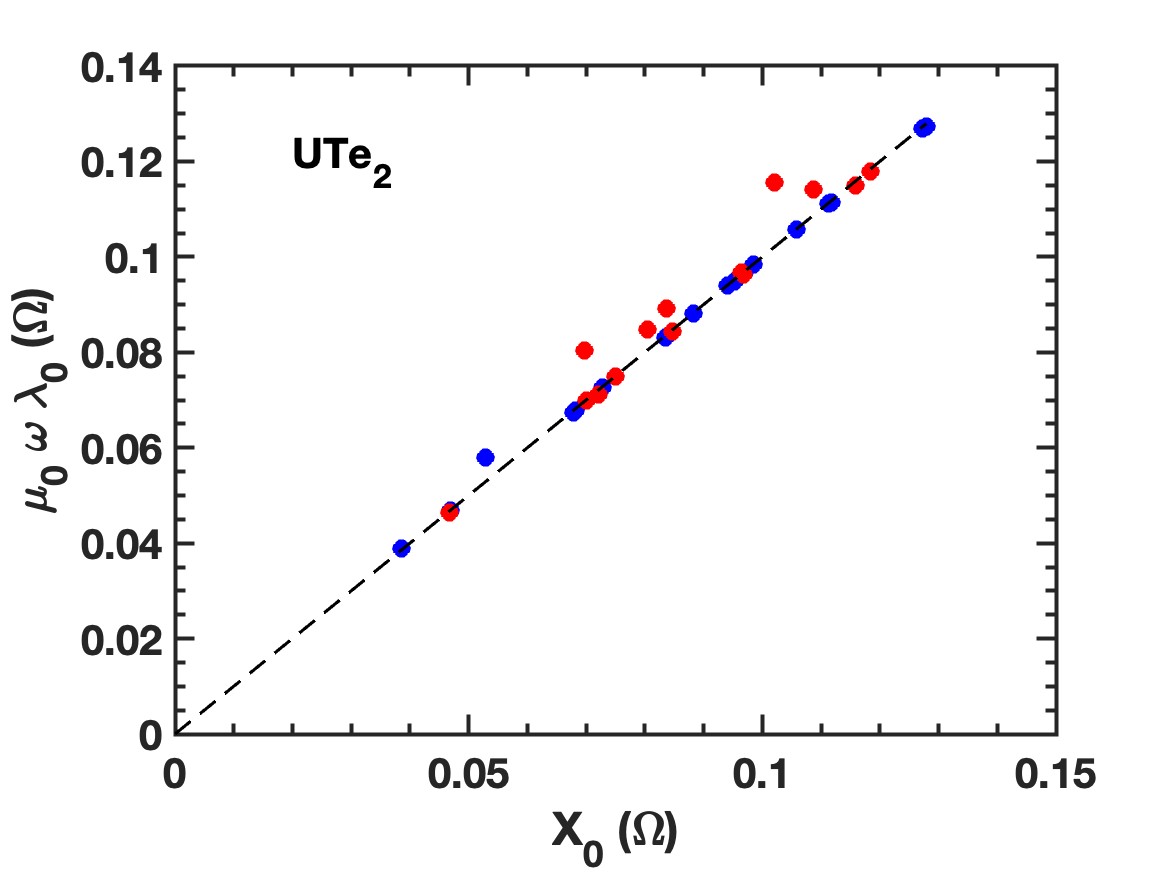}
\label{fig:lambda_0_vs_X0_sup_mat}
\end{subfigure}
\begin{subfigure}[b]{0.5\textwidth}
\caption{}
\hspace*{0cm}\includegraphics[scale=0.23,width=1.0\textwidth]{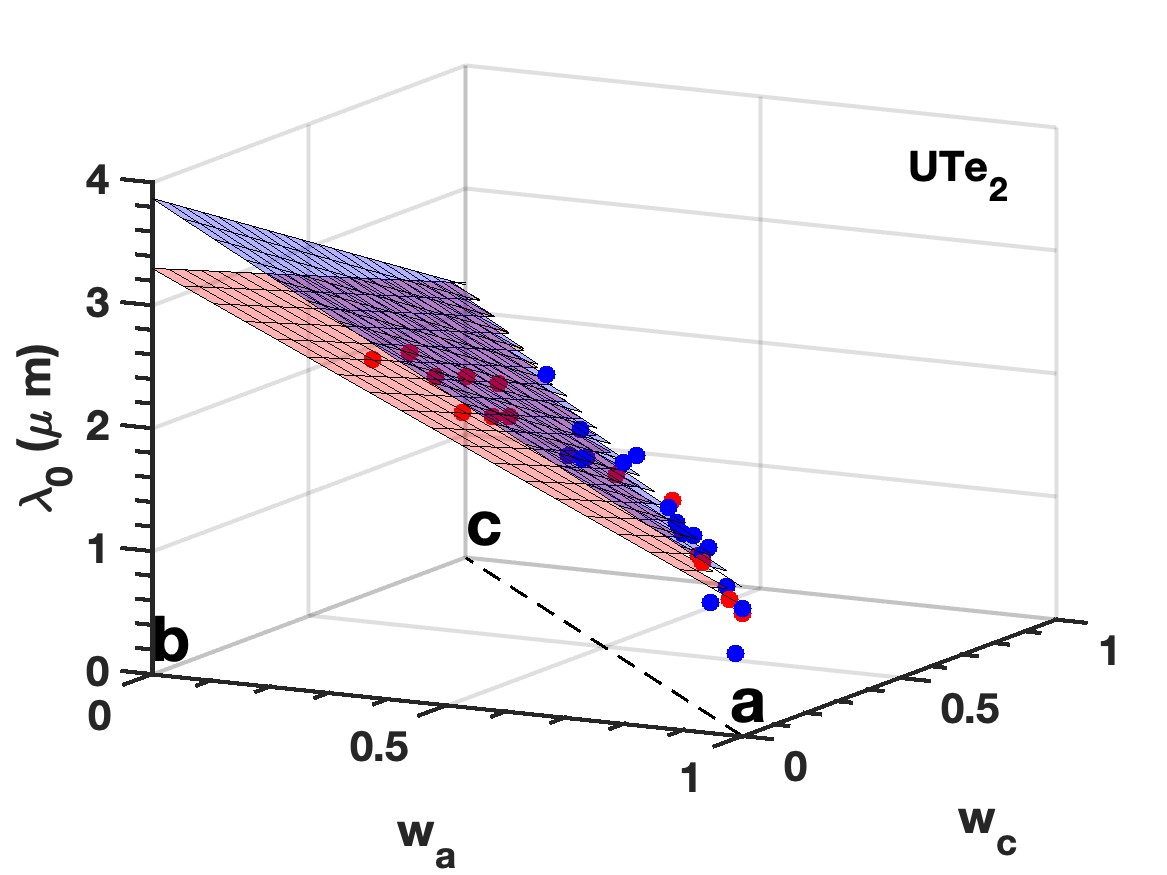}
\label{fig:lambda_0_vs_w_sup_mat}
\end{subfigure}
\caption{Plots of composite $\lambda_0$ for two UTe$_2$ samples vs (a) the minimum temperature reactance and (b) the crystal induced-current weightings. UTe$_2$ B39 is shown in blue and UTe$_2$ B40 is shown in red. For (a), the dashed line indicates $\mu_0\omega\lambda_0=X_0$. For (b), planar fits for each data set are included in the corresponding colors.}
\label{fig:lambda_plane_sup_mat}
\end{figure}

The $\Delta\lambda(T)/G$ data demonstrates an independence on $X_0/G$ for $T<T_c/3$, so the effect of $X_0$ is then only to uniformly shift the value of $\lambda$ at low temperatures. Assuming $\sigma_2 \gg \sigma_1$, this would be expected since $\lambda\approx X_s/(\mu_0\omega)$ at low temperatures; however, 
%as discussed in \S\ref{sec:SM_sigma}, 
we do not find $\sigma_2 \gg \sigma_n$ in general for UTe$_2$. Despite this, we find a very good correspondence between $\mu_0\omega\lambda_0$ and $X_0$ as shown in Fig.~\ref{fig:lambda_0_vs_X0_sup_mat}.

We estimate the axis-resolved minimum temperature penetration depths using a planar fit to the composite penetration depth data, shown in Fig.~\ref{fig:lambda_0_vs_w_sup_mat}. This is justified by the fact that $\mu_0\omega\lambda\approx X_s$ for $T\ll T_c$ assuming $\sigma_2\gg \sigma_1$. We observe that this relationship holds for the composite $\lambda_0$ and $X_0$, as shown in Fig.~\ref{fig:lambda_0_vs_X0_sup_mat}. We find $\lambda_{0,a}=1.2\ (1.1)\ \mu m$, $\lambda_{0,b}=3.9\ (3.3)\ \mu m$, and $\lambda_{0,c}=2.2\ (2.2)\ \mu m$ for UTe$_2$ B39 (B40).

\begin{figure}	%---------------------------------
\begin{subfigure}[b]{0.4\textwidth}
\caption{}
\vspace*{-0.75cm}
\hspace*{-2cm}
\includegraphics[scale=0.23,width=1.0\textwidth]{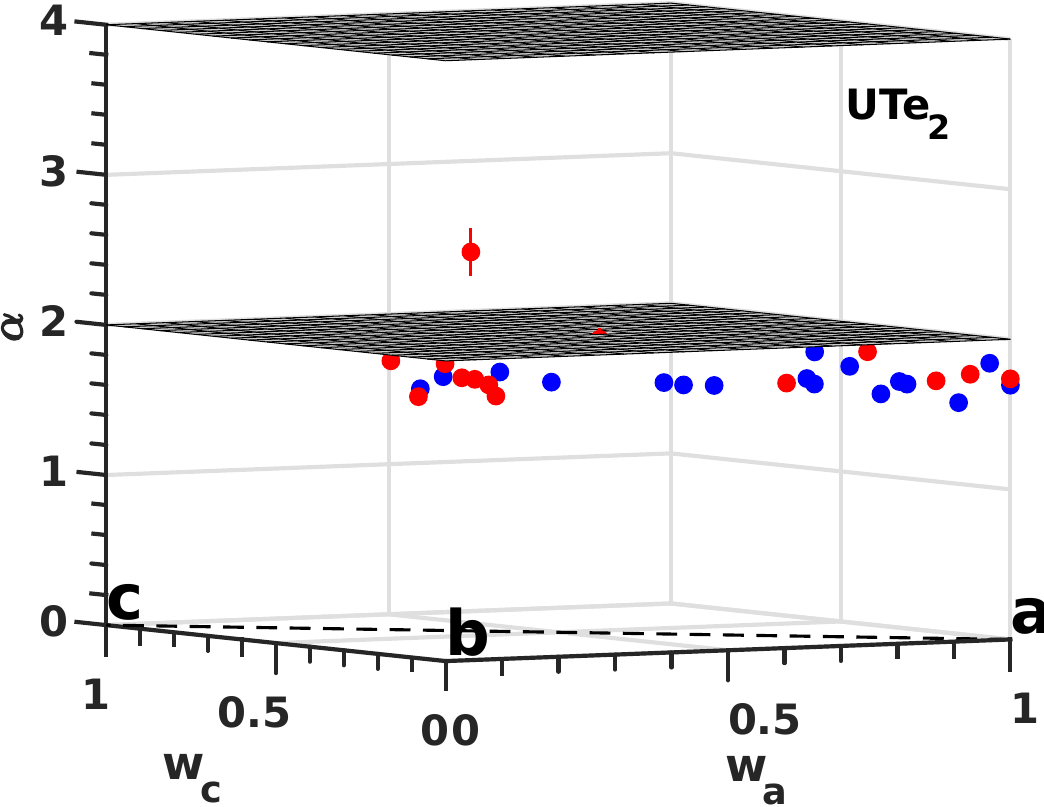}
\label{fig:alpha_vs_w_sup_mat}
\end{subfigure}
\begin{subfigure}[b]{0.4\textwidth}
\vspace*{0.6cm}
\caption{}
\vspace*{-1.2cm}
\hspace*{-2cm}
\includegraphics[scale=0.23,width=1.0\textwidth]{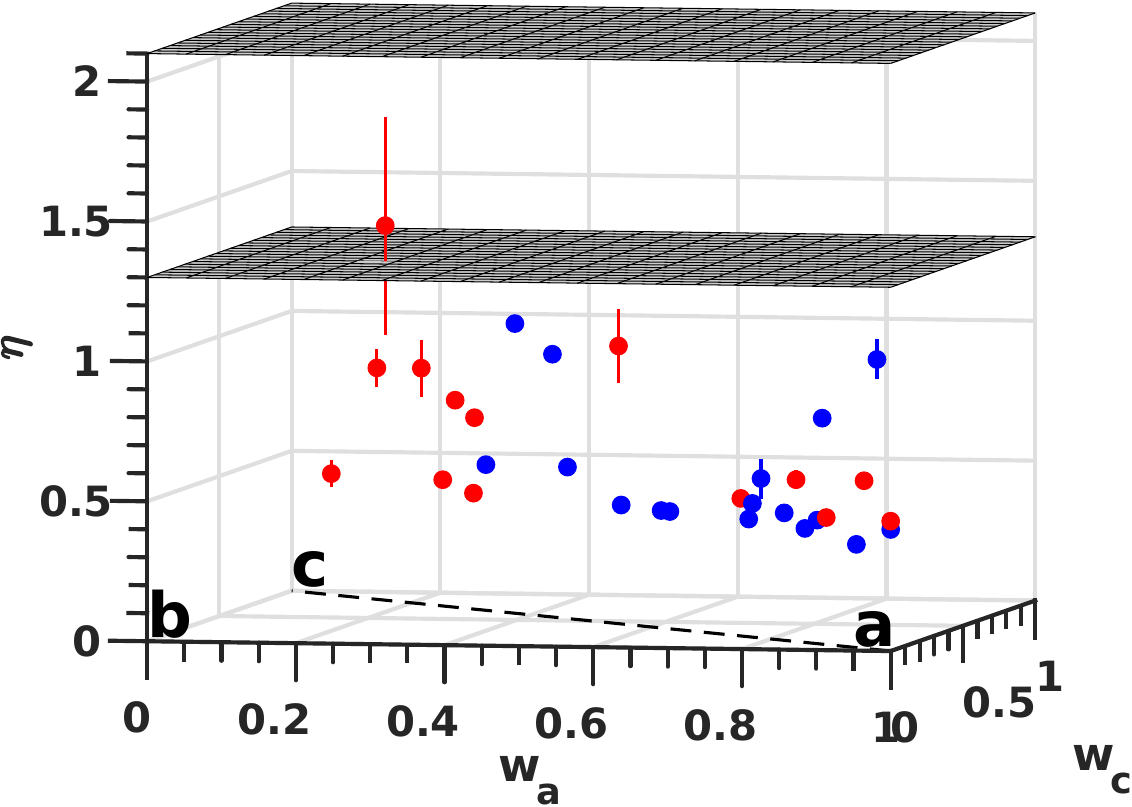}
\label{fig:eta_vs_w_sup_mat}
\end{subfigure}
\begin{subfigure}[b]{0.4\textwidth}
\vspace*{-0.3cm}
\caption{}
\vspace*{-0.4cm}
\hspace*{-2cm}
\includegraphics[scale=0.23,width=1.0\textwidth]{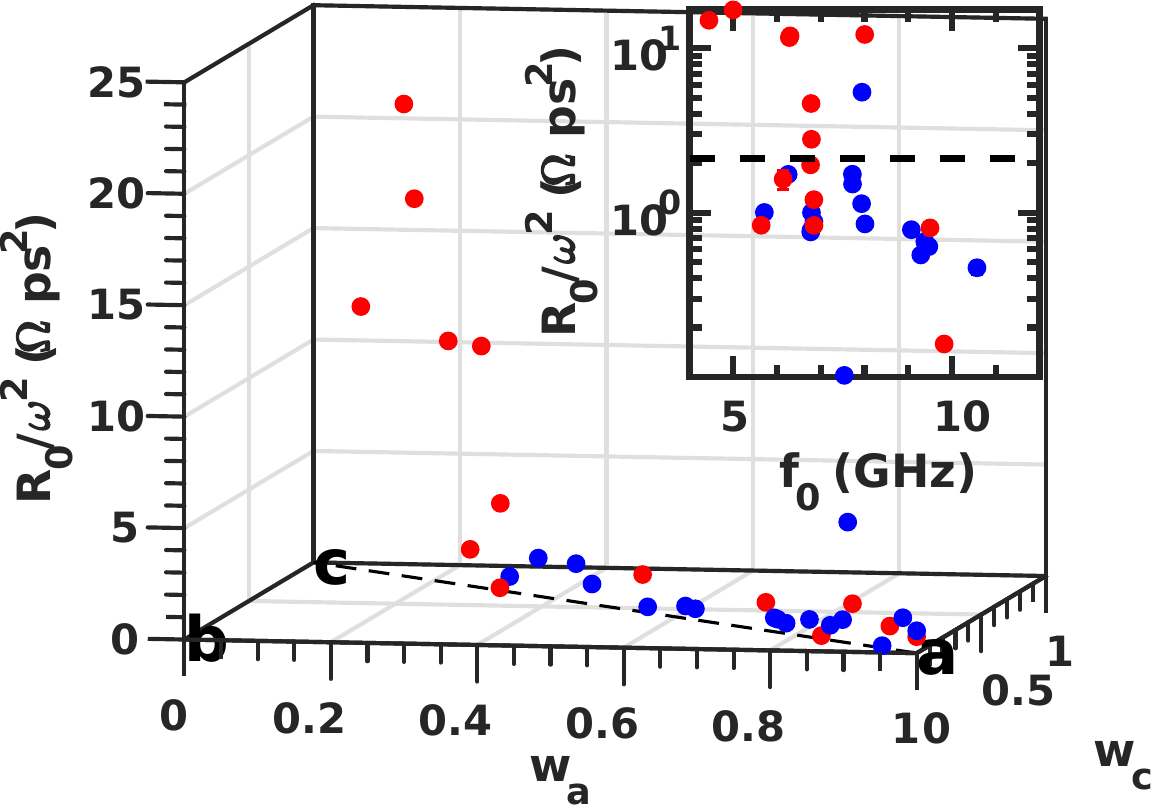}
\label{fig:Rs_0_vs_w_sup_mat}
\end{subfigure}
\caption{Plots of power-law fit parameters and residual loss for two UTe$_2$ samples vs contribution of current flow in each crystallographic axis. UTe$_2$ B39 is shown in blue and UTe$_2$ B40 is shown in red. (a) shows the penetration depth power-law exponent $\alpha$ and the p-wave predictions of $\alpha=2$ parallel and $\alpha=4$ perpendicular to the point nodes as planes. (b) shows the penetration depth power-law coefficient $\eta$ and the p-wave predictions of $\eta=1.3$ parallel and $\eta=2.1$ perpendicular to the point nodes as planes. (c) shows the minimum temperature surface resistance divided by the square of the angular frequency, and its inset shows the same quantity vs frequency with our estimate of the universal d-wave loss $R_0^{line-nodal}/\omega^2 = \frac{1}{2}\mu_0^2\lambda_L^3\sigma_{00}$ shown as the black dashed line. Residual loss divided by frequency squared is expected to be roughly frequency independent.}
\label{fig:w_dep_sup_mat}
\end{figure}

Figure \ref{fig:w_dep_sup_mat} repeats much of the same information that is shown in Fig.~\ref{fig:w_dep} of the main text, but in a different format.  In addition to the $\alpha=2, \eta=1.3$ prediction for currents parallel to point nodes, Figs.~\ref{fig:alpha_vs_w_sup_mat}, \ref{fig:eta_vs_w_sup_mat} show the perpendicular-to-nodes prediction of $\alpha=4, \eta=2.1$ \cite{Klemm88}.  The residual loss values $R_0/\omega^2$ are shown in three-dimensional format in Fig.~\ref{fig:Rs_0_vs_w_sup_mat}, and the frequency dependence of the $R_0/\omega^2$ values is shown as an inset.  The inset also shows the estimated value of $R_0/\omega^2$ (dashed line) for a line-nodal superconductor, as estimated in Supp. Mat. Section \ref{sec:SM_res_loss}. Note that there is one mode ($w_a=0.25$, $w_b=0.40$, $w_a=0.35$) with $\alpha=2.61\pm0.16$, which is much larger than that of the other modes, even considering its larger uncertainty. While this point is near the peak in $R_0/\omega^2$ shown in Fig.~\ref{fig:Rs_0_vs_w_sup_mat}, the modes nearby in the induced-current weights do not show increased values of $\alpha$ as they do $R_0/\omega^2$. This mode similarly has a larger $\eta=1.43\pm0.39$.

	%---------------------------------

	%---------------------------------

\section{Testing the Accuracy of Data Analysis Methods with Synthetic Data}
\label{sec:App_synth_data}
Here we discuss a method with which we generate synthetic data to test the validity of our data analysis and the assumptions that we make about composite vs.\ axis-resolved electrodynamic properties.
\subsection{Testing the Data Analysis Method with Synthesized Data}
\label{sec:Synth_data}
Our objective is to generate synthetic frequency shift $\Delta f(T)$ and $Q(T)$ data that has the full physics of the electrodynamic response of an anisotropic superconducting (and normal metal) crystal, and to use it to test the accuracy of our data analysis techniques.  More specifically, the goal of this exercise is to put the correct physics of anisotropic normal metal and superconducting response into synthetic data, and see if the data analysis methods correctly extract the anisotropic properties (i.e. penetration depth, conductivity, etc.) that were put into the calculation.

\begin{figure}  %---------------------------------
% \centering
\hspace*{0cm}\includegraphics[scale=0.36]{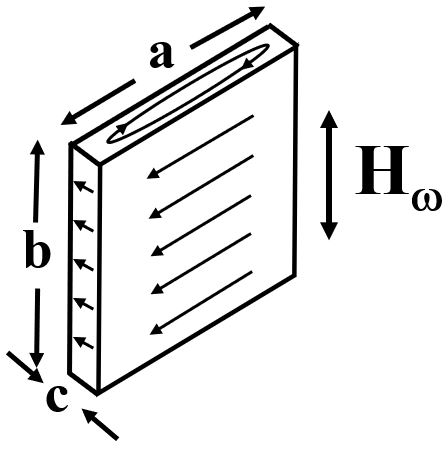}
\caption{Picture of a rectangular crystal sample with anisotropic electrodynamic properties with respect to an rf magnetic field direction, denoted as $H_{\omega}$.  The arrows indicate the screening current directions on the surfaces, and the letters denote the crystal dimensions in each of the three crystallographic directions in the orthorhombic structure. 
 Following Fig. 1 of Ref. \cite{Trunin01}.}
\label{fig:orient}
\end{figure}

We rely on analytical models of the electrodynamics of anisotropic metals.  The first such model that we used was developed by Trunin and collaborators \cite{Trunin01}, based on earlier work by Gough and Exon \cite{Gough94}.  Their complex frequency shift of the cavity due to the presence of the sample at some temperature $T$ is given by,\cite{Nef03}
\begin{equation} 
 \Tilde{\Delta}\left( \frac{1}{Q(T)} \right) -2i\frac{\Tilde{\delta} f(T)}{f(T)} = \frac{i\mu_0 \mu(T) v H^2_0}{2W}   
\label{eq:TruninCmplxFreqShift}
\end{equation}
where $\mu(T)$ is Trunin's expression for the complex relative permeability of the anisotropic sample \cite{Trunin01}, $v$ is the volume of the crystal, and $W$ is the stored energy in the entire cavity on resonance.  Here $H_0$ is the magnetic field at the surface of the crystal, which is assumed to be not enhanced from the value in the absence of the sample.  In other words, this expression ignores de-magnetization effects which create enhancement of fields at edges and corners created by geometrical blocking and re-direction of the fields that exist in the empty cavity.  As is, this expression is most appropriate for a long thin crystal with the long axis parallel to the field.  This is basically the orientation shown in Fig.~\ref{fig:orient} with crystal side-lengths $b \gg a,c$.

	%---------------------------------
The definitions of $\Tilde{\Delta}$ and {$\Tilde{\delta}$ are as follows.  First, $\Tilde{\Delta} f$ is the frequency shift of the cavity with sample minus that of the empty cavity, $\Tilde{\Delta} f(T) =f_{sample}(T) - f_{empty}(T)$, in other words, the background-subtracted frequency shift due to the sample alone.  Likewise $\Tilde{\Delta} (\frac{1}{Q})$ is the difference of inverse-$Q$ between the cavity with sample vs. the empty cavity.  In other words $\Tilde{\Delta} (\frac{1}{Q}) = \frac{1}{Q_{sample}} - \frac{1}{Q_{empty}}$ is the background-subtracted inverse $Q$ of the sample.  Now $\Tilde{\delta} f$ is the frequency shift of the sample in the cavity minus the case of the sample having perfect conductivity (zero penetration of the field).  This latter case is when the sample creates an excluded volume inside the cavity defined by its geometrical boundaries.  Trunin assumes that there is a constant offset, $f_{off}$, between $\Tilde{\Delta} f$ and $\Tilde{\delta} f$, namely $f_{off} = \Tilde{\Delta} f(T) - \Tilde{\delta} f(T)$.  Hence we can write $\Tilde{\delta} f(T) = \Tilde{\Delta} f(T) - f_{off}$.  In turn, we can write the imaginary part of Eq.~\ref{eq:TruninCmplxFreqShift} as $\frac{\Tilde{\delta} f(T)}{f(T)} = \frac{\Tilde{\Delta} f(T)}{f(T)} - \frac{f_{off}}{f(T)}$.  The last term on the right is related to our parameter $X_0$.

The expression for the complex relative permeability $\mu(T)$ presented by the sample depends on what type of anisotropy is assumed.  Trunin considered the case of a bi-anisotropic sample with a-axis and b-axis properties being identical, and the c-axis properties being different,\cite{Trunin01}
\begin{equation}  \label{eq:Mu}
    \mu = \frac{8}{\pi^2} \sum_{\textrm{Odd }n>0} \frac{1}{n^2} \left( \frac{\tan(\alpha_n)}{\alpha_n} + \frac{\tan(\beta_n)}{\beta_n}  \right)
\end{equation}
with
\begin{equation}  \label{eq:Coeffs}
\begin{split}
    \alpha_n^2 = -\frac{a^2}{\delta_c^2} \left( \frac{i}{2} + \frac{\pi^2}{4} \frac{\delta_{ab}^2}{c^2} n^2  \right) \\
    \beta_n^2 = -\frac{c^2}{\delta_{ab}^2} \left( \frac{i}{2} + \frac{\pi^2}{4} \frac{\delta_{c}^2}{a^2} n^2  \right)
    \end{split}
\end{equation}
where $n$ is a positive integer, $a$ is the length of the crystal in its a-direction, and $c$ is the length of the crystal in its c-direction, and $\delta_{ab}$ is the skin depth for currents flowing in the ab-plane, while $\delta_{c}$ is the skin depth for currents flowing in the c-direction.  Note that not all authors \cite{Gough94} agree on these definitions for $a$, $c$, $\delta_{ab}$, and  $\delta_{c}$!

An important feature of Eq.~\ref{eq:Mu} is the fact that it explicitly takes into account the finite size of the crystal, through the parameters $(a, c)$.  In other words, as the temperature approaches T$_c$, and the skin depths approach the size of the crystal, this expression correctly captures the resulting finite-size effects.

The temperature dependent quantities here are the \textit{complex} skin depths $\delta_{ab}$, and  $\delta_{c}$, ignoring thermal expansion of the crystal dimensions.  The skin depths are defined as,

\begin{equation} \label{eq:SkinDepths}
\begin{split}
    \delta_{ab}(T) = \sqrt{\frac{2}{\omega \mu_0 \sigma_{ab}(T)}} \\
    \delta_{c}(T) = \sqrt{\frac{2}{\omega \mu_0 \sigma_{c}(T)}}
    \end{split}
\end{equation}
where $\sigma_{ab}$ and $\sigma_{c}$ are the \textit{complex} conductivities of the crystal for currents flowing in the $ab$-plane and $c$-direction, respectively.

Now one must create a model for the complex conductivities $\sigma_{ab}(T)$ and $\sigma_{c}(T)$ of the crystal, and their temperature dependence.  
%We have been using various versions of the two-fluid model for this task.  There is a Mathematica notebook named ``Trunin Complex Frequency Shift of Anisotropic Crystal" that creates the phony frequency shift and $Q$ data.  
The model assumes temperature-dependent momentum-relaxation lifetimes (in the normal state) and quasiparticle lifetimes (in the superconducting state) that are continuous at T$_c$ and given by,
\begin{equation}  \label{eq:Taus}
\begin{split}
  \tau_{ab}(T) =
    \begin{cases}
      \frac{\mu_0 \lambda_{Lab}^2}{\rho_{ab}(T)} & \text{if $T>T_c$}\\
      \frac{\mu_0 \lambda_{Lab}^2}{\rho_{ab}(T_c)} \frac{T_c+T_{sat}}{T+T_{sat}} & \text{if $T<T_c$}\\
    \end{cases}  \\     
  \tau_{c}(T) =
    \begin{cases}
      \frac{\mu_0 \lambda_{Lc}^2}{\rho_{c}(T)} & \text{if $T>T_c$}\\
      \frac{\mu_0 \lambda_{Lc}^2}{\rho_{c}(T_c)} \frac{T_c+T_{sat}}{T+T_{sat}} & \text{if $T<T_c$}\\
    \end{cases}     
    \end{split}
\end{equation}
where the temperature-dependent normal-state resistivities $\rho_{ab}(T)$ and $\rho_c(T)$ are incorporated into the definitions of the temperature-dependent momentum relaxation times above T$_c$.  Here $\lambda_{Lab}$ and $\lambda_{Lc}$ are the axis-dependent London penetration depths of the crystal.  The temperature $T_{sat}>0$ represents the (negative) temperature at which the quasiparticle lifetime diverges.  The two-fluid model complex conductivity for each axis is defined as,
\begin{equation}
\begin{split} \label{eq:Conducs}
  \sigma_{ab}(T) = \frac{1}{\mu_0 \lambda_{Lab}^2} \left( \frac{f_n(T) \tau_{ab}(T)}{1+i\omega \tau_{ab}(T)} + \frac{f_s(T)}{i\omega} \right)\\
  \sigma_c(T) = \frac{1}{\mu_0 \lambda_{Lc}^2} \left( \frac{f_n(T) \tau_{c}(T)}{1+i\omega \tau_{c}(T)} + \frac{f_s(T)}{i\omega} \right)
  \end{split}
\end{equation}
where $f_s(T)$ and $f_n(T)$ are the temperature-dependent dimensionless superfluid and normal fluid fractions, $0 \le f_s, f_n \le 1$.

One can then calculate the complex surface impedances associated with currents flowing either in the $ab$-plane, or the $c$-direction, as follows,
$$Z_{s,ab}(T) = \sqrt{\frac{i \mu_0 \omega}{\sigma_{ab}(T)}}, $$
$$Z_{s,c}(T) = \sqrt{\frac{i \mu_0 \omega}{\sigma_{c}(T)}}, $$
The two-fluid conductivities, Eq. \ref{eq:Conducs}, can also be used to calculate the complex skin depths in Eq. \ref{eq:SkinDepths}.  These are used to calculate the coefficients in Eq. \ref{eq:Coeffs}, as well as the complex relative permeability of the crystal in Eq. \ref{eq:Mu}.  Finally one can calculate the complex frequency shift of the cavity due to the crystal, using Eq. \ref{eq:TruninCmplxFreqShift}.  For our numerical calculations we take $v=(1\ mm)^3$, $W=10^{-12}\ W$ and $H_0 = 1\ A/m$.

%The treatments of Gough and Exon \cite{Gough94} and Trunin \cite{Trunin01} assume that the crystal presents two different in-plane conductivities to the rf magnetic field.   In our case, we believe that the full three-axis anisotropy of real $UTe_2$ crystals is involved in the screening. We are not aware of any treatment of 3-axis anisotropic screening.  

Note that our data analysis approach operates under the assumption that ``impedances add" (i.e. as defined by Eq. \ref{eq:Zs_comp}) when analysing the screening of anisotropic materials.  A justification of this statement in the bi-anisotropic case is given in Eq. (55) of Ref.\ \cite{MaoThesis95}, which includes the weighting depending on the exposed area of each facet of the crystal.  This is basically the same as the assumption made by Kitano, \textit{et al}. \cite{Kitano95} in their Eq. (2), also showing the relative weighting.  One goal of this exercise is to test whether or not this ``impedances add" data analysis approach is valid in the parameter regime of our experiments on UTe$_2$ crystals.

\subsubsection{Synthetic Data Analysis Results}
Synthetic data generated by the method outlined in Section \ref{sec:Synth_data} was analyzed by the same data analysis methods applied to the measured UTe$_2$ crystals.  A set of 5 modes with a variety of different scenarios were created, covering a significant range of $w_a$, $w_b$ and $w_c$ values.   The deduced crystal weights $w_a$, $w_b$ and $w_c$ were recovered to within 2\% in all cases tested. The expected composite $\lambda_0$ was recovered to within 4\% in all cases.  The extracted London penetration depths were within 2\% of the assumed values, except the b-axis, which had an error of 10\%.  An isotropic superfluid density temperature dependence of $f_s(T) =1-(T/T_c)^2$ was employed, and the resulting penetration depth temperature dependence at low temperatures $\Delta \lambda /\lambda \sim (T/T_c)^{\alpha}$ yielded an axis-average of $\alpha = 1.97 \pm 0.01$. These results are a rigorous test of the basic assumptions of our data analysis approach and give us confidence that our determinations of axis-resolved weightings and screening length scales should be accurate.

\bibliography{UTe2_EM_bib.bib}

\end{document}